\documentclass[aps, prc, reprint, amsmath, amssymb, superscriptaddress, floatfix, nofootinbib]{revtex4-1}

\usepackage[pdftex]{graphicx}% Include figure files
\DeclareGraphicsExtensions{.jpg,.png,.pdf}

\usepackage[utf8]{inputenc}
\usepackage{hyperref}
\usepackage{amsmath}
\usepackage{amssymb}
\usepackage{amsfonts}
\usepackage{tabularx}
\usepackage{booktabs}
\usepackage{graphicx}
\usepackage{color}
\usepackage{multirow}
\usepackage[perpage,symbol]{footmisc}
\usepackage{lineno}% linenumbers
\usepackage{verbatim}
\usepackage{dcolumn}% Align table columns on decimal point
\usepackage{bm}% bold math
\usepackage[inline]{enumitem}
\graphicspath{{Fig/}}
\definecolor{theblue}{RGB}{0,50,230}
\hypersetup{
  colorlinks=true,
  linkcolor=theblue,
  citecolor=theblue,
  urlcolor=theblue
} 

\newcommand{\pt}{\ensuremath{p}_{\rm T}}
\newcommand{\raa}{\ensuremath{R}_{\rm AA}}
\newcommand{\vtwo}{\ensuremath{v}_{\rm 2}}
\newcommand{\snn}{\sqrt{s_{\rm NN}}}
\newcommand{\s}{\sqrt{s}}
\begin{document}

%\preprint{APS/123-QED}

\title{Production of open-charm mesons in relativistic heavy-ion collisions}% Force line breaks with \\

\author{Shuang~Li}
\email{lish@ctgu.edu.cn}
\affiliation{%
College of Science, China Three Gorges University, Yichang 443002, China\\
}%
\affiliation{%
Key Laboratory of Quark and Lepton Physics (MOE), Central China Normal University, Wuhan 430079, China\\
}%
\author{Chaowen~Wang}%
\affiliation{%
College of Science, China Three Gorges University, Yichang 443002, China\\
}%
\author{Xianbao~Yuan}%
\email{ztsbaby@163.com}
\affiliation{%
College of Science, China Three Gorges University, Yichang 443002, China\\
}%
\author{Shengqin Feng}%
\affiliation{%
College of Science, China Three Gorges University, Yichang 443002, China\\
}%
\affiliation{%
Key Laboratory of Quark and Lepton Physics (MOE), Central China Normal University, Wuhan 430079, China\\
}%
\affiliation{%
School of Physics and Technology, Wuhan University, Wuhan 430072, China\\
}%

\date{\today}%

\begin{abstract}
We present a theoretical framework to study open charm production in relativistic heavy-ion collisions.
Charm quarks are regarded as an effective probes of the deconfined medium formed in these collisions,
the Quark-Gluon Plasma (QGP).
The initial conditions of such collisions are simulated with a transverse profile described via a Glauber-based model,
and a longitudinal behaviour modeled by a data-inspired parameterization.
The space-time evolution of the temperature and the flow velocity field of the medium,
is quantified by means of a 3+1 dimensional relativistic viscous hydrodynamics.
The Brownian motion of charm quarks propagating through the QGP, is described by utilizing the Langevin Transport Equation.
The subsequent hadronization is implemented via a ``dual'' model, including fragmentation
and heavy-light coalescence mechanisms. In particular, in the coalescence also the contribution from higher hadronic states
components is considered.
The parameters of the model are tuned based on comparison to data.
The coupling strength between the charm quarks and the QGP constituents, quantified by the spatial diffusion coefficient $2\pi TD_{s}$,
is obtained by performing a phenomenological fit analysis to the lattice QCD calculations,
resulting in $2\pi TD_{s}=const.$ (\textbf{Model-A}) and $2\pi TD_{s}=1.3 + (T/T_{c})^2$ (\textbf{Model-B}).
We find that the relative azimuthal distribution of the initially back-to-back generated $c\bar{c}$ pairs
presents a broadening behaviour, which is more pronounced for $c\bar{c}$ pairs with small initial $\pt$,
and when the Model-B approach is adopted.
The competition between the initial drag and the subsequent collective effects tends to restrict
the time dependence of charm quark $\raa$.
Concerning the theoretical uncertainty on final D-meson nuclear modification,
the nuclear shadowing and pp baseline components are dominat at high
and low $\pt$ ($\pt\lesssim3~{\rm GeV/{\it c}}$), respectively.
The measured D-meson $\raa(\pt)$ favors Model-A assumption for the diffusion coefficient both at RHIC and LHC,
while their $\vtwo(\pt)$ prefer Model-B at moderate $\pt$.
These results confirm the necessity to consider the temperature- and/or momentum-dependence of $2\pi TD_{s}$ to
describe well the D-meson $\raa$ and $\vtwo$ simultaneously.
\end{abstract}

\maketitle

%\tableofcontents

%%==============================================
\section{INTRODUCTION}\label{sec:Introduction}
Relativistic heavy-ion collisions provide a unique opportunity to create
and investigate the properties of strongly interacting matter
in extreme conditions of temperature and energy density,
where the formation of a deconfined medium, the Quark-Gluon Plasma (QGP), is expected~\cite{Gyulassy05, Shuryak05}.
Experiments with heavy-ion collisions have been carried at the Relativistic Heavy Ion Collider (RHIC) at BNL
and at the Large Hadron Collider (LHC) at CERN~\cite{Muller12, Jurgen17, Shuryak17} in the last two decades.
Among the various probes of the QGP, heavy quarks (HQ), i.e. charm and bottom quarks, are of particular
interest~\cite{HQQGPBraaten91, HQQGPRapp10, HFSummaryGROUP16, HFSummaryAarts17, HFQM17Greco, CUJET3QM17} since,
due to their large mass, they are mainly produced in hard
scattering processes at the early stages of the heavy-ion collisions.
Subsequently, they interact with the QGP constituents and experience the full evolution of the QGP medium.
Thermal production of $c\bar{c}$ pairs in the QGP medium is expected to be negligible
at the temperatures reached in heavy-ions at the RHIC and at the LHC.
Interactions with the QGP constituents do not change the flavour,
which make charm quark ideal probes of the medium properties.

HQ interact with the medium constituents in two main scenarios~\cite{RalfSummary16}:
inelastic interactions via the exchange of color charge, resulting in the gluon radiation;
multiple elastic collisions with small momentum transfer.
Both of them cause energy loss for the HQ, usually referred as radiative and collisional energy loss, respectively.
Therefore, HQ allow one to probe the mechanisms of multiple interactions with the medium,
together with the strength of the collective expansion of the fireball.
Considering the fact that HQ fragmentation function is quite hard~\cite{PYTHIA64},
its properties can be inherited well enough by the corresponding
open heavy-flavour hadrons (having charm or bottom quarks among these valence quarks)
such as D mesons ($D^{0}$, $D^{+}$, $D^{*+}$ and $D^{+}_{s}$~\cite{ShuangHP13,ShuangQM14})
and B mesons ($B^{0}$, $B^{+}$ and $B_{\rm s}$~\cite{BMesonCMS17}).

Experimentally, the mentioned energy loss effects are studied by measuring
the open heavy-flavour hadron nuclear modification factor $\raa$,
which is defined as the ratio of the binary-scaled particle production cross section in nucleus-nucleus collisions
to that in nucleon-nucleon collisions at the same energy,
as well as the collective effects are investigated by measuring the 
elliptic flow coefficient $\vtwo$, which is the second order coefficient
of the Fourier expansion of particle azimuthal distributions.
A strong suppression ($\raa<1$) of high $p_{\rm T}$ D-meson was observed at mid-rapidity
in central nucleus-nucleus collisions at the BNL-RHIC and CERN-LHC~\cite{HFSummaryDong17,HFSQM17},
indicating that charm quark energy loss effect is significant.
Meanwhile, a positive $\vtwo$ was measured in semi-central collisions and intermediate $\pt$,
suggesting that charm quark participages in the collective expansion of the medium.
Theoretically, models were developed~\cite{Akamatsu09,HFModelHee08,CaoPRC13,MCatHQsPRC16,PHSDPRC16,POWLANGJHEP18}
to describe the available measurements.
It was realized~\cite{JFLPRL09,Das15,JFLCPL15,CUJET3JHEP16} that the simultaneous description of $\raa$ and $\vtwo$
of open charmed meson at low and intermediate $\pt$ is sensitive to the temperature-dependence of the interaction strength,
which can be quantified by the spatial diffusion coefficient $2\pi TD_{s}$.
%Moreover, concerning the strange charmed meson ($D_{s}^{+}$) productions at intermediate transverse momentum~\cite{HFSummaryQM17},
%it is still an open question about the coexistence/competition between the strangeness enhancement effect and the heavy-light coalescence effect.
Also, as pointed out in Ref.~\cite{HFSummaryAarts17}, one should explore the propagation of theoretical uncertainties
in $\raa$ calculations, including these due to the pp baseline calculation and the (anti-)shadowing parameterization.

In this work, we try to address these questions by taking into account different models for the temperature-dependence
of $2\pi TD_{s}$, which are phenomenologically extracted from lattice QCD calculations,
and then investigate its effect on the observables ($\raa$ and $\vtwo$ ) for both charm quarks and open charmed mesons.
Additionally, based on an instantaneous approach, the typical heavy-light coalescence model for charm quark
is adopted with the additional feature of including the contribution from higher states of
the harmonic oscillator wave function (see Sec.~\ref{subsubsec:CoalMod} for details).

The paper is organized as follows:
Sec.~\ref{sec:LangevinTransport} is dedicated to the description of the HQ transport model
which is implemented via Langevin approach, as well as to the discussion of the temperature dependence of $2\pi TD_{s}$.
In order to simulate as completely as possible the evolution of HQ in heavy-ion collisions,
Sec.~\ref{sec:hybridModel} presents the additional components used to build our hybrid model,
including the initial conditions, the hydrodynamics expansion of the underlying thermal medium and
the hadronization mechanisms for both the medium constituent and the charm quarks.
Sec.~\ref{sec:numercialResults} shows the results about $\raa$ and $\vtwo$ at parton and hadron level.
A summary section can be finally found.

%%==============================================
\section{Langevin transport approach for heavy quark}\label{sec:LangevinTransport}
In this section, we summarize the kinetic transport theory, including the Langevin approach that we use,
to describe the heavy quark space-time evolution in a thermal medium.
We also illustrate the development to include the recoil force induced by the radiated gluon.
Moreover, we introduce a phenomenological model for the temperature-dependence of the drag and diffusion coefficients.
The different parton in medium energy loss mechanisms are discussed as well.

%%-----------------------------
%%-----------------------------
\subsection{Heavy quark diffusion as Brownian motion with the Langevin approach}\label{subsec:LangevinTransport_Diffusion}
While traversing the QGP,
HQ experience multiple elastic scatterings with its constituents and
propagate with a Brownian motion, which can be quantified by a Boltzmann Transport Equation~\cite{Benjamin88}.
For large quark masses and moderate medium temperatures, the typical momentum transfers
in interactions between HQ and the medium are small and the Boltzmann Transport Equation
reduces to the Fokker-Plank Transport Equation~\cite{RalfSummary08}.
In the framework of Fokker-Plank Transport,
the interactions between HQ and the medium constituents are conveniently encoded in the drag and diffusion coefficients,
which are related to each other via the relevant dissipation-fluctuation relation (or Einstein relation).
Consequently, the phase-space distribution of HQ behaves according to the Boltzmann-J$\ddot{\rm u}$ttner approach~\cite{Juttner}
and reaches the thermodynamic equilibrium in the infinite time limit.

In ultra-relativistic heavy-ion collisions, the Fokker-Plank Transport is equivalent to the Langevin approach,
which consists of a ``deterministic'' drag term $F^{\rm Drag}$  and ``stochastic'' diffusion term $F^{\rm Diff}$,
expressed in terms of HQ momentum and its position as~\cite{RalfSummary16}
\begin{equation}
\begin{aligned}\label{eq:TransprotLVEP}
&dx_{\rm i}=\frac{p_{\rm i}}{E_{\rm i}}dt
& \\
&dp_{\rm i}=(F_{\rm i}^{\rm Drag} + F_{\rm i}^{\rm Diff})dt
\end{aligned}
\end{equation}
where, $dx_{\rm i}$ and $dp_{\rm i}$ are the HQ position and momentum changes in the $i^{th}$ time-step $dt$.
The drag force reads
\begin{equation}\label{eq:DragForce}
F_{\rm i}^{\rm Drag}=-\Gamma(p_{\rm i}) \cdot p_{\rm i},
\end{equation}
where $\Gamma(p_{\rm i})$ is the drag coefficient.
The thermal random force which acts on the HQ is expressed as
\begin{equation}\label{eq:ThermalForce}
F_{\rm i}^{\rm Diff}=\frac{1}{\sqrt{dt}}C_{\rm ij}(p_{\rm i})\rho_{j}.
\end{equation}
In the so-called post-point scheme,
the strength of the thermal noise $C_{\rm ij}(p_{\rm i})$ can be associated
to the momentum diffusion coefficient $\kappa$ via~\cite{CaoThesis} $C_{\rm ik}C^{\rm k}_{\rm j}=\kappa(p)\delta_{\rm ij}$
by assuming a isotropic momentum-dependence of the diffusion coefficient.
As mentioned above, $\Gamma(p)$ and $\kappa(p)$ are bridged via the dissipation-fluctuation relation~\cite{CaoThesis}
\begin{equation}\label{eq:LTEFDR}
\Gamma(p)=\frac{\kappa(p)}{2TE}.
\end{equation}
As shown in Eq.~\ref{eq:ThermalForce}, $C_{\rm ij}$ is weighted by a random variable
$\vec{\rho}=(\rho_{1},\rho_{2},\rho_{3})$ which folloes the Gaussian-normal distribution.

%%-----------------------------
%%-----------------------------
\subsection{Temperature-dependence of the drag and diffusion coefficients}\label{subsec:TemperatureDependenceOfInteraction}
The spatial diffusion coefficient~\cite{Moore04},
defined as $D_{s}=\lim_{p\rightarrow0}\frac{T}{m_{\rm Q} \cdot \Gamma(p)}$ in the non-relativistic limit,
is usually employed to characterize the coupling strength in HQ transport calculations.
The spatial diffusion coefficient is related to the momentum diffusion coefficient $\kappa$ via~\cite{Moore04} $D_{s}=2T^{2}/\kappa$.
In addition, $D_{s}$ is usually scaled by the thermal wavelength $\lambda_{\rm th}=1/(2\pi T)$, namely $D_{s}/\lambda_{\rm th}=2\pi TD_{s}$.
The main features concerning its temperature and momentum dependence have been developed in many models~\cite{LQCDbanerjee12, Tolos13, Berrehrah14, BerrehrahMC}.
The drag and diffusion coefficients can be represented in terms of $2\pi TD_{s}$ as
\begin{equation}\label{eq:Gamma2Ds}
\Gamma=\frac{1}{(2\pi TD_{s})} \cdot \frac{2\pi T^{2}}{E},
\end{equation}
\vspace{-1.5em}
\begin{equation}\label{eq:Kappa2Ds}
\kappa=\frac{1}{(2\pi TD_{s})} \cdot {4\pi T^{3}}.
\end{equation}
Note that, (1) the spatial diffusion coefficient $D_{s}$ is defined in the zero-momentum limit,
while the notation $D_{s}$, as shown in Eq.~\ref{eq:Gamma2Ds} and \ref{eq:Kappa2Ds},
refers to the definition of spatial diffusion coefficient extended to larger momentum values;
(2) in this work, the HQ transport coefficient $\hat{q}_{\rm Q}$ is related to
the momentum space diffusion coefficient $\kappa$ via $\hat{q}_{\rm Q}=2\kappa$~\cite{CaoThesis}.

We discuss below two approaches to model the temperature and momentum dependence of the spatial diffusion coefficient $2\pi TD_{s}(T,p)$.
%in order to extract the physics message in model-to-data comparison.

\textbf{Model-A}: Following the discussion in Ref.~\cite{Moore04, Akamatsu09},
one may neglect both the $T$- and $p$-dependence of $2\pi TD_{s}$
and simplify it as
\begin{equation}\label{eq:ModelA} 
2\pi TD_{s}=const.
\end{equation}
In this case, there is only one parameter $2\pi TD_{s}$,
which characterizes the coupling strength of HQ with the medium.
It can be adjusted according to model-to-data comparisons.
A remarkable feature of this approach is that the drag coefficient behaves as $\Gamma \propto T^{2}$,
which is similar to AdS/CFT and pQCD calculations~\cite{Akamatsu09}.
%\newline\textbf{=== Add the thermalization time? ===}

\textbf{Model-B}: alternatively, one can neglect the momentum dependence of $2\pi TD_{s}$,
as mentioned in Ref.~\cite{HFSummaryAarts17}, and parameterize its temperature dependence as
\begin{equation}\label{eq:ModelB} 
2\pi TD_{s}=a + b \cdot (\frac{T}{T_{c}})^2
\end{equation}
where, $a$ and $b$ are the adjustable parameters, and $T_{c}$ is the critical temperature
for the transition from the deconfined QGP to a hadron gas.
In this approach, the drag coefficient shows a weak dependence on the temperature
which is consistent with the results presented in Ref.~\cite{Das15, GrecoSQM16}.

\begin{figure}[!htbp]
\begin{center}
\vspace{-1.0em}
\setlength{\abovecaptionskip}{-0.1mm}
\setlength{\belowcaptionskip}{-1.5em}
\includegraphics[width=.42\textwidth]{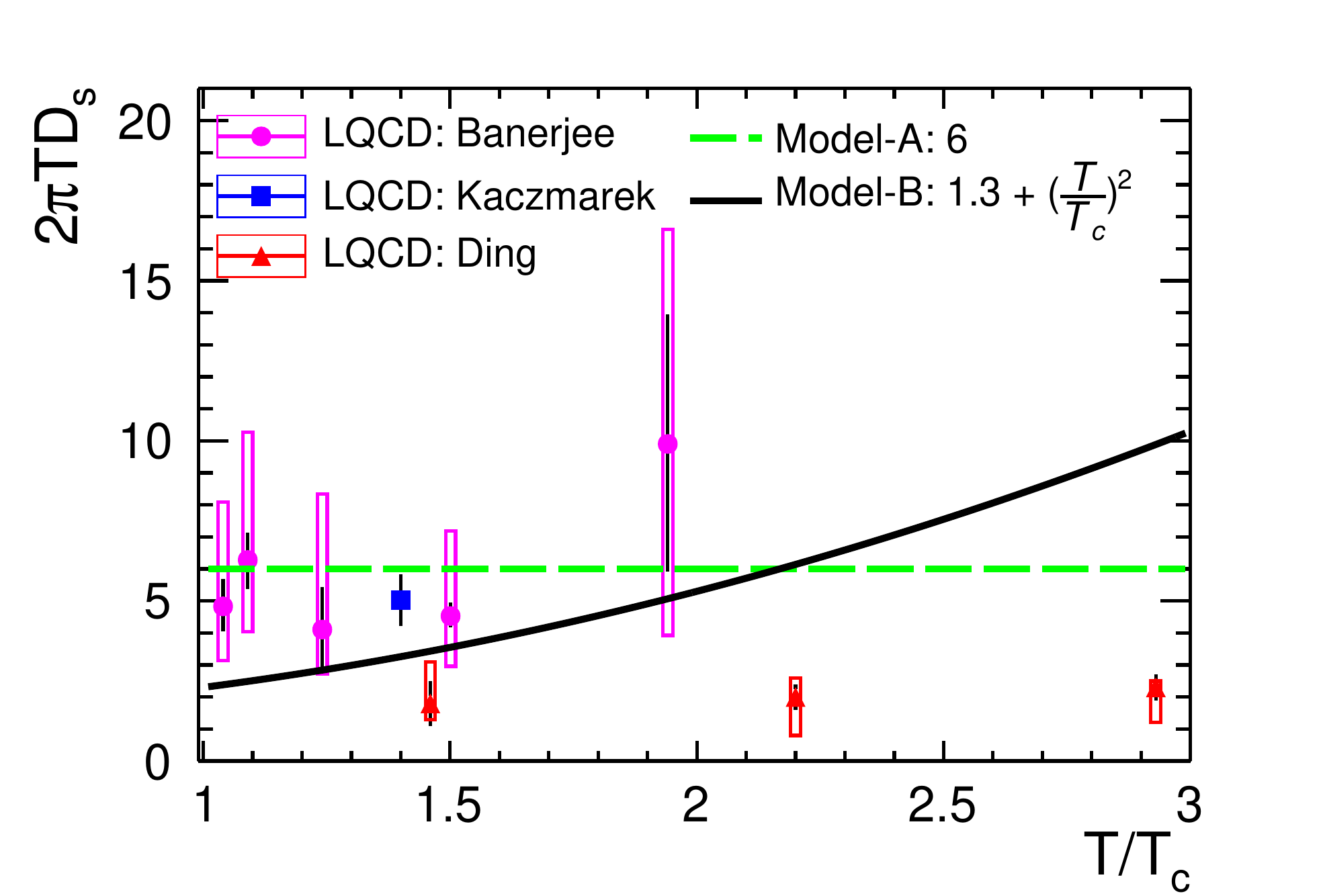}
\caption{(Color online) Spatial diffusion coefficient $2\pi TD_{s}$ of charm quarks ($m_{\rm c}=1.5$ GeV)
from lattice QCD calculations (pink circle~\cite{LQCDbanerjee12}, blue square~\cite{LQCDolaf14} and red triangle~\cite{LQCDding12} symbols)
at zero momentum. The phenomenological approaches (dashed green and solid black curves) are displayed as well.}
\label{fig:DsVsT}
\end{center}
\end{figure}
Figure~\ref{fig:DsVsT} presents the temperature dependence of the spatial diffusion coefficient $2\pi TD_{s}$
as obtained from lattice QCD calculations, i.e. Banerjee (pink circles~\cite{LQCDbanerjee12}),
Kaczmarek (blue square~\cite{LQCDolaf14}) and Ding (red triangles~\cite{LQCDding12}),
as well as the results from the two approaches,
i.e. Model-A (dashed green curve; Eq.~\ref{eq:ModelA}) and Model-B (solid black curve; Eq.~\ref{eq:ModelB}) described above.
The model parameters were tuned to fit the lattice QCD results and their values are summarized in Tab.~\ref{tab:Coefficients}.
It is interesting to note that the values of ${\kappa}/{T^{3}}$ and ${\hat{q}_{\rm Q}}/{T^{3}}$ obtained at certain values of ${T}/{T_{c}}$
fall within the ranges, reported in Refs.~\cite{LQCDFrancis15, JETCoef14}.
\begin{table}[!htbp]
\centering
\begin{tabular}{|c|c|c|c|c|}
\hline
\multicolumn{2}{|c}{\multirow{2}{*}{\centering }}
 & \multicolumn{1}{|c}{\multirow{2}{*}{\centering Model-A}}
 & \multicolumn{1}{|c}{\multirow{2}{*}{\centering Model-B}}
 & \multicolumn{1}{|c|}{\multirow{4}{*}{\centering Reference}}
 \\
\multicolumn{2}{|c}{\centering }
 & \multicolumn{1}{|c}{\centering}
 & \multicolumn{1}{|c}{\centering}
 & \multicolumn{1}{|c|}{}
 \\
\cline{1-4}
\multicolumn{2}{|c}{\multirow{2}{*}{\centering $2\pi TD_{s}$}}
 & \multicolumn{1}{|c}{\multirow{2}{*}{\centering 6}}
 & \multicolumn{1}{|c}{\multirow{2}{*}{\centering $1.3+(\frac{T}{T_{c}})^2$}}
 & \multicolumn{1}{|c|}{\multirow{2}{*}{}}
  \\
\multicolumn{2}{|c}{}
 & \multicolumn{1}{|c}{}
 & \multicolumn{1}{|c}{}
 & \multicolumn{1}{|c|}{}
  \\
\hline
\multicolumn{2}{|c}{\multirow{2}{*}{\centering $\frac{\kappa}{T^{3}}(\frac{T}{T_{c}}=1.5)$}}
 & \multicolumn{1}{|c}{\multirow{2}{*}{\centering 2.09}}
 & \multicolumn{1}{|c}{\multirow{2}{*}{\centering 3.53}}
 & \multicolumn{1}{|c|}{\multirow{2}{*}{\centering $1.8\sim3.4$~\cite{LQCDFrancis15}}}
  \\
\multicolumn{2}{|c}{}
 & \multicolumn{1}{|c}{}
 & \multicolumn{1}{|c}{}
 & \multicolumn{1}{|c|}{}
  \\
\hline
\multicolumn{2}{|c}{\multirow{2}{*}{\centering $\frac{\hat{q}_{\rm Q}}{T^{3}}(\frac{T}{T_{c}}=1.88)$}}
 & \multicolumn{1}{|c}{\multirow{2}{*}{\centering 4.19}}
 & \multicolumn{1}{|c}{\multirow{2}{*}{\centering 5.20}}
 & \multicolumn{1}{|c|}{\multirow{2}{*}{\centering $3.4\sim5.8$~\cite{JETCoef14}}}
  \\
\multicolumn{2}{|c}{}
 & \multicolumn{1}{|c}{}
 & \multicolumn{1}{|c}{}
 & \multicolumn{1}{|c|}{}
  \\
\hline
\multicolumn{2}{|c}{\multirow{2}{*}{\centering $\frac{\hat{q}_{\rm Q}}{T^{3}}(\frac{T}{T_{c}}=2.61)$}}
 & \multicolumn{1}{|c}{\multirow{2}{*}{\centering 4.19}}
 & \multicolumn{1}{|c}{\multirow{2}{*}{\centering 3.11}}
 & \multicolumn{1}{|c|}{\multirow{2}{*}{\centering $2.3\sim5.1$~\cite{JETCoef14}}}
  \\
\multicolumn{2}{|c}{}
 & \multicolumn{1}{|c}{}
 & \multicolumn{1}{|c}{}
 & \multicolumn{1}{|c|}{}
  \\
\hline
\end{tabular}
\caption{Summary of the two different models for $2\pi TD_{s}$ (see Fig.~\ref{fig:DsVsT}),
as well as values obtained for ${\kappa}/{T^{3}}$ and ${\hat{q}_{\rm Q}}/{T^{3}}$.
The values for other predictions are shown for comparison.}
\label{tab:Coefficients}
\end{table}

%%-----------------------------
%%-----------------------------
\subsection{Heavy quark in-medium energy loss}\label{subsec:HQEloss}
As introduced in the previous sub-sections,
the multiple scattering of heavy quarks (HQ) off the thermal partons inside a hot and dense QCD medium results in a Brownian motion,
which can be described by the Langevin Transport Equation in the small momentum transfer limit.
This accounts to the so-called collisional energy loss of the HQ.
However, after traversing the QGP, heavy quarks can interact with the medium constituents
via inelastic scattering, resulting in gluon radiation~\cite{EnergyLossMovivateJDB82, GluonRadiation94}.
This medium-induced gluon radiation leads to the so-called radiative energy loss.
In this analysis, we follow the strategy proposed in Ref.~\cite{CaoQM12, CaoPRC13}
to incorporate in the Langevin Equation both the collisional and the radiative energy loss of HQ propagating through the QGP medium.
Equation~\ref{eq:TransprotLVEP} is therefore modified as
\begin{equation}\label{eq:TransprotLVEP_Update}
dp_{\rm i}=(F_{\rm i}^{\rm Drag} + F_{\rm i}^{\rm Diff} + F_{\rm i}^{\rm Gluon})dt,
\end{equation}
with
\begin{equation}\label{eq:RecoilForce}
F_{\rm i}^{\rm Gluon}=-\frac{dp_{\rm ij}^{\rm Gluon}}{dt}.
\end{equation}
where, $F_{\rm i}^{\rm Gluon}$ is the recoil force which acts on the HQ,
and $p_{\rm ij}$ indicates the momentum of the radiated gluon.
The transverse momentum and radiation time dependence of the radiated gluon
is quantified by pQCD Higher-Twist calculations~\cite{HTPRL04}.

It should be noticed that the Langevin Equation (Eq.~\ref{eq:TransprotLVEP})
is modified to include the recoil force induced by the emitting gluon (Eq.~\ref{eq:TransprotLVEP_Update} and~\ref{eq:RecoilForce}),
resulting in the possible violation of the fluctuation-dissipation relation (Eq.~\ref{eq:LTEFDR}) by a certain amount:
moreover, in this approach, the collisional and radiative energy loss effects are treated as independent,
while, as pointed in Ref.~\cite{Gossiaux06, Das10},
they are not entirely independent since the transport coefficients for collisional
and radiative processes are correlated, which is not taken into account in this work.
Finally, note that a lower cut-off is imposed on the gluon energy ($\omega\geqslant\pi T$~\cite{CaoThesis})
to balance the gluon radiation and the inverse absorption,
and to constrain the evolution of low-energy heavy quarks to follow the soft multiple
scattering scenario, where the detailed balance is well defined.
We follow Ref.~\cite{CaoThesis} by assuming that
the fluctuation-dissipation relation (Eq.~\ref{eq:LTEFDR}) is still valid between the drag and the diffusion terms
of the modified Langevin Transport Equation (Eq.~\ref{eq:TransprotLVEP_Update}).

%%==============================================
\section{Hybrid Modeling of Heavy Quark Evolution}\label{sec:hybridModel}
In order to simulate open charm hadron production in heavy-ion collisions,
one needs to employ a hybrid model including the initial conditions,
the hydrodynamics expansion of the underlying medium
and the hadronization mechanisms for both the medium and the charm quarks.
%%-----------------------------
%%-----------------------------
\subsection{Initial distribution of heavy quarks}\label{subsec:hybridModel_initial}
\subsubsection{Spatial-space initialization via Glauber model}\label{subsubsec:initialSpatial}
The initial spatial distributions of heavy quark pairs are determined
by simulating the initial entropy density distributions in
heavy-ion collisions\footnote[5]{Exactly, the spatial distributions of heavy quark pairs are sampled
according to a event-averaged smooth initial transverse profile,
which will be discussed in detail in Fig.~\ref{fig:EvtWgt} (Sec.~\ref{subsubsec:HudroSim}).}.
The relevent transverse profile, i.e. perpendicular to the beam direction, is modeled by the MC-Glauber model ($SuperMC$~\cite{iEBE})
which allows one to sample randomly the position of each nucleon
inside the projectile and the target nuclei according to their Woods-Saxon distributions,
while the longitudinal profile, i.e. parallel to the beam direction, is described by a data-inspired phenomenological function.

At the initial time of the collision, $\tau=\sqrt{t^{2}+z^{2}}\equiv0$,
the entropy density, $s(\tau=0, \vec{r}_{\perp}, \eta_{s})$, can be factorized as
\begin{equation}\label{eq:entropyDis3D}
s(\tau=0, \vec{r}_{\perp}, \eta_{s}) \equiv s(\tau=0, \vec{r}_{\perp})\cdot\rho(\eta_{s})
\end{equation}
where, $s(\tau=0, \vec{r}_{\perp})$ is the initial entropy density deposited in the transverse plane~\cite{iEBE}.
The function $\rho(\eta_{s})$ (Eq.~\ref{eq:entropyDis3D}) allows one to quantify the longitudinal profile of
initial entropy density as a function of the spatial pseudorapidity $\eta_{s}=0.5 ln(t+z)/(t-z)$.
Experimentally, charged particle pseudorapidity distributions exhibit a plateau behaviour in the central region ($\eta\sim0$),
followed by a rapid drop-off toward forward/backward regions (i.e. at large $\eta$)~\cite{PHOBOS}.
It was argued~\cite{TH01} that this observation can be reproduced by composing the initial
entropy density into two regions: the initial entropy density is flat near $\eta_{s}\sim0$
and smoothly fall-off as a half part of a Gaussian approach in the forward/backward space-time rapidity regions.
Therefore, we parameterize the longitudinal distribution $\rho(\eta_{s})$ as
\begin{equation}\label{eq:rhos}
\rho(\eta_{s})=H(y_{\rm beam}-|\eta_{s}|) \cdot e^{-\frac{(|\eta_{s}|-\Delta\eta)^{2}}{2\sigma_{\eta}^{2}} \cdot H(|\eta_{s}|-\Delta\eta)}
\end{equation}
where, $y_{\rm beam}$ is the beam rapidity;
$\Delta\eta$ and $\sigma_{\eta}$ describe the plateau and Gaussian fall-off behaviour, respectively; $H$ is the Heaviside step function.
Using the typical parameters such as the initial time scale $\tau_{0}=0.6~{\rm fm/{\it c}}$,
the shear viscosity $\eta/s=1/(4\pi)$ corresponding to the predicted low-limit
and the critical temperature $T_{c}=165 ~{\rm MeV}$ at both RHIC and LHC energies,
we can compare the calculated charge particle multiplicity with the available measurements,
and fix the parameters of Eq.~\ref{eq:rhos}.
For instance, we take $\Delta\eta=0.5$ and $\sigma_{\eta}=0.7$ in central ($0-10\%$) Au--Au collisions at $\snn=200~{\rm GeV}$,
and the resulting mean multiplicity per participant pair is $<2N_{ch}/N_{part}>_{model}=3.745$, which is consistent with the available measurements
$<2N_{ch}/N_{part}>_{data}=3.64\sim3.82$~\cite{ChgRHIC200}.

\subsubsection{Momentum-space initialization via pQCD-based calculation}\label{subsubsec:initialMomentum}
The initial momentum distributions of heavy quarks are determined according to
FONLL (Fixed Order Next-to-Leading Logarithms~\cite{FONLL98, FONLL01, FONLL12}) calculations in the desired rapidity intervals,
considering also the related systematic uncertainties on the calculations.

\begin{figure}[!htbp]
\begin{center}
\vspace{-1.0em}
\setlength{\abovecaptionskip}{-0.1mm}
\setlength{\belowcaptionskip}{-1.5em}
\includegraphics[width=.42\textwidth]{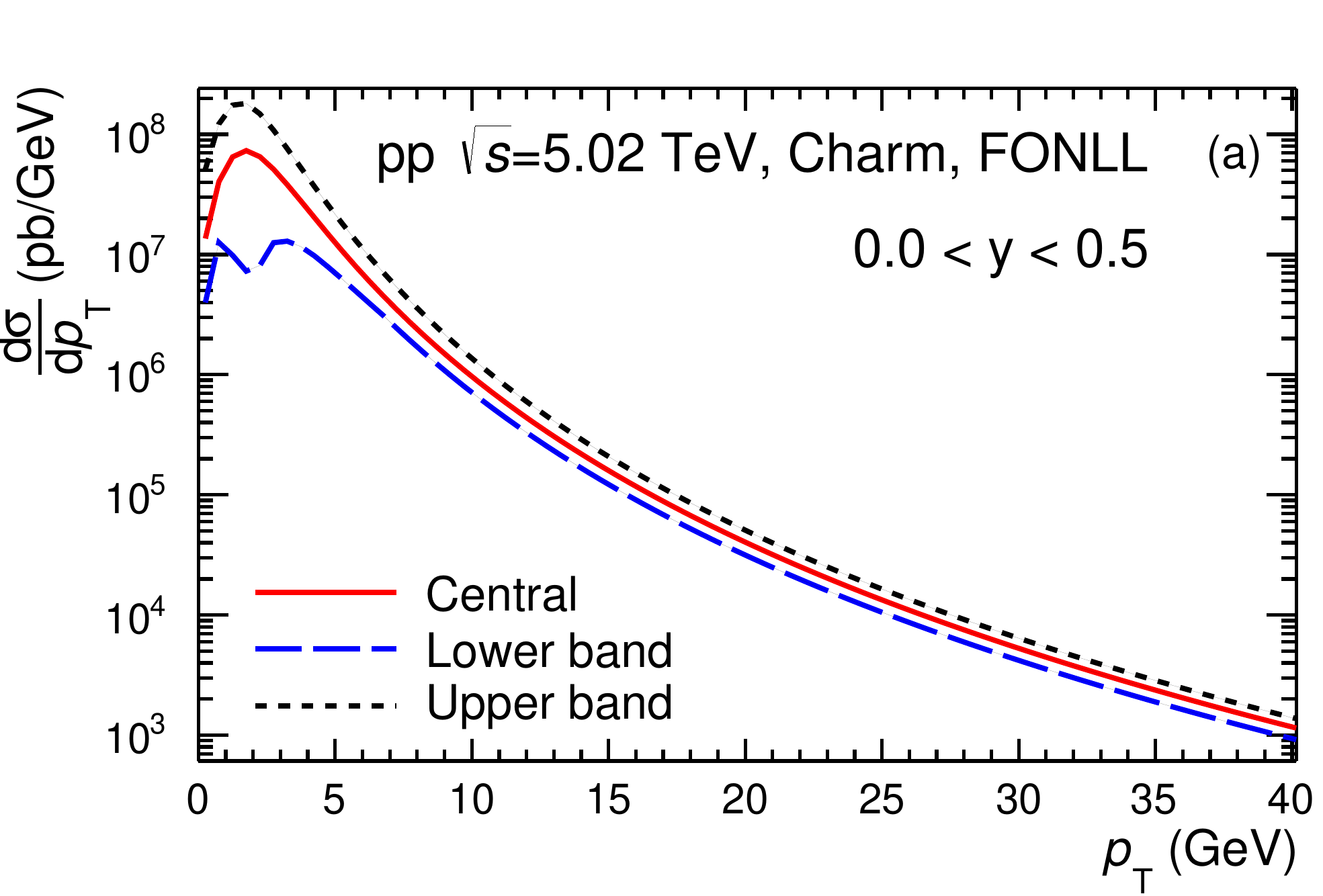}
\includegraphics[width=.42\textwidth]{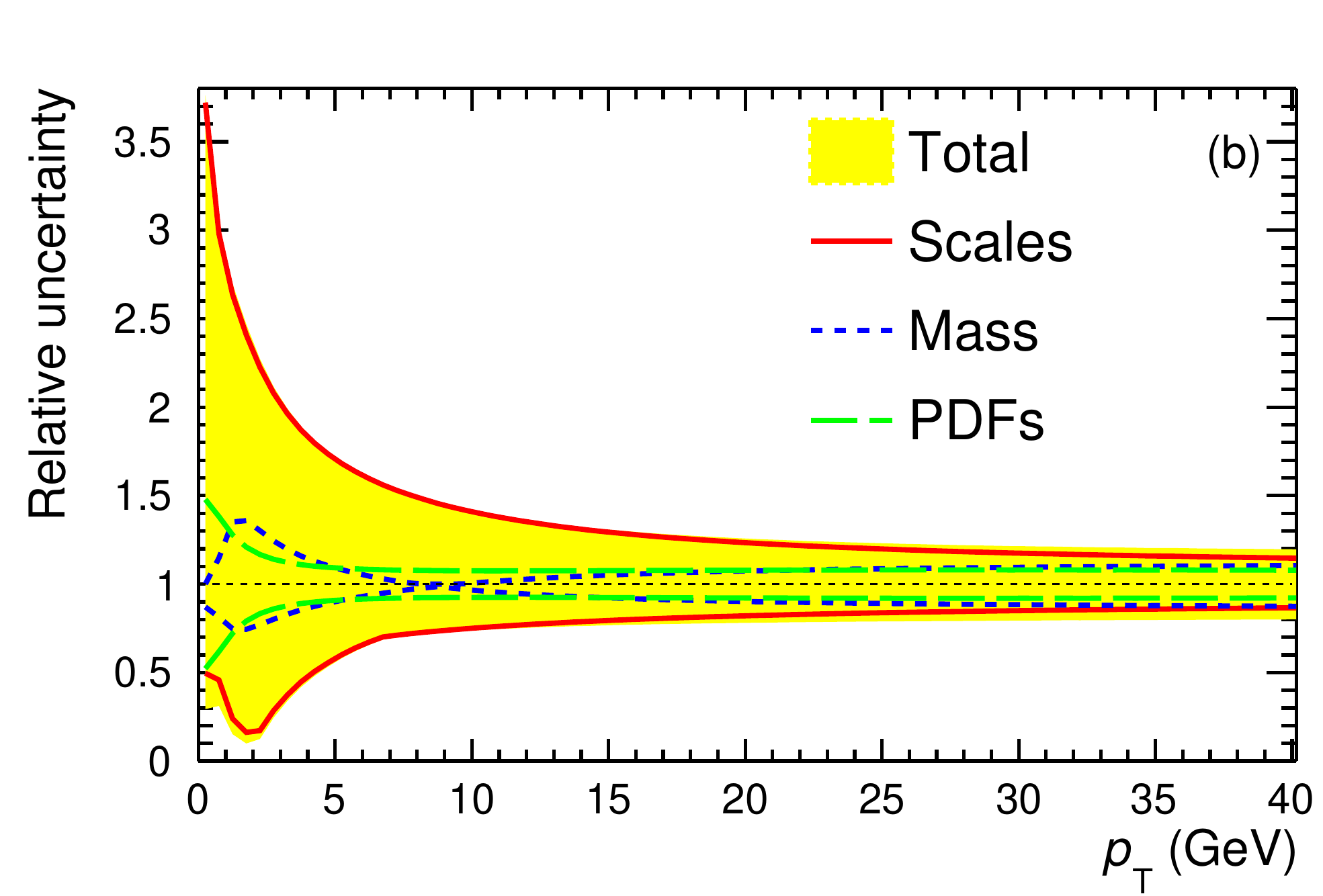}
\caption{(Color online) The (a) production cross section of charm quarks in $0<y<0.5$ in pp collisions at $\s=5.02$ TeV,
as well as the (b) relative uncertainty due to various sources (see legend and text for details).}
\label{fig:FONLLFig}
\end{center}
\end{figure}
The differential production cross section of charm quarks calculated
in the range $0<y<0.5$ for pp collisions at $\s=5.02$ TeV
is shown in the panel-a of Fig.~\ref{fig:FONLLFig}.
The corresponding central values (solid red curve) of FONLL calculations are obtained with~\cite{FONLL98}
\begin{equation}\label{eq:FONLLCentForm}
\mu_{\rm R}=\mu_{\rm F}=\mu_{0}\equiv\sqrt{\pt^{2}+m_{\rm Q}^{2}} %\nonumber
\end{equation}
where $\mu_{\rm R}$ and $\mu_{\rm F}$ are the renormalization and
factorization scales, respectively; $m_{\rm Q}$ is the heavy quark mass,
and its central value is $m_{\rm c}=1.5~{\rm GeV}$ and $m_{\rm b}=4.75~{\rm GeV}$
for charm and bottom, respectively.
The upper (dashed black) and lower (dotted blue) curves represent the systematic uncertainties which are
estimated by varying the renormalization and factorization scales and the quark mass in a conservative approach.
The common variations are~\cite{PDG2016, FONLLErr}
\begin{equation}
\begin{aligned}\label{eq:FONLLErr}
&\frac{1}{2}\mu_{0}<\mu_{\rm R},~\mu_{\rm F}<2\mu_{0}; \qquad \frac{1}{2}\mu_{\rm R}<\mu_{\rm F}<2\mu_{\rm R}; \\
& \\
&1.3 < m_{c} < 1.7~{\rm GeV}; \qquad 4.5 < m_{b} < 5.0~{\rm GeV}.
\end{aligned}
\end{equation}
The ratios to the central values are shown in the panel-b of Fig.~\ref{fig:FONLLFig}.
The uncertainty on parton distribution functions (PDFs) is given by different sets of inputs from CTEQ6~\cite{CTEQ6M}.
One can see that the uncertainty on QCD scales (solid red curve) dominates in the considered $\pt$ region,
while the one on PDFs (long dashed green curve) is negligible for $2<\pt<15~{\rm GeV/{\it c}}$.
Note that the different sources mentioned above are considered in this analysis.

The charm quantum numbers are conserved in strong interactions, therefore,
the charm quark $c$ is always created together with its anti-quark $\bar{c}$.
Then we assume the back-to-back azimuthal correlations,
\begin{equation}\label{eq:HQSpatial}
{r}_{c,i}={r}_{\bar{c},i},
\end{equation}
\vspace{-2.5em}
\begin{equation}\label{eq:HQMomentum}
{p}_{c,i}=-{p}_{\bar{c},i}
\end{equation}
where, $i=x,~y,~z$.
Consequently, the $\pt$- and $y$-dependence of the $c\bar{c}$ pair yields are sampled
according to the FONLL calculations (e.g. panel-a in Fig.~\ref{fig:FONLLFig}) via Monte-Carlo,
and then, they are restricted to satisfy the above conditions.

\subsubsection{Shadowing effect in nucleus--nucleus collisions}\label{subsubsec:initialCNM}
The nuclear modification of the parton distribution functions (nPDFs) should be taken into account
in nucleus--nucleus collisions since the nucleons are bound in a nucleus.
The most relevant effect at RHIC and LHC energy is a depletion at small Bjorken-$x$,
usually called shadowing~\cite{Shadowing}, which reduces the production cross section of charm quarks at low $\pt$.
At large Bjorken-$x$, shadowing is replaced by an enhancement of the PDF,
usually called anti-shadowing in the literature~\cite{Shadowing}.

\begin{figure}[!htbp]
\begin{center}
\vspace{-1.0em}
\setlength{\abovecaptionskip}{-0.1mm}
\setlength{\belowcaptionskip}{-1.5em}
\includegraphics[width=.42\textwidth]{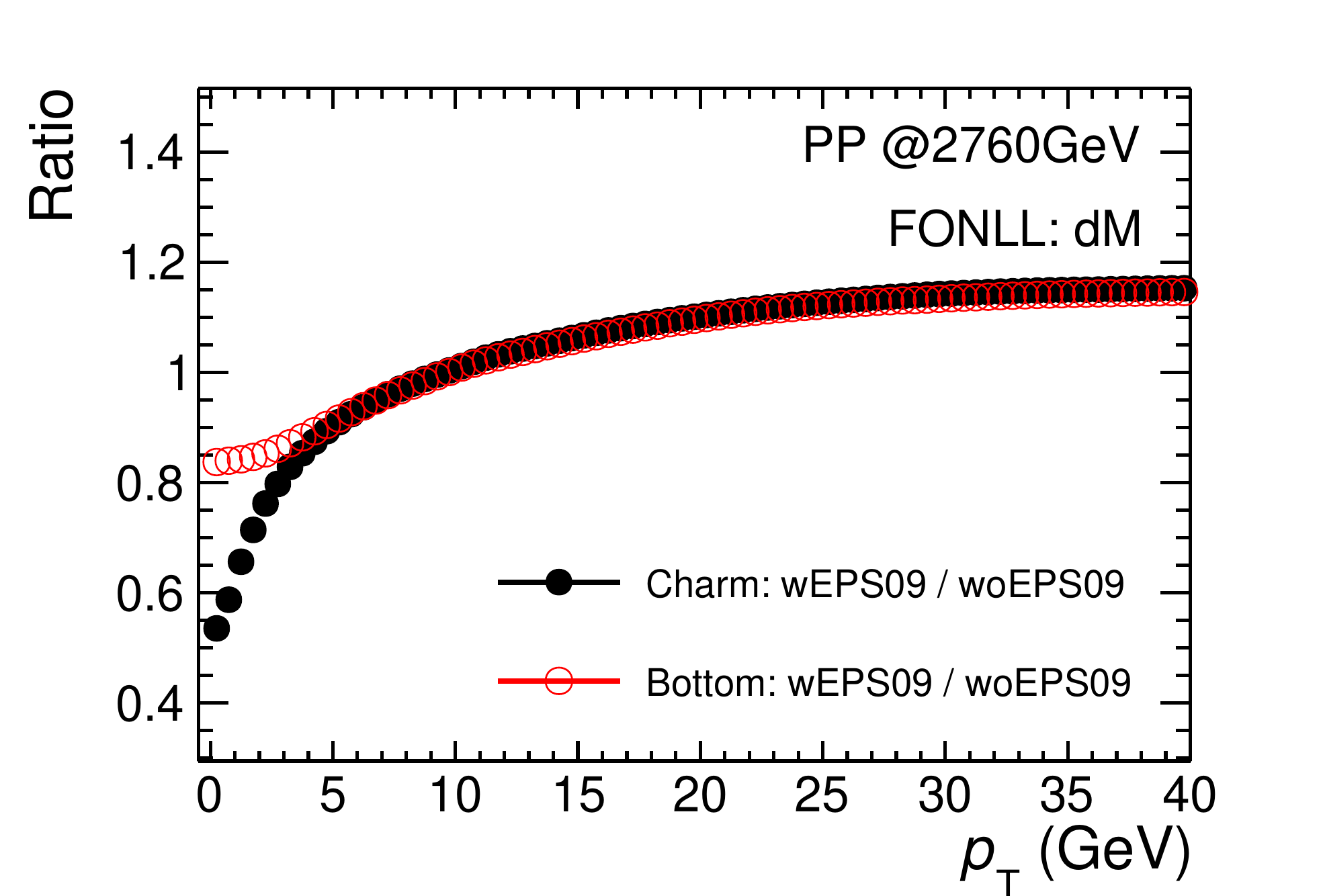}
\caption{(Color online) Ratio of the production cross sections with and without including EPS09
modification of lead (Pb) nuclear PDFs in the rapidity interval $-1<y<1$ in collisions at $\s=2.76~{\rm TeV}$.
Results for charm and bottom quark pairs are shown as filled black and open red circles, respectively.}
\label{fig:EPS09Comparison}
\end{center}
\end{figure}
In this work, we employ EPS09 NLO parameterization~\cite{EPS09} for the gold (Au) and lead (Pb) nucleus PDFs.
Figure~\ref{fig:EPS09Comparison} shows the effect on HQ production
as the ratio of the production cross sections with and without EPS09 modification on lead (Pb) nuclear PDFs at $\s=2.76~{\rm TeV}$.
It is found that the effect of shadowing is more pronounced for charm (filled black circle) than that for bottom quarks (open red circle),
due to the smaller Bjorken-$x$ ($\propto m_{\rm T}=\sqrt{m_{\rm Q}^{2}+{\pt}^2}$ at a given rapidity) for charm in this region
probed by charm.
The effect becomes similar towards high $\pt$, induced by the similar Bjorken-$x$ values probed when $\pt \gg m_{\rm Q}$.
It is found that the ratio is slightly larger than unity ($\sim15\%$ at maximum) in the range $10\lesssim\pt\lesssim40~{\rm GeV}$,
which will enhance the heavy-flavor production at high $\pt$.

The corresponding systematic uncertainties on the EPS09 NLO parameterization
are defined baed on various nPDFs sets which are obtained by tuning
fit parameters\footnote[5]{see Eq.~2.12 and 2.13 in Ref.~\cite{EPS09} for details.
We use the nPDFs sets up to $k=7$  in this analysis.}.

%%-----------------------------
%%-----------------------------
\subsection{Underlying QGP medium}\label{subsec:hybridModel_Hydro}
The hot and dense strongly-interacting medium produced in heavy-ion collisions
is in pre-equilibrium state until it reaches local thermalization.
We assume that the QCD medium undergoes a rapid thermalization and forms a QGP in equilibrium
at $\tau_{0}=0.6~{\rm fm}/{\it c}$, at which the hydrodynamical evolution commences.
The thermalization time scale is much shorter than the total life time of the QGP.
Therefore, we neglect the pre-equilibrium evolution and thermalization in this analysis,
assuming $s(\tau=0)\approx s(\tau_{0}=0.6)$ hereafter.
As discussed below, we utilize the initial conditions described above to model
the initial entropy density distribution at the starting time scale of the hydrodynamical evolution,
as well as an Equation of State (EoS) obtained via the lattice QCD calculations
to describe the phase transition from the deconfined partons to the hadronic state.

\subsubsection{Hydrodynamic description}\label{subsubsec:HudroSim}
The description of the QGP medium evolution is implemented by means of a 3+1
dimensional relativistic viscous hydrodynamics based on the HLLE algorithm~{\cite{vhlle}},
with $\tau_{0}=0.6~{\rm fm}/{\it c}$, shear viscosity ${\eta/s=1/(4\pi)}$
and critical temperature $T_{c}=165~{\rm MeV}$ in Au--Au and Pb--Pb collisions.
It provides the space-time evolution of the temperature and the flow velocity field.

\begin{figure}[!htbp]
\begin{center}
\vspace{-1.0em}
\setlength{\abovecaptionskip}{-0.1mm}
\setlength{\belowcaptionskip}{-1.5em}
\includegraphics[width=.50\textwidth]{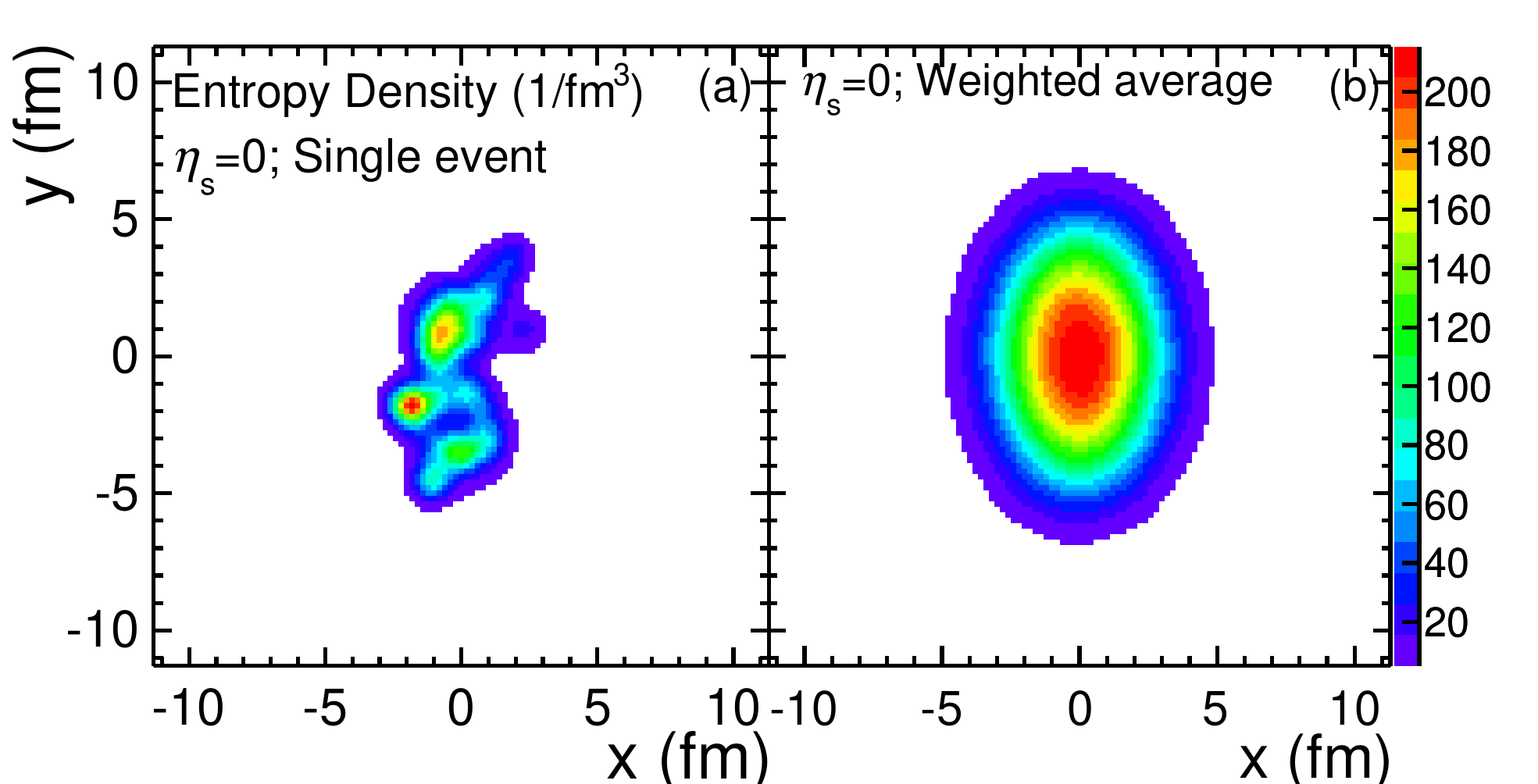}
\caption{(Color online) Left (a): initial entropy density distribution deposited in the transverse plane for a single event
in semi-central ($30-50\%$) Pb--Pb collisions at $\snn=2.76~{\rm TeV}$.
Right (b): results after weighting all the events in the $30-50\%$ centrality.}
\label{fig:EvtWgt}
\end{center}
\end{figure}
Concerning the initial state simulation for the hydrodynamic medium evolution,
we rely on the Glauber-based model introduced in Sec.~\ref{subsubsec:initialSpatial} (see Eq.~\ref{eq:entropyDis3D}).
However, by considering that the full event-by-event hydrodynamic simulation requires a huge
computational time and disk space, we overcome these issues by utilizing a weighting approach
to have an event-averaged smooth initial transverse profile of the entropy density distribution.
Figure~\ref{fig:EvtWgt} shows the results obtained for the centrality interval $30-50\%$ for Pb--Pb collisions at $\snn=2.76$ TeV.
The initial entropy density distribution for a single event is presented in the panel-a (left),
while the result after averageing all the events belonging to $30-50\%$ is displayed in the panel-b (right).
As expected, event-by-event fluctuations are largely suppressed.

\subsubsection{Isothermal freeze-out}\label{subsubsec:FreZout}
\iffalse
The QGP medium expands and cools down, and the (local) temperature
drops below the critical one $T_{c}$, resulting in the
transition from QGP phase to hadrons gas, namely hadronization.
After the transition, the hadron gas can in principle continue to interact inelastically until the chemical freeze-out, subsequently, the hadronic system
continues to expand and interact elastically until the kinetic freeze-out.
\fi
In this work, we neglect the chemical freeze-out procedure and consider, only, the kinetic freeze-out (or freeze-out since now)
occurs at $T_{c}=165~{\rm MeV}$.

To model the freeze-out of the QGP medium, we utilize an instantaneous
approach across a hypersurface of constant temperature,
namely isothermal freeze-out~\cite{isothermalFrzOut}.
We employ a widely used model, cornelius~\cite{cornelius},
to reconstruct the isothermal particlization hypersurface,
on which the momentum distributions of the different hadron species are evaluated
using the Cooper-Frye formalism~\cite{cooperFrye}.

%%-----------------------------
%%-----------------------------
\subsection{Simulation of heavy quark Brownian motion}\label{subsec:hybridModel_Simu}
In this sub-section, we describe the numerical framework
utilized for the Langevin evolution of heavy quarks (HQ) coupled with the expanding underlying hydrodynamic medium.
Generally, in the local rest frame (LRF) of the fluid cell,
the HQ motion follows the modified Langevin Transport Equation,
and the local temperature and the local flow velocity at the considered cell position
are provided by the relativistic hydrodynamics simulations.
The steps of this numerical procedure are
\begin{enumerate}
\item[(1)] Sample the HQ pairs at the position $x^{\mu}$ and momentum $p^{\mu}$, in the laboratory frame (LAB),
according to the initial phase space configurations ($\tau\sim0$);

\item[(2)] Move all the HQ from $\tau\sim0$ to $\tau_{0}=0.6~{\rm fm/{\it c}}$ as free streaming particles,
and modify the positions $x^{\mu}$ correspondingly;

\item[(3)] Search the fluid cell at $x^{\mu}$,
and extract its temperature $T$ and velocity $u^{\mu}$ from the hydrodynamic simulations;
then, boost the HQ to the LRF of the fluid cell and get the HQ momentum in this frame;

\item[(4)] Make a discrete time-step $\Delta t=0.01~{\rm fm/{\it c}}$ for the HQ in order to update its momentum $p^{\mu}$
        \begin{equation}
        p_{\rm i}(t+\Delta t)-p_{\rm i}(t)=(F_{\rm i}^{\rm Drag} + F_{\rm i}^{\rm Diff} + F_{\rm i}^{\rm Gluon}){\Delta t} \nonumber
        \end{equation}
        where the three terms in the right hand side are driven by
        \begin{itemize}
        \item the drag force term $F_{\rm i}^{\rm Drag}$: drag coefficient $\Gamma$, which is determined
                by substituting Eq.~\ref{eq:ModelA} or Eq.~\ref{eq:ModelB} into Eq.~\ref{eq:Gamma2Ds}, with the fluid cell temperature $T$ obtained in the previous step;
        \item the thermal force term $F_{\rm i}^{\rm Diff}$: the relevant time correlation profile behaves as:
                \begin{equation}
                \langle F_{\rm i}^{\rm Diff}(t) \cdot F_{\rm j}^{\rm Diff}(t+n\Delta t)\rangle \equiv \frac{\kappa}{\Delta t}\delta_{\rm ij}\delta_{0n} \nonumber
                \end{equation}
                where, the momentum diffusion coefficient $\kappa$ is given by Eq.~\ref{eq:Kappa2Ds}.
                The above correlation is implemented by applying a momentum deflection sampled according to a Gaussian distribution
                with the width $\sqrt{\kappa/\Delta t}$;
        \item the recoil force term $F_{\rm i}^{\rm Gluon}$: during each time step $\Delta t$,
                the Higher-Twist model gives the average number of radiated gluons,
                which is assumed to follow the Poisson distributions. The resulting total probability to
                radiate at least one gluon is used to determine whether or not the radiation process is triggered;
        \end{itemize}

\item[(5)] Update the HQ position after the time-step $\Delta t$
        \begin{equation}
        x_{\rm i}(t+\Delta t)-x_{\rm i}(t)=\frac{p_{\rm i}(t)}{E_{p_{\rm i}}(t)}{\Delta t} \nonumber
        \end{equation}
        with the four-momentum $p_{\rm i}$ obtained in the previous step,
        and boost back the HQ from the LRF to the LAB reference frame;

\item[(6)] Repeat the above steps (3)-(5) until hadronization conditions are reached,
           i.e. until the temperature in the local fluid cell satisfies $T\geqslant T_{c}$.
\end{enumerate}

%%-----------------------------
%%-----------------------------
\subsection{``Dual" hadronization model of heavy quarks}\label{subsec:hybridModel_Had}
As discussed above, the QGP medium hadronizes in our model when the local temperature reaches the critical one $T_{c}=165~{\rm GeV}$.
When the temperature $T$ reaches $T_{c}$, the heavy quark (HQ) will hadronize into the relevant heavy-flavor hadrons.
It is known that the hadronization is an intrinsically non-perturbative process,
which is treated as phenomenological models.
Two approaches are usually employed to describe the HQ hadronization processes,
namely ``fragmentation"~\cite{FragOriginalPythia87} and ``heavy-light coalescence"~\cite{CoalOriginalDover91}.
%It is found that the former one is validated in all the momentum region in nucleon-nucleon collisions,
%moreover, it is dominated at high momentum in nucleus-nucleus collisions,
%while the latter one is significant at moderate momentum region~\cite{GKLhadPRL, FMNBhadPRL}.
In this work, we adopt a ``dual" approach~\cite{CoalCMK09, CoalGRECO18}, including both fragmentation and coalescence,
to model the HQ hadronization in heavy-ion collisions.

\subsubsection{Fragmentation model}\label{subsubsec:FragMod}
The HQ fragmentation can be implemented by using the
``Lund symmetric fragmentation function'' (PYTHIA 6.4~\cite{PYTHIA64}) with all the defaults parameters.
Alternatively, in this analysis, we utlize two other phenomenological models:
\begin{itemize}
\item Peterson fragmentation function~\cite{FragOriginalPeterson83}:
with the parameter $\epsilon$ fixed to $\epsilon_{\rm c}=0.06$ and $\epsilon_{\rm b}=0.006$ for charm and bottom, respectively;
\item Braaten fragmentation functions~\cite{FragBraaten93}:
with the parameter $r=0.1$ for charm quarks with $m_{\rm c}=1.5~{\rm GeV}$~\cite{FragFONLLPRL}.
\end{itemize}

\begin{figure}[!tbp]
\begin{center}
\vspace{-1.8em}
\setlength{\abovecaptionskip}{-0.1mm}
\setlength{\belowcaptionskip}{-1.5em}
\includegraphics[width=.42\textwidth]{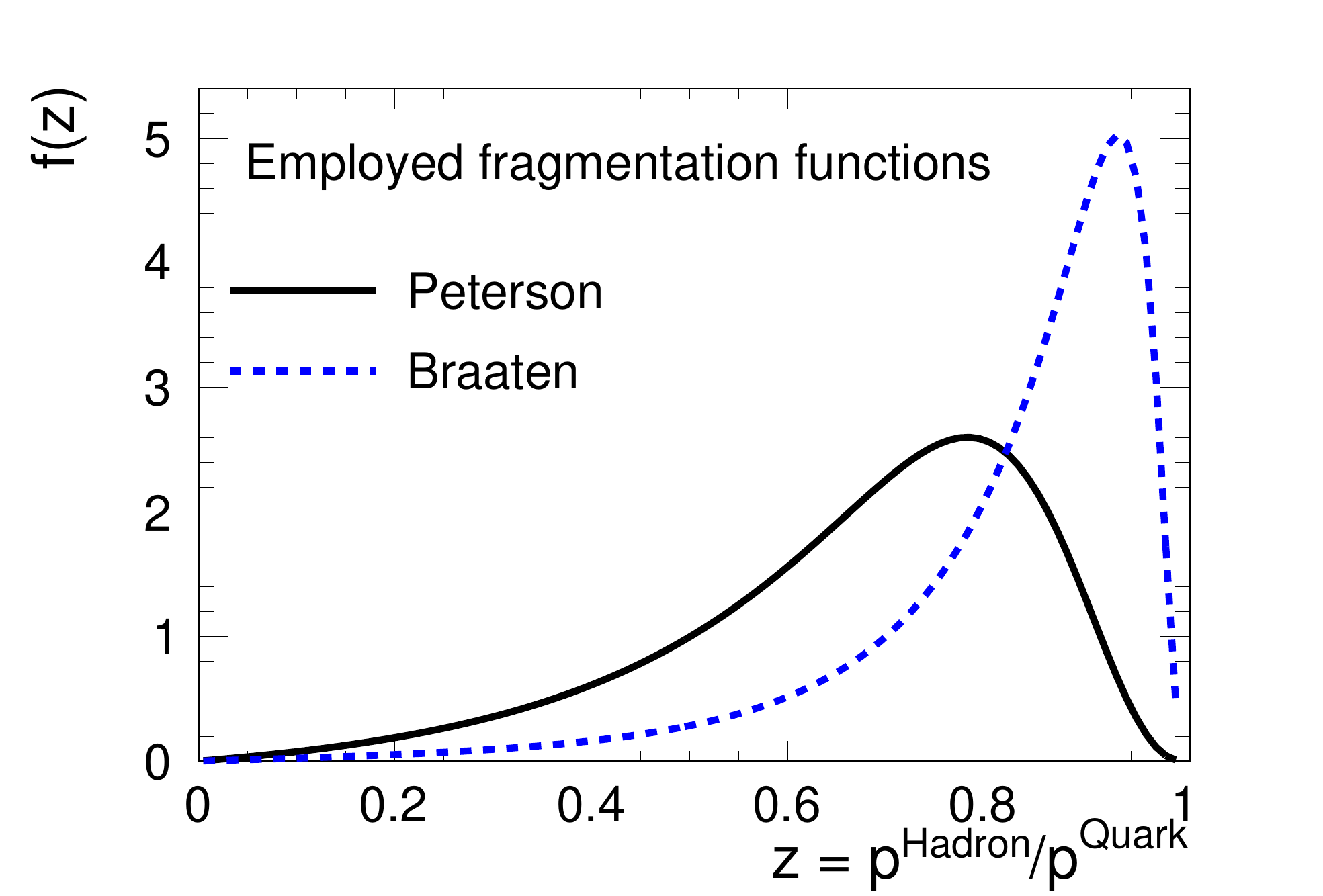}
\caption{(Color online) Normalized fragmentation functions obtained in the two phenomenological models considered in this work:
Peterson (solid black curve) and Braaten for vector mesons (dashed blue curve).}
\label{fig:FragFun}
\end{center}
\end{figure}
Figure~\ref{fig:FragFun} shows the normalized fragmentation functions as a function of the fragmentation fraction $z$,
which is defined as the momentum fraction taken away by the fragmented heavy-flavor hadron with respect to that of its mother HQ.
The average value of $z$ is larger for Braaten model (dashed blue curve) as compared to Peterson model (solid black curve),
resulting in a harder transverse momentum distribution at hadron level for the former one.
Note that the fragmentation functions are assumed to be universal, i.e. to be the same, for different colliding systems
and at different colliding energies.

To model the hadronization, we need also to provide the fragmentation fractions for the
various hadron species, i.e. the fraction of charm quarks hadronizing in the different hadron species,
except for the Lund-PYTHIA approach.
Reference~\cite{FragFracZEUS13} reforms the fragmentation fractions for the open charmed hadrons,
$D^{0}$, $D^{+}$, $D^{\ast +}$ and $D_{s}^{+}$, measured by different DIS, $\gamma p$ and $e^{+}e^{-}$ experiments (see references therein).
We took the weighted average of the fragmentation fractions reported in this work,
using as weights the total uncertainties in the measurements,
including the systematic and statistical components which are added in quadrature.
The resulting fractions are $f(c\rightarrow D^{0})=0.566$, $f(c\rightarrow D^{+})=0.227$, $f(c\rightarrow D^{\ast +})=0.230$
and $f(c\rightarrow D_{s}^{+})=0.081$.

\subsubsection{Heavy-light coalescence model}\label{subsubsec:CoalMod}
Within the instantaneous hadronization approach~\cite{GKLhadPRL, FMNBhadPRL},
the heavy-light quark coalescence is commonly modeled in terms of the overlap among the Wigner functions,
which are based on Gaussian wave packets for the heavy quark and the light anti-quark,
and the harmonic oscillator wave function the for charm hadron.
However, in the calculations, some groups~\cite{CaoPRC15, PHSDPRC16, DasPRC17} consider only
the harmonic oscillator wave functions of the ground state.
The heavy-light coalescence probability for the excited charm hadron,
such as $c\rightarrow D_{1}(2420)^{0}$, is then obtained with some artificial assumptions.
This was recently updated, for light quarks~\cite{NewCoal16}, by including the contribution from higher states
of the harmonic oscillator wave functions.
We follow this strategy and further extend it to charm quarks in this analysis.

According to the heavy-light coalescence model,
the momentum distributions of heavy-flavor mesons ($M$) composed of a heavy quark ($Q$) and a light anti-quark ($\bar{q}$) are given as
\begin{equation}
\begin{aligned}\label{eq:MesonCoal}
\frac{dN_{\rm M}}{d^{3}\vec{p}_{\rm M}}=&g_{\rm M}\int d^{3}\vec{x}_{\rm Q}d^{3}\vec{p}_{\rm Q} d^{3}\vec{x}_{\rm\bar{q}}d^{3}\vec{p}_{\rm\bar{q}} f_{\rm Q}(\vec{x}_{\rm Q},\vec{p}_{\rm Q}) f_{\rm\bar{q}}(\vec{x}_{\rm\bar{q}},\vec{p}_{\rm\bar{q}}) \\
&{\overline W}_{\rm M}^{\rm (n)}(\vec{y}_{\rm M},\vec{k}_{\rm M}) \delta^{(3)}(\vec{p}_{\rm M}-\vec{p}_{\rm Q}-\vec{p}_{\rm\bar{q}})
\end{aligned}
\end{equation}
where, $g_{\rm M}$ is the degeneracy factor accounting for the spin-color degrees of freedom;
$f_{\rm Q}(\vec{x}_{\rm Q},\vec{p}_{\rm Q})$ and $f_{\rm\bar{q}}(\vec{x}_{\rm\bar{q}},\vec{p}_{\rm\bar{q}})$
are the phase-space distributions of heavy quark and light anti-quark, respectively.
For the heavy quark, $f_{\rm Q}(\vec{x}_{\rm Q},\vec{p}_{\rm Q})$
can be obtained after the HQ propagate through the underlying QGP medium.
For the thermal light anti-quark, $f_{\rm\bar{q}}(\vec{x}_{\rm\bar{q}},\vec{p}_{\rm\bar{q}})$
follows the Boltzmann-J$\ddot{\rm u}$ttner distribution in the momentum space
and it is spatially distributed on the freeze-out hypersurface
\footnote[5]{To-be discussed in more detail in the following (Sec.~\ref{subsubsec:HQHad}).}.
The coalescence probability is quantified by ${\overline W}_{\rm M}^{\rm (n)}(\vec{y}_{\rm M},\vec{k}_{\rm M})$,
which is the overlap integral of the Wigner function of the meson and of the $Q\bar{q}$ pair,
\begin{equation}
\begin{aligned}\label{eq:InteWig}
{\overline W}_{\rm M}^{\rm (n)}(\vec{y}_{\rm M},\vec{k}_{\rm M})=&\int \frac{d^{3}\vec{x}^{\;\prime}_{\rm Q}d^{3}\vec{p}^{\;\prime}_{\rm Q}}{(2\pi)^{3}} \frac{d^{3}\vec{x}^{\;\prime}_{\rm\bar{q}}d^{3}\vec{p}^{\;\prime}_{\rm\bar{q}}}{(2\pi)^{3}} W_{\rm Q}(\vec{x}^{\;\prime}_{\rm Q}, \vec{p}^{\;\prime}_{\rm Q}) \\
&W_{\rm\bar{q}}(\vec{x}^{\;\prime}_{\rm\bar{q}}, \vec{p}^{\;\prime}_{\rm\bar{q}}) W_{\rm M}^{\rm (n)}(\vec{y}^{\;\prime}_{\rm M}, \vec{k}^{\;\prime}_{\rm M})
\end{aligned}
\end{equation}
where,
\begin{equation}
\begin{aligned}\label{eq:RelativeXP}
&\vec{y}_{\rm M}\equiv\vec{y}_{\rm M}(\vec{x}_{\rm Q},\vec{x}_{\rm\bar{q}})=(\vec{x}_{\rm Q}-\vec{x}_{\rm\bar{q}}) \\
\\
&\vec{k}_{\rm M}\equiv\vec{k}_{\rm M}(\vec{p}_{\rm Q},\vec{p}_{\rm\bar{q}})=(m_{\rm\bar{q}}\vec{p}_{\rm Q}-m_{\rm Q}\vec{p}_{\rm\bar{q}})/(m_{\rm Q}+m_{\rm\bar{q}})
\end{aligned}
\end{equation}
are the relative coordinate and the relative momentum, respectively, in the center-of-mass (CMS) frame of the $Q\bar{q}$ pair;
$W_{\rm Q}(\vec{x}^{\;\prime}_{\rm Q}, \vec{p}^{\;\prime}_{\rm Q})$ and $W_{\rm\bar{q}}(\vec{x}^{\;\prime}_{\rm\bar{q}}, \vec{p}^{\;\prime}_{\rm\bar{q}})$
are, respetively, the Wigner functions of heavy quark and light anti-quark
with their centroids at $(\vec{x}_{\rm Q},\vec{p}_{\rm Q})$ and $(\vec{x}_{\rm\bar{q}},\vec{p}_{\rm\bar{q}})$,
and they are both defined by taking the relevant wave function to be a Gaussian wave packet~\cite{CoalOriginalDover91}.
$W_{\rm M}^{\rm (n)}(\vec{y}^{\;\prime}_{\rm M}, \vec{k}^{\;\prime}_{\rm M})$
denotes the Wigner function of heavy-flavor meson, which is based on the well-known harmonic oscillator~\cite{CoalOriginalDover91},
resulting in
\begin{eqnarray}\label{eq:WignerMeson}
W_{\rm M}^{\rm (n)}(\vec{y}, \vec{k}) &&= \left\{ \begin{array}{ll}
8e^{-\frac{\vec{y}^{\;2}}{\sigma_{\rm M}^{2}} - \sigma_{\rm M}^{2}\vec{k}^{\;2}} & \textrm{(n=0)} \\
\\
\frac{16}{3}(\frac{\vec{y}^{\;2}}{\sigma_{\rm M}^{2}}-\frac{3}{2}+\sigma_{\rm M}^{2}\vec{k}^{\;2}) e^{-\frac{\vec{y}^{\;2}}{\sigma_{\rm M}^{2}} - \sigma_{\rm M}^{2}\vec{k}^{\;2}} & \textrm{(n=1)}
\end{array} \right.
\end{eqnarray}
Finally, the overlap integral function for a heavy-flavor meson (Eq.~\ref{eq:InteWig})
in the $n^{th}$ excited state in the CMS of the $Q\bar{q}$ pair is re-written as~\cite{NewCoal16}
\begin{equation}\label{eq:InteWigner2}
{\overline W}_{\rm M}^{\rm (n)}(\vec{y},\vec{k})=\frac{\upsilon^{n}}{n!} e^{-\upsilon}, \qquad
\upsilon=\frac{1}{2}\biggr(\frac{\vec{y}^{\;2}}{\sigma_{\rm M}^{2}}+\sigma_{\rm M}^{2}\vec{k}^{\;2}\biggr).
\end{equation}
Note that, in this work, we just consider the open charmed mesons
up to their first excited states ($n\leqslant1$) according to the PDG data~\cite{PDG2016}.
The width parameter $\sigma_{\rm M}$ in the harmonic oscillator wave function
is determined by the radius of the formed heavy-flavor meson.
The charge radius of the $Q\bar{q}$ system is given by~\cite{CaoThesis}
\begin{equation}%\label{ChgR2R}
\langle r_{\rm M}^{2}\rangle=\frac{e_{\rm Q}m_{\rm\bar{q}}^{2} + e_{\rm\bar{q}}m_{\rm Q}^{2}}{(e_{\rm Q}+e_{\rm\bar{q}})(m_{\rm Q}+m_{\rm\bar{q}})^{2}} \cdot \langle r^{2}\rangle,
\end{equation}
%in the limit of non-relativistic,
where, $e_{\rm Q}$ and $e_{\rm\bar{q}}$ are the absolute values of the heavy quark and light anti-quark charges, respectively.
$\langle r^{2}\rangle$ denotes the average squared distance, and it can be calculated from the Wigner function
\begin{equation}\label{eq:R2Wigner}
\langle r^{2}\rangle=\frac{\int d^{3}\vec{r}d^{3}\vec{p} \cdot r^{2} \cdot W_{\rm M}^{\rm (n)}(\vec{r}, \vec{p})}{\int d^{3}\vec{r}d^{3}\vec{p} \cdot W_{\rm M}^{\rm (n)}(\vec{r}, \vec{p})}.
\end{equation}
By substituting Eq.~\ref{eq:WignerMeson} into Eq.~\ref{eq:R2Wigner},
we can relate $\sigma_{\rm M}^{2}$ to $\langle r_{\rm M}^{2} \rangle$ via
\begin{eqnarray}
\sigma_{\rm M}^{2}~&&= \left\{ \begin{array}{ll}
\frac{2}{3} \frac{(e_{\rm Q}+e_{\rm\bar{q}})(m_{\rm Q}+m_{\rm\bar{q}})^{2}}{e_{\rm Q}m_{\rm\bar{q}}^{2} + e_{\rm\bar{q}}m_{\rm Q}^{2}} \cdot \langle r_{\rm M}^{2} \rangle & \textrm{\qquad (n=0)} \\
\\
\frac{2}{5} \frac{(e_{\rm Q}+e_{\rm\bar{q}})(m_{\rm Q}+m_{\rm\bar{q}})^{2}}{e_{\rm Q}m_{\rm\bar{q}}^{2} + e_{\rm\bar{q}}m_{\rm Q}^{2}} \cdot \langle r_{\rm M}^{2} \rangle & \textrm{\qquad (n=1)}
\end{array} \right.
\end{eqnarray}
where, the light (anti-)quark masses take the values
$m_{\rm u}=m_{\rm\bar{u}}=m_{\rm d}=m_{\rm\bar{d}}=300~{\rm MeV}$ and $m_{\rm s}=m_{\rm\bar{s}}=475~{\rm MeV}$.
In this analysis, we adopt the assumption proposed in Ref.~\cite{PHSDPRC16}
and we set the charge radius of open charmed mesons to be equal to the one of the proton,
i.e. $\langle r_{\rm M}^{2}\rangle \approx \langle r_{\rm p}\rangle^{2}\approx (0.9~{\rm fm)^{2}}$.

Various species of open charmed mesons are considered up to their first excited states ($n\leqslant1$),
which are listed in Tab.~\ref{tab:DMesonStates}.
Note that, (1) the further decay of the D-meson produced in the decay of the excited state is not shown in this table;
(2) the branching ratios for $D_{1}(2420)^{0}\rightarrow D^{0}\pi^{+}\pi^{-}$ and $D_{1}(2420)^{0}\rightarrow D^{*}(2010)^{+}\pi^{-}$
are estimated according to the spin-color degeneracy factors of $D^{0}$ and $D^{*}(2010)^{+}$ ($g_{\rm M}$ in Eq.~\ref{eq:MesonCoal}),
i.e. $g_{D^{0}} / g_{D^{*}(2010)^{+}} = 1/3$. Similar case for $D^{*}_{2}(2460)^{0}$ and $D^{*}_{2}(2460)^{+}$.
\begin{table}[!htbp]
\centering
\begin{tabular}{|c|c|c|c|c|}
 \toprule[1.5pt]
\multicolumn{1}{c}{\multirow{1}{*}{\centering Species}}
 & \multicolumn{1}{c}{\multirow{1}{*}{\centering $^{2s+1}L_{J}$}}
 & \multicolumn{1}{c}{\multirow{1}{*}{\centering Mass (GeV)}}
 & \multicolumn{1}{c}{\multirow{1}{*}{\centering Decay Modes}}
 & \multicolumn{1}{c}{\multirow{1}{*}{\centering BR ($\%$)}}
  \\
 \hline
\multicolumn{1}{c}{\multirow{1}{*}{\centering $D^{0}$}}
 & \multicolumn{1}{c}{\multirow{1}{*}{\centering $^{1}S_{0}$}}
 & \multicolumn{1}{c}{\multirow{1}{*}{\centering 1.86}}
 & \multicolumn{1}{c}{\multirow{1}{*}{\centering }}
 & \multicolumn{1}{c}{\multirow{1}{*}{\centering }}
  \\
\hline
\multicolumn{1}{c}{\multirow{1}{*}{\centering $D^{*}(2007)^{0}$}}
 & \multicolumn{1}{c}{\multirow{1}{*}{\centering $^{3}S_{1}$}}
 & \multicolumn{1}{c}{\multirow{1}{*}{\centering 2.01}}
 & \multicolumn{1}{c}{\multirow{1}{*}{\centering $D^{0}\pi^{0}$}}
 & \multicolumn{1}{c}{\multirow{1}{*}{\centering 64.7}}
  \\
\multicolumn{1}{c}{\multirow{1}{*}{\centering }}
 & \multicolumn{1}{c}{\multirow{1}{*}{\centering }}
 & \multicolumn{1}{c}{\multirow{1}{*}{\centering }}
 & \multicolumn{1}{c}{\multirow{1}{*}{\centering $D^{0}\gamma$}}
 & \multicolumn{1}{c}{\multirow{1}{*}{\centering 35.3}}
  \\
\hline
\multicolumn{1}{c}{\multirow{1}{*}{\centering $D^{*}_{0}(2400)^{0}$}}
 & \multicolumn{1}{c}{\multirow{1}{*}{\centering $^{3}P_{0}$}}
 & \multicolumn{1}{c}{\multirow{1}{*}{\centering 2.32}}
 & \multicolumn{1}{c}{\multirow{1}{*}{\centering $D^{+}\pi^{-}$}}
 & \multicolumn{1}{c}{\multirow{1}{*}{\centering 1}}
  \\
\hline
\multicolumn{1}{c}{\multirow{1}{*}{\centering $D_{1}(2420)^{0}$}}
 & \multicolumn{1}{c}{\multirow{1}{*}{\centering $^{1}P_{1}$}}
 & \multicolumn{1}{c}{\multirow{1}{*}{\centering 2.42}}
 & \multicolumn{1}{c}{\multirow{1}{*}{\centering $D^{0}\pi^{+}\pi^{-}$}}
 & \multicolumn{1}{c}{\multirow{1}{*}{\centering 25}}
  \\
\multicolumn{1}{c}{\multirow{1}{*}{\centering }}
 & \multicolumn{1}{c}{\multirow{1}{*}{\centering }}
 & \multicolumn{1}{c}{\multirow{1}{*}{\centering }}
 & \multicolumn{1}{c}{\multirow{1}{*}{\centering $D^{*}(2010)^{+}\pi^{-}$}}
 & \multicolumn{1}{c}{\multirow{1}{*}{\centering 75}}
 \\
\hline
\multicolumn{1}{c}{\multirow{1}{*}{\centering $D_{2}^{*}(2460)^{0}$}}
 & \multicolumn{1}{c}{\multirow{1}{*}{\centering $^{3}P_{2}$}}
 & \multicolumn{1}{c}{\multirow{1}{*}{\centering 2.46}}
 & \multicolumn{1}{c}{\multirow{1}{*}{\centering $D^{+}\pi^{-}$}}
 & \multicolumn{1}{c}{\multirow{1}{*}{\centering 25}}
  \\
\multicolumn{1}{c}{\multirow{1}{*}{\centering }}
 & \multicolumn{1}{c}{\multirow{1}{*}{\centering }}
 & \multicolumn{1}{c}{\multirow{1}{*}{\centering }}
 & \multicolumn{1}{c}{\multirow{1}{*}{\centering $D^{*}(2010)^{+}\pi^{-}$}}
 & \multicolumn{1}{c}{\multirow{1}{*}{\centering 75}}
 \\
\midrule[1pt]
\multicolumn{1}{c}{\multirow{1}{*}{\centering $D^{+}$}}
 & \multicolumn{1}{c}{\multirow{1}{*}{\centering $^{1}S_{0}$}}
 & \multicolumn{1}{c}{\multirow{1}{*}{\centering 1.87}}
 & \multicolumn{1}{c}{\multirow{1}{*}{\centering }}
 & \multicolumn{1}{c}{\multirow{1}{*}{\centering }}
  \\
\hline
\multicolumn{1}{c}{\multirow{1}{*}{\centering $D^{*}(2010)^{+}$}}
 & \multicolumn{1}{c}{\multirow{1}{*}{\centering $^{3}S_{1}$}}
 & \multicolumn{1}{c}{\multirow{1}{*}{\centering 2.01}}
 & \multicolumn{1}{c}{\multirow{1}{*}{\centering $D^{0}\pi^{+}$}}
 & \multicolumn{1}{c}{\multirow{1}{*}{\centering 67.7}}
  \\
\multicolumn{1}{c}{\multirow{1}{*}{\centering }}
 & \multicolumn{1}{c}{\multirow{1}{*}{\centering }}
 & \multicolumn{1}{c}{\multirow{1}{*}{\centering }}
 & \multicolumn{1}{c}{\multirow{1}{*}{\centering $D^{+}\pi^{0}$}}
 & \multicolumn{1}{c}{\multirow{1}{*}{\centering 30.7}}
  \\
\multicolumn{1}{c}{\multirow{1}{*}{\centering }}
 & \multicolumn{1}{c}{\multirow{1}{*}{\centering }}
 & \multicolumn{1}{c}{\multirow{1}{*}{\centering }}
 & \multicolumn{1}{c}{\multirow{1}{*}{\centering $D^{+}\gamma$}}
 & \multicolumn{1}{c}{\multirow{1}{*}{\centering 1.6}}
  \\
\hline
\multicolumn{1}{c}{\multirow{1}{*}{\centering $D^{*}_{2}(2460)^{+}$}}
 & \multicolumn{1}{c}{\multirow{1}{*}{\centering $^{3}P_{2}$}}
 & \multicolumn{1}{c}{\multirow{1}{*}{\centering 2.47}}
 & \multicolumn{1}{c}{\multirow{1}{*}{\centering $D^{0}\pi^{+}$}}
 & \multicolumn{1}{c}{\multirow{1}{*}{\centering 25}}
  \\
\multicolumn{1}{c}{\multirow{1}{*}{\centering }}
 & \multicolumn{1}{c}{\multirow{1}{*}{\centering }}
 & \multicolumn{1}{c}{\multirow{1}{*}{\centering }}
 & \multicolumn{1}{c}{\multirow{1}{*}{\centering $D^{*}(2007)^{0}\pi^{+}$}}
 & \multicolumn{1}{c}{\multirow{1}{*}{\centering 75}}
 \\
\midrule[1pt]
\multicolumn{1}{c}{\multirow{1}{*}{\centering $D^{+}_{s}$}}
 & \multicolumn{1}{c}{\multirow{1}{*}{\centering $^{1}S_{0}$}}
 & \multicolumn{1}{c}{\multirow{1}{*}{\centering 1.97}}
 & \multicolumn{1}{c}{\multirow{1}{*}{\centering }}
 & \multicolumn{1}{c}{\multirow{1}{*}{\centering }}
  \\
\hline
\multicolumn{1}{c}{\multirow{1}{*}{\centering $D^{*+}_{s}$}}
 & \multicolumn{1}{c}{\multirow{1}{*}{\centering $^{3}S_{1}$}}
 & \multicolumn{1}{c}{\multirow{1}{*}{\centering 2.11}}
 & \multicolumn{1}{c}{\multirow{1}{*}{\centering $D^{+}_{s}\gamma$}}
 & \multicolumn{1}{c}{\multirow{1}{*}{\centering 93.5}}
  \\
\multicolumn{1}{c}{\multirow{1}{*}{\centering }}
 & \multicolumn{1}{c}{\multirow{1}{*}{\centering }}
 & \multicolumn{1}{c}{\multirow{1}{*}{\centering }}
 & \multicolumn{1}{c}{\multirow{1}{*}{\centering $D^{+}_{s}\pi^{0}$}}
 & \multicolumn{1}{c}{\multirow{1}{*}{\centering 6.5}}
  \\
\hline
\multicolumn{1}{c}{\multirow{1}{*}{\centering $D^{*}_{s0}(2317)^{+}$}}
 & \multicolumn{1}{c}{\multirow{1}{*}{\centering $^{3}P_{0}$}}
 & \multicolumn{1}{c}{\multirow{1}{*}{\centering 2.32}}
 & \multicolumn{1}{c}{\multirow{1}{*}{\centering $D^{+}_{s}\pi^{0}$}}
 & \multicolumn{1}{c}{\multirow{1}{*}{\centering 1}}
  \\
\hline
\multicolumn{1}{c}{\multirow{1}{*}{\centering $D_{s1}(2460)^{+}$}}
 & \multicolumn{1}{c}{\multirow{1}{*}{\centering $^{3}P_{1}$}}
 & \multicolumn{1}{c}{\multirow{1}{*}{\centering 2.46}}
 & \multicolumn{1}{c}{\multirow{1}{*}{\centering $D_{s}^{*+}\pi^{0}$}}
 & \multicolumn{1}{c}{\multirow{1}{*}{\centering 48}}
  \\
\multicolumn{1}{c}{\multirow{1}{*}{\centering }}
 & \multicolumn{1}{c}{\multirow{1}{*}{\centering }}
 & \multicolumn{1}{c}{\multirow{1}{*}{\centering }}
 & \multicolumn{1}{c}{\multirow{1}{*}{\centering $D_{s}^{+}\gamma$}}
 & \multicolumn{1}{c}{\multirow{1}{*}{\centering 18}}
  \\
\multicolumn{1}{c}{\multirow{1}{*}{\centering }}
 & \multicolumn{1}{c}{\multirow{1}{*}{\centering }}
 & \multicolumn{1}{c}{\multirow{1}{*}{\centering }}
 & \multicolumn{1}{c}{\multirow{1}{*}{\centering $D_{s}^{+}\pi^{+}\pi^{-}$}}
 & \multicolumn{1}{c}{\multirow{1}{*}{\centering 4.3}}
  \\
\multicolumn{1}{c}{\multirow{1}{*}{\centering }}
 & \multicolumn{1}{c}{\multirow{1}{*}{\centering }}
 & \multicolumn{1}{c}{\multirow{1}{*}{\centering }}
 & \multicolumn{1}{c}{\multirow{1}{*}{\centering $D_{s}^{*+}\gamma$}}
 & \multicolumn{1}{c}{\multirow{1}{*}{\centering 8}}
  \\
\multicolumn{1}{c}{\multirow{1}{*}{\centering }}
 & \multicolumn{1}{c}{\multirow{1}{*}{\centering }}
 & \multicolumn{1}{c}{\multirow{1}{*}{\centering }}
 & \multicolumn{1}{c}{\multirow{1}{*}{\centering $D^{*}_{s0}(2317)^{+}\gamma$}}
 & \multicolumn{1}{c}{\multirow{1}{*}{\centering 3.7}}
 \\
\hline
\multicolumn{1}{c}{\multirow{1}{*}{\centering $D_{s1}(2536)^{+}$}}
 & \multicolumn{1}{c}{\multirow{1}{*}{\centering $^{1}P_{1}$}}
 & \multicolumn{1}{c}{\multirow{1}{*}{\centering 2.54}}
 & \multicolumn{1}{c}{\multirow{1}{*}{\centering $D^{*}(2010)^{+}K^{0}$}}
 & \multicolumn{1}{c}{\multirow{1}{*}{\centering 85}}
  \\
\multicolumn{1}{c}{\multirow{1}{*}{\centering }}
 & \multicolumn{1}{c}{\multirow{1}{*}{\centering }}
 & \multicolumn{1}{c}{\multirow{1}{*}{\centering }}
 & \multicolumn{1}{c}{\multirow{1}{*}{\centering $D^{+}K^{0}$}}
 & \multicolumn{1}{c}{\multirow{1}{*}{\centering 15}}
  \\
\hline
\multicolumn{1}{c}{\multirow{1}{*}{\centering $D^{*}_{s2}(2573)$}}
 & \multicolumn{1}{c}{\multirow{1}{*}{\centering $^{3}P_{2}$}}
 & \multicolumn{1}{c}{\multirow{1}{*}{\centering 2.57}}
 & \multicolumn{1}{c}{\multirow{1}{*}{\centering $D^{0}K^{+}$}}
 & \multicolumn{1}{c}{\multirow{1}{*}{\centering 1}}
  \\
\bottomrule[1.5pt]
\end{tabular}
\caption{Open charmed meson species taken into account in this analysis. Results adopted from Ref.~\cite{PDG2016}.}
\label{tab:DMesonStates}
\end{table}
%%%%

\subsubsection{Implementation of the ``dual'' hadronization model}\label{subsubsec:HQHad}
Since we focus on the open charmed meson production, composed of a $c$ ($\bar{c}$) and its partner $\bar{q}$ ($q$),
in this work, a ``dual'' hadronization approach is implemented as described in the following.
We will take as example the $c\bar{q}$ combination.
\begin{enumerate}
\item[(1)] Extract the three-vectors $\vec{r}_{\rm i,c}$ and $\vec{p}_{\rm i,c}$ for the $i^{th}$ charm quark
position and momentum at $T_{c}$, after the propagating through the QCD medium;

\item[(2)] Sample a number of associated $\bar{q}$ candidates, $N_{\rm i,partners}$, for the considered charm quark,
and initialize their positions and momentum according to:
        \begin{itemize}
        \item position: set the three-vector for the $j^{th}$ partner, $\vec{r}_{\rm ij,\bar{q}}$,
                according to the coordinate of the hypersurface cells
                which the considered current charm quark is located;
        \item momentum initialization: sample the partner momentum, $\vec{p}_{\rm ij,\bar{q}}$, in the LRF of the fluid cell,
                according to the Boltzmann-J$\ddot{\rm u}$ttner distribution;
                boost the partner to the LAB frame;
        \end{itemize}

\item[(3)] Calculate coalescence probabilities up to the first excited state,
i.e. $\overline{W}_{\rm M}^{\rm (0)}(\vec{y}_{\rm ij},\vec{k}_{\rm ij})$ and $\overline{W}_{\rm M}^{\rm (1)}(\vec{y}_{\rm ij},\vec{k}_{\rm ij})$,
via Eq.~\ref{eq:InteWigner2}, in the CMS of each $c\bar{q}$ pair,
with $\vec{y}_{\rm ij}(\vec{r}_{\rm i,c},\vec{r}_{\rm ij,\bar{q}})$ and $\vec{k}_{\rm ij}(\vec{p}_{\rm i,c},\vec{p}_{\rm ij,\bar{q}})$ defined in Eq.~\ref{eq:RelativeXP};
get the relevant total coalescence probability:
\begin{equation}
P_{\rm ij}^{\rm Total}=\overline{W}_{\rm M}^{\rm (0)}(\vec{y}_{\rm ij},\vec{k}_{\rm ij}) + \overline{W}_{\rm M}^{\rm (1)}(\vec{y}_{\rm ij},\vec{k}_{\rm ij}); \nonumber
\end{equation}
search the target $c\bar{q}$ pair giving the maximum value
\begin{equation}
P_{\rm i}^{\rm Max}=MAX\biggr\{P_{\rm i1}^{\rm Total},P_{\rm i2}^{\rm Total}~...~P_{\rm iN_{\rm i,partners}}^{\rm Total}\biggr\} \nonumber
\end{equation}

\item[(4)] Generate a random number, $rdm$, with flat distribution between zero and one and compare it to $P_{\rm i}^{\rm Max}$
        \begin{itemize}
        \item $rdm>P_{\rm i}^{\rm Max}$: the fragmentation process will be triggered for the considered charm quark;
        \item $rdm<P_{\rm i}^{\rm Max}$: the coalescence approach will be implemented
                \begin{itemize}
                \item[*] $rdm<\overline{W}_{\rm M}^{\rm (0)}$: the $c$ and $\bar{q}$ quarks are combined via coalescence to form open charmed meson with ground state;
                \item[*] $\overline{W}_{\rm M}^{\rm (0)}<rdm<\overline{W}_{\rm M}^{\rm (1)}$: an open charm meson in the first excited state is produced;
                \end{itemize}
        \end{itemize}

\item[(5)] Repeat the above steps for all the charm quarks.
\end{enumerate}
%%%%%%%%%%%%%%%%%
\iffalse
Assuming that the various light (anti-)quarks are thermalized inside QGP,
and their density distributions are therefore quantified by the Boltzmann-J$\ddot{\rm u}$ttner function.
Then, the relevant flavor species ($u/\bar{u}$, $d/\bar{d}$ and $s/\bar{s}$) can be determined
according to the Boltzmann-J$\ddot{\rm u}$ttner integrated over the whole momentum range.
The obtained parton densities are about $0.18~{\rm fm}^{-3}$ for $u$ and $d$, and $0.10~{\rm fm}^{-3}$ for $s$ quark in the case of $T_{c}=165~{\rm MeV}$.
The relative ratio is, therefore, about $u:d:s\approx1:1:0.5$ which is kept during the determination of the flavors of the light (anti-)quarks.
\fi
%%%%%%%%%%%%%%%%%

In the following, we will show the coalescence probability
for $c$ quarks into D-meson, including $D^{0}$, $D^{\ast 0}$, $D^{+}$, $D^{\ast +}$,
$D_{s}^{+}$ and their first excited states listed in Tab.~\ref{tab:DMesonStates},
for different centrality classes and at different energies.

\begin{figure}[!htbp]
\begin{center}
\vspace{-1.0em}
\setlength{\abovecaptionskip}{-0.1mm}
\setlength{\belowcaptionskip}{-1.5em}
\includegraphics[width=.42\textwidth]{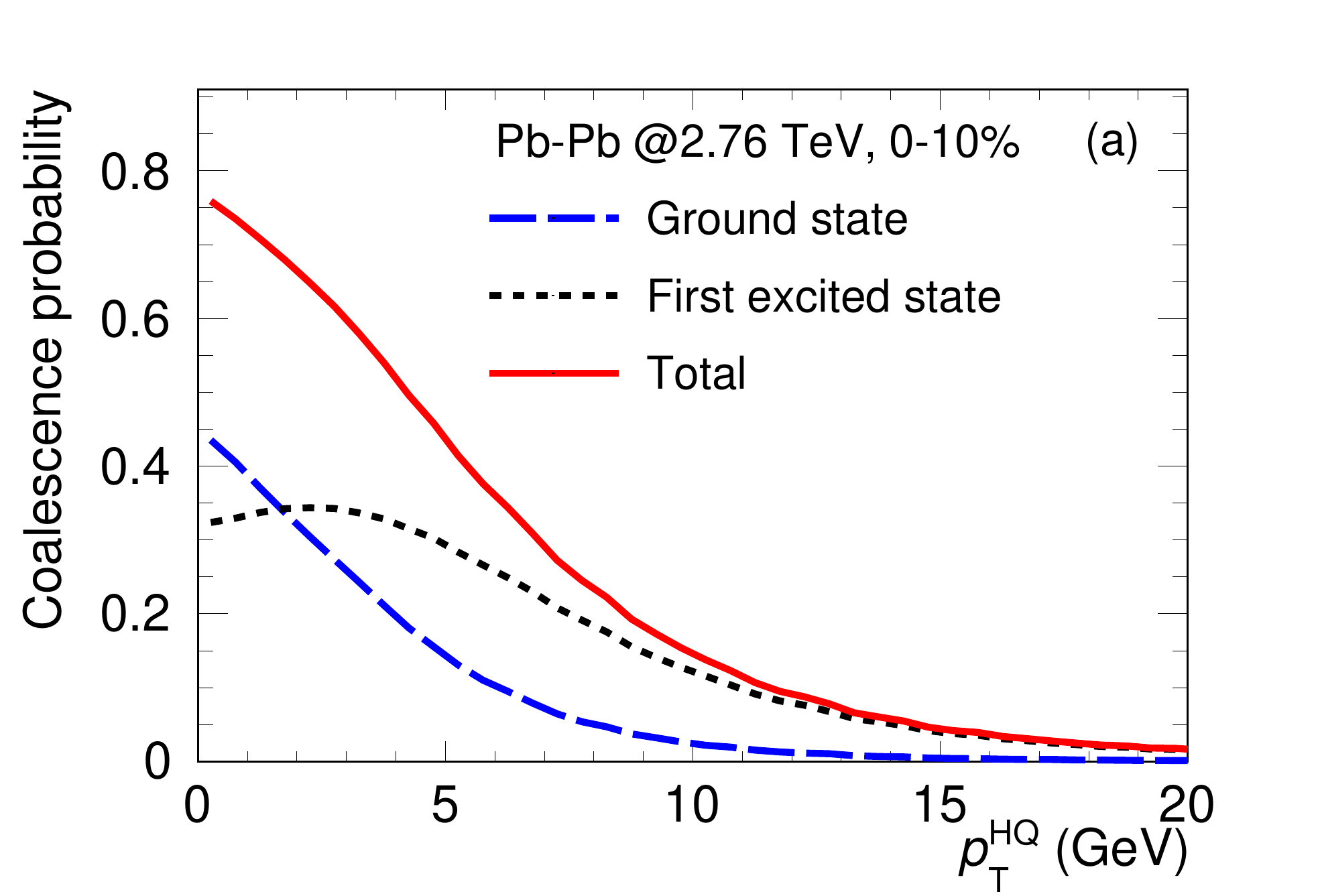}
\includegraphics[width=.42\textwidth]{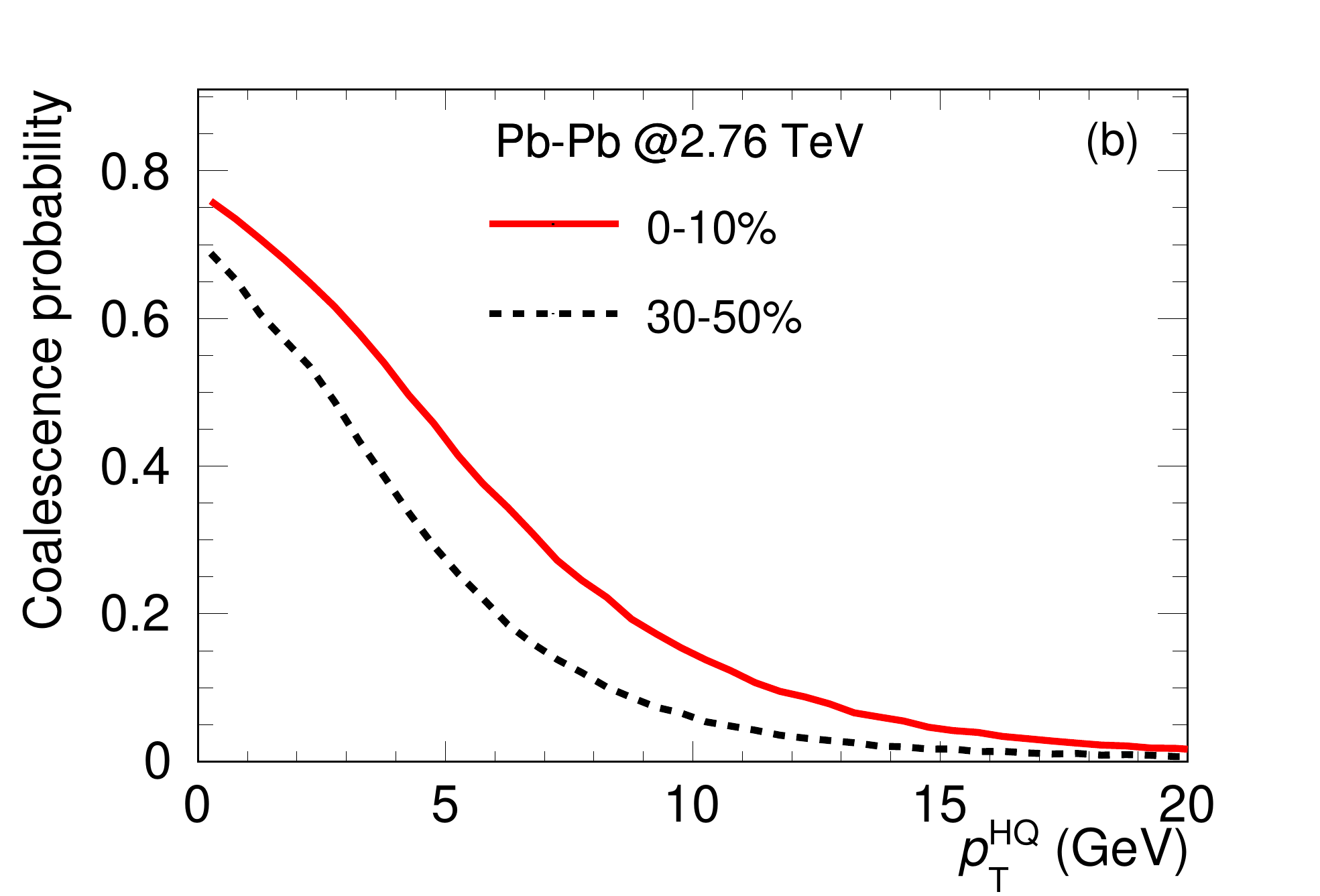}
\caption{(Color online) Upper (a): comparison of the coalescence probability, for $c\rightarrow$ D-meson in central ($0-10\%$) Pb--Pb collisions at $\snn=2.76~{\rm TeV}$,
contributed by (a) the ground states (long dashed blue curve) and the first excited states (dashed black curve).
The combined results (solid red curve) are presented as well.
Bottom (b): comparison of the coalescence probability, for $c\rightarrow$ D-meson in Pb--Pb collisions at $\snn=2.76~{\rm TeV}$,
obtained in central ($0-10\%$, solid red curve) and semi-central ($30-50\%$, dashed blue curve) regions.}
\label{fig:CoalProbPbPb2760}
\end{center}
\end{figure}

In the panel-a (upper) of Fig.~\ref{fig:CoalProbPbPb2760} the coalescence probabilities obtained
in central ($0-10\%$) Pb--Pb collisions at $\snn=2.76~{\rm TeV}$,
are presented as a function of the charm quark transverse momentum.
The contributions of the ground states and the first excited states
are shown separatedly as the long dashed blue and short dashed black curves, respectively.
It is found that the coalescence into a ground state has maximum probability at $\pt^{\rm HQ}\sim0$,
and it decreases towards high $\pt$,
due to the difficulty to find a coalescence partner in this region.
On the other case the coalescence probability into the first excited states
shows a slightly increasing behaviour in the range $\pt^{\rm HQ}\lesssim3~{\rm GeV}$,
followed by a decreasing trend at higher $\pt^{\rm HQ}$.
This behaviour may be induced by the fact that energetic charm (anti-)quarks are needed to form D mesons in the highly excited states.
The total coalescence probability is shown as a solid red curve,
which decreases from $\sim0.75$ at $\pt^{\rm HQ}\sim0$ to $0.15$ at $\pt^{\rm HQ}=10~{\rm GeV}$.
Moreover, the total coalescence probability is larger than $0.5$ in the range $\pt^{\rm HQ}\lesssim4~{\rm GeV}$,
reflecting its dominance in this region.
Similar behaviour was found for Pb--Pb and Au--Au collisions in different centrality classes.

The panel-b (bottom) of Fig.~\ref{fig:CoalProbPbPb2760} shows the results calculated
for Pb--Pb collisions at $\snn=2.76~{\rm TeV}$ in the $0-10\%$ (solid curve) and $30-50\%$ (dashed curve).
The coalescence probability is systematically larger for more central collisions.
This is because the parton density is larger in the $0-10\%$ than in the $30-50\%$,
resulting in a larger probability to form heavy-light combinations.

%%-----------------------------
%%-----------------------------
\subsection{Experimental observables}\label{subsec:observables}
We investigate the nuclear modification factor $\raa$ which is defined as the ratio
of the binary-scaled particle production cross section in nucleus-nucleus collisions
to that in nucleon-nucleon collisions at the same energy,
\begin{equation}\label{eq:RAA}
\raa(\pt,y)=\frac{{\rm d}^{2}\sigma_{\rm AA}/{\rm d}\pt {\rm d}y}{{\rm d}^{2}\sigma_{\rm pp}/{\rm d}\pt {\rm d}y},
\end{equation}
where, ${\rm d}^{2}\sigma_{\rm AA}/{\rm d}\pt {\rm d}y$ is the $\pt$ and $y$ double-differential production cross section
in nucleus-nucleus collisions, scaled by the number of binary nucleon-nucleon collisions;
${\rm d}^{2}\sigma_{\rm pp}/{\rm d}\pt {\rm d}y$ is the double-differential result in nucleon-nucleon collisions.
The deviation of $\raa$ from unity is sensitive to the effects such
as initial (anti-)shadowing and the in-medium energy loss,
consequently, it can theorefore be used to quantify the nuclear effects in heavy-ion collisions.

The elliptic flow coefficient $\vtwo$ is defined as the second harmonic
when representing the particle azimuthal distributions via a Fourier expansion:
\begin{equation}\label{eq:v2}
\vtwo=\langle cos(2\phi) \rangle = \biggr \langle \frac{p_{x}^{2}-p_{y}^{2}}{p_{x}^{2}+p_{y}^{2}} \biggr \rangle.
\end{equation}
Therefore, $\vtwo$ allows to describe the anisotropy of the transverse momentum distribution of the produced particles.
It is sensitive to the EoS and to the initial conditions in the low $\pt$ region,
while at hight $\pt$ it originates for the path-length-dependence of in-medium energy loss.

As discussed in Sec.~\ref{subsubsec:initialMomentum} (Eq.~\ref{eq:HQSpatial} and \ref{eq:HQMomentum}),
the $c\bar{c}$ pairs are initially back-to-back generated before including the nuclear matter effects
such as (anti-)shadowing and in-medium energy loss.
Therefore, the initial relative azimuthal distribution $\rm dN^{c\bar{c}~pair}/d|\Delta\phi|$
can be described by a delta function at $|\Delta\phi|=\pi$,
with the relative azimuthal angle $|\Delta\phi|$ defined as,
\begin{eqnarray}
|\Delta\phi|=&&\left\{ \begin{array}{ll}
|\phi_{\rm c}-\phi_{\rm\bar{c}}| & \textrm{($|\phi_{\rm c}-\phi_{\rm\bar{c}}|<\pi$)} \\
\\
2\pi-|\phi_{\rm c}-\phi_{\rm\bar{c}}| & \textrm{($|\phi_{\rm c}-\phi_{\rm\bar{c}}|>\pi$)}
\end{array} \right.
\end{eqnarray}
where, $\phi_{\rm c}$ ($\phi_{\rm\bar{c}}$) denotes the azimuthal angle of the $c$ ($\bar{c}$) quark.
However, the relative azimuthal distribution will be broadened by a certain amount
after the propagation of the $c$ quarks through the medium
and this behaviour can be inherited by the corresponding heavy-flavour hadrons $\rm dN^{D\bar{D}}/d|\Delta\phi|$.

Note that, in this work, we neglect the hadronic rescatterings in the late stages,
which can slightly reduce the open charmed meson $\raa$ at high $\pt$ and enhance its $\vtwo$ at moderate $\pt$~\cite{CaoPRC15},
but, it is not expected to significantly affect the azimuthal correlation of $D\bar{D}$ pairs~\cite{DDbarCorrelationZPF07}.

%%==============================================
\section{Numerical Results}\label{sec:numercialResults}
In this section, we summarize the results obtained at parton and hadron level, respectively.
The comparisons with available measurements are discussed as well.

%%-----------------------------
%%-----------------------------
\subsection{Results for heavy quarks}\label{subsec:HQSpectra}
\subsubsection{Profile of heavy quark energy loss}\label{subsubsec:HQProfile}
\begin{figure}[!tbp]
\begin{center}
\vspace{-1.0em}
\setlength{\abovecaptionskip}{-0.1mm}
\setlength{\belowcaptionskip}{-1.5em}
\includegraphics[width=.42\textwidth]{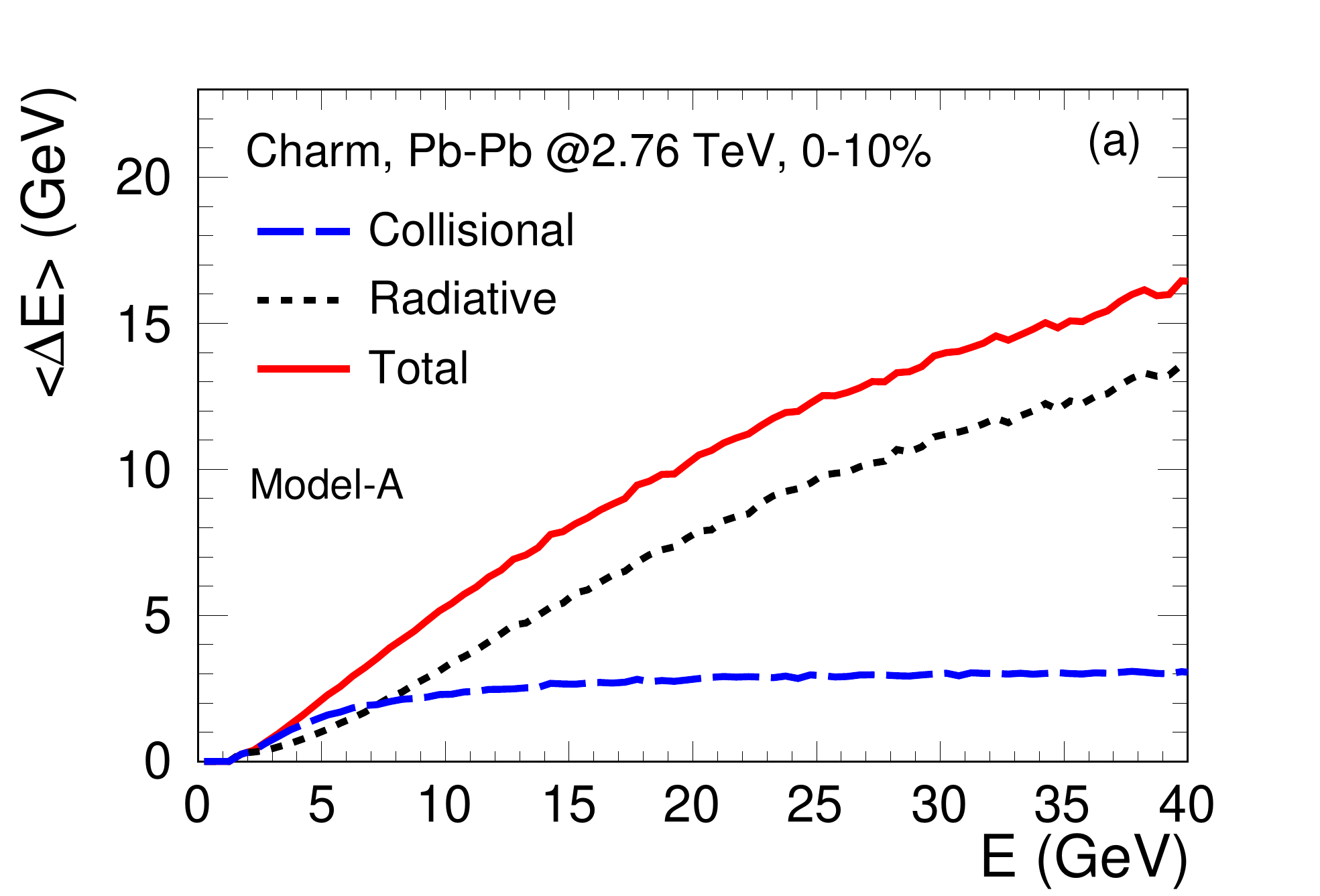}
\includegraphics[width=.42\textwidth]{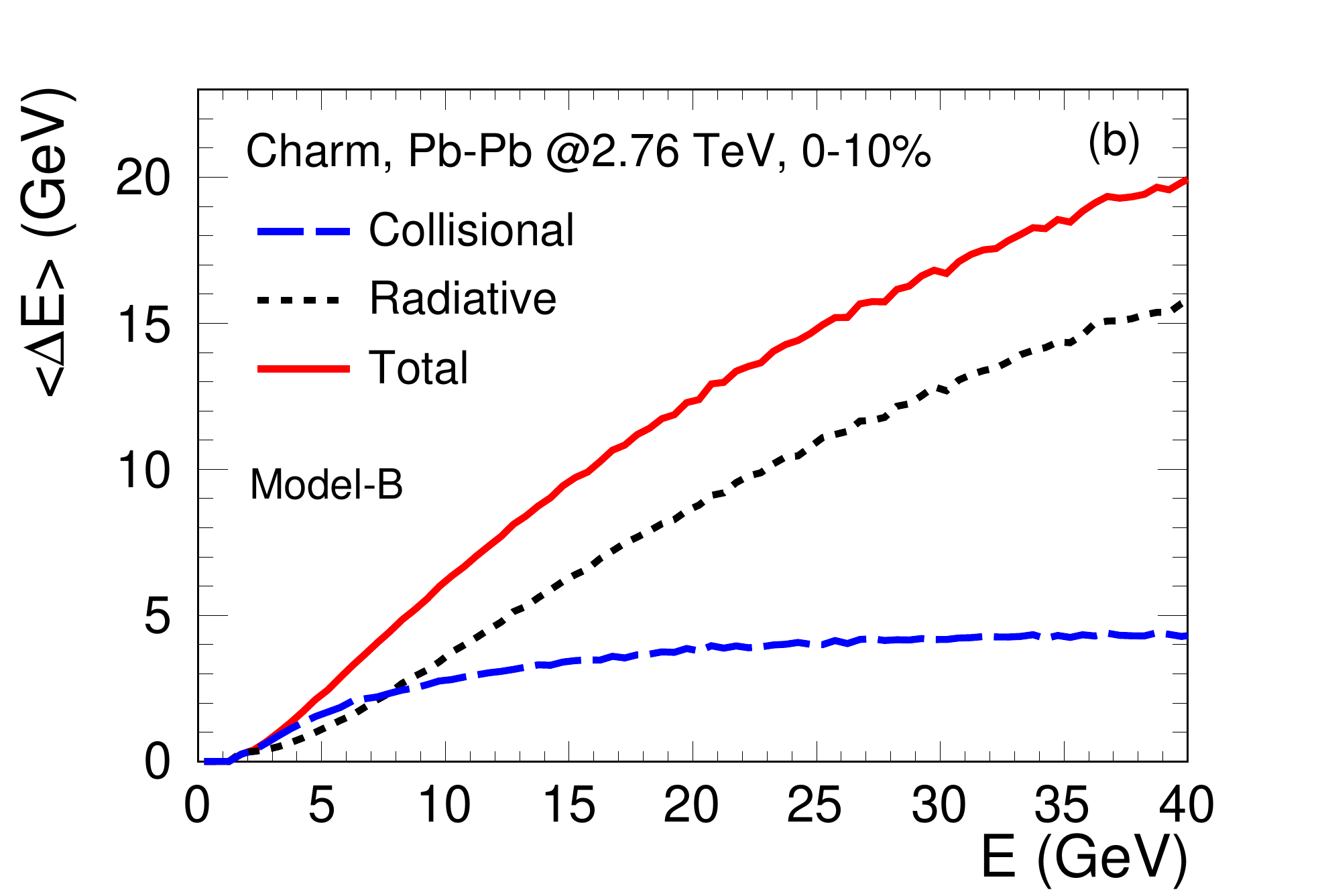}
\caption{(Color online) Energy loss of charm quarks obtained via (a) Model-A
and (b) Model-B: collisional and radiative contributions are shown separately
as long dashed blue and dashed black curves,
respectively, in each panel.
The combined results are shown as solid red curve.}
\label{fig:ElossPbPb2760Cent1ModelAB}
\end{center}
\end{figure}

The average in-medium energy loss of charm quarks is shown in Fig.~\ref{fig:ElossPbPb2760Cent1ModelAB}
as a function of the initial energy,
displaying separately the contributions of collisional (long dashed blue curve) and radiative (dashed black curve) mechanisms.
The results based on Model-A (Eq.~\ref{eq:ModelA}) are shown in the panel-a (upper).
It can be seen that collisional energy loss is significant at low energy,
while radiative energy loss is the dominant mechanism at high energy.
The crossing point between collisional and radiative contributions is around ${\rm E}=7\sim8~{\rm GeV}$.
In the panel-b (bottom) of Fig.~\ref{fig:ElossPbPb2760Cent1ModelAB},
the results based on Model-B (Eq.~\ref{eq:ModelB}) are shown.
The energy loss with Model-B is slightly larger than that with Model-A;
the crossing point between collisional and radiative contributions is around ${\rm E}=8\sim9~{\rm GeV}$.
This is caused by the fact that, (1) the initial transverse momentum spectrum of HQ
is much more harder that of medium constituent, hence, the multiple elastic scatterings are dominated by the drag term;
(2) the drag coefficient with Model-B is larger than Model-A around $T_{c}$ (see Fig.~\ref{fig:DsVsT} together with Eq.~\ref{eq:Gamma2Ds}),
resulting in a stronger interaction strength between the HQ and the incident medium constituents.
Consequently, the HQ lose more energy with Model-B approach.

\subsubsection{Correlation in relative azimuthal angle}\label{subsubsec:DeltaPhi}
\begin{figure}[!tbp]
\begin{center}
\vspace{-1.0em}
\setlength{\abovecaptionskip}{-0.1mm}
\setlength{\belowcaptionskip}{-1.5em}
\includegraphics[width=.42\textwidth]{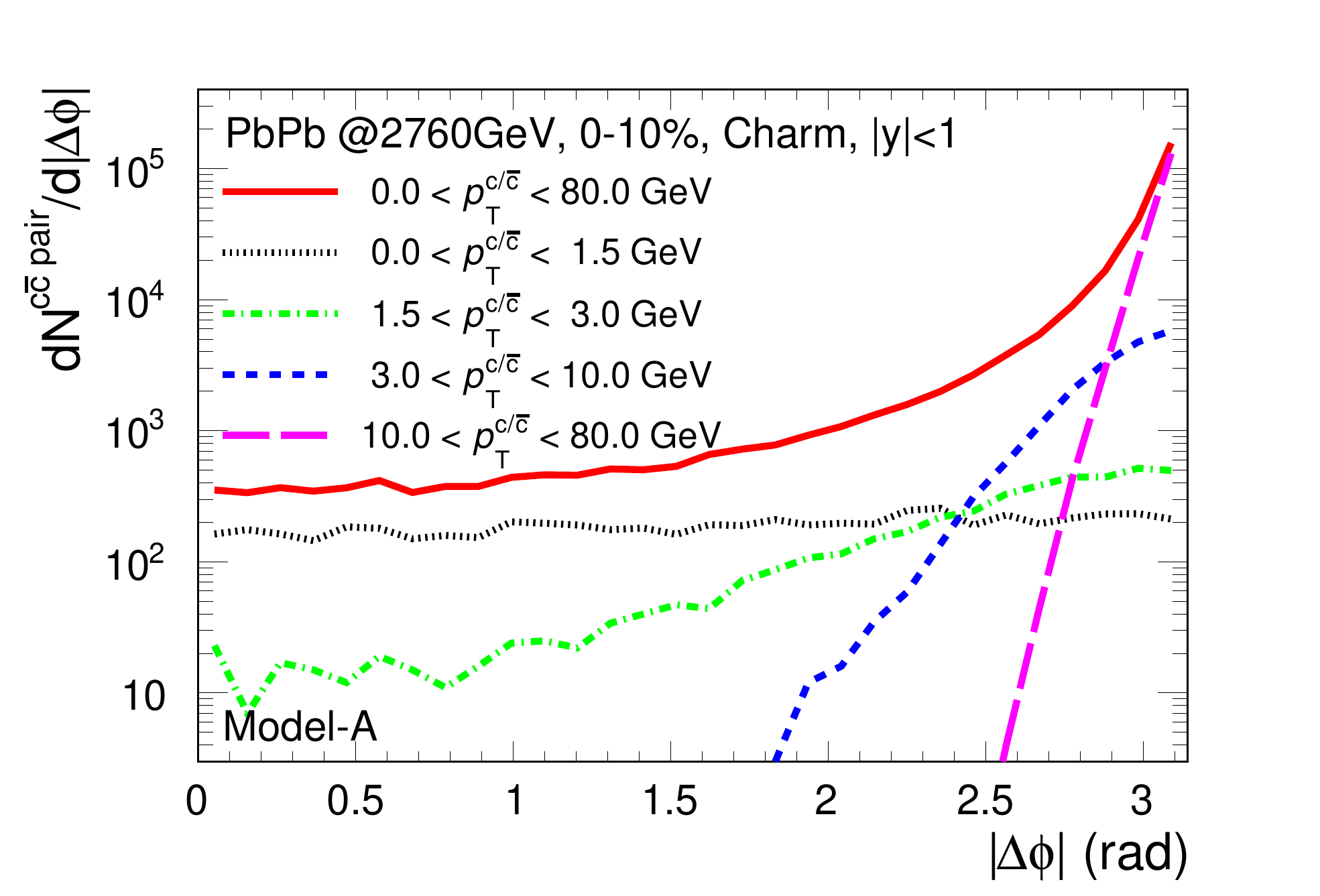}
\caption{(Color online) Relative azimuthal angle between $c$ and $\bar{c}$ quarks
with Model-A approach in central ($0-10\%$) Pb--Pb collisions at $\snn=2.76~{\rm TeV}$.
$c\bar{c}$ pairs were generated back-to-back ($|\Delta\phi|=\pi$).
The in-medium energy loss effects are included, while the nuclear (anti-)shadowing is neglected.
The curves in different styles indicate the results within different $\pt$ intervals (see legend for details).}
\label{fig:CharmDeltaPhi}
\end{center}
\end{figure}

Figure~\ref{fig:CharmDeltaPhi} shows the (raw) yields of the initially back-to-back generated $c\bar{c}$ pairs,
after propagating through the medium,
with Model-A approach (Eq.~\ref{eq:ModelA}) in central ($0-10\%$) Pb--Pb collisions at $\snn=2.76~{\rm TeV}$.
The different dashed curves refer to different intervals of $c$ quark $\pt$.
Note that the (anti-)shadowing effects are not included in this plot.
It is clearly observed that an almost flat $|\Delta\phi|$ distribution
with the lower initial transverse momentum interval $\pt^{\rm c/\bar{c}}<1.5~{\rm GeV}$ (dotted black curve),
indicating the initially back-to-back properties are largely washed out
throughout the interactions with the surrounding medium constituents.
The broadening of the distributions tends to decrease with increasing $\pt^{\rm c/\bar{c}}$,
reflecting a larger survival probability, for the initial back-to-back correlation, towards high $\pt^{\rm c/\bar{c}}$.
The results in the whole momentum range $\pt^{\rm c/\bar{c}}<80~{\rm GeV}$ are shown as solid red curve.
Similar conclusions are obtained with Model-B (Eq.~\ref{eq:ModelB}),
which the broadening is more pronounced as compared to Model-A.
As explained above, the larger initial drag term cases, the stronger interactions in Model-B,
which are more powerful to pull the $c\bar{c}$ pairs from high momentum to low momentum.

\begin{figure}[!htbp]
\begin{center}
\vspace{-1.0em}
\setlength{\abovecaptionskip}{-0.1mm}
\setlength{\belowcaptionskip}{-1.5em}
\includegraphics[width=.42\textwidth]{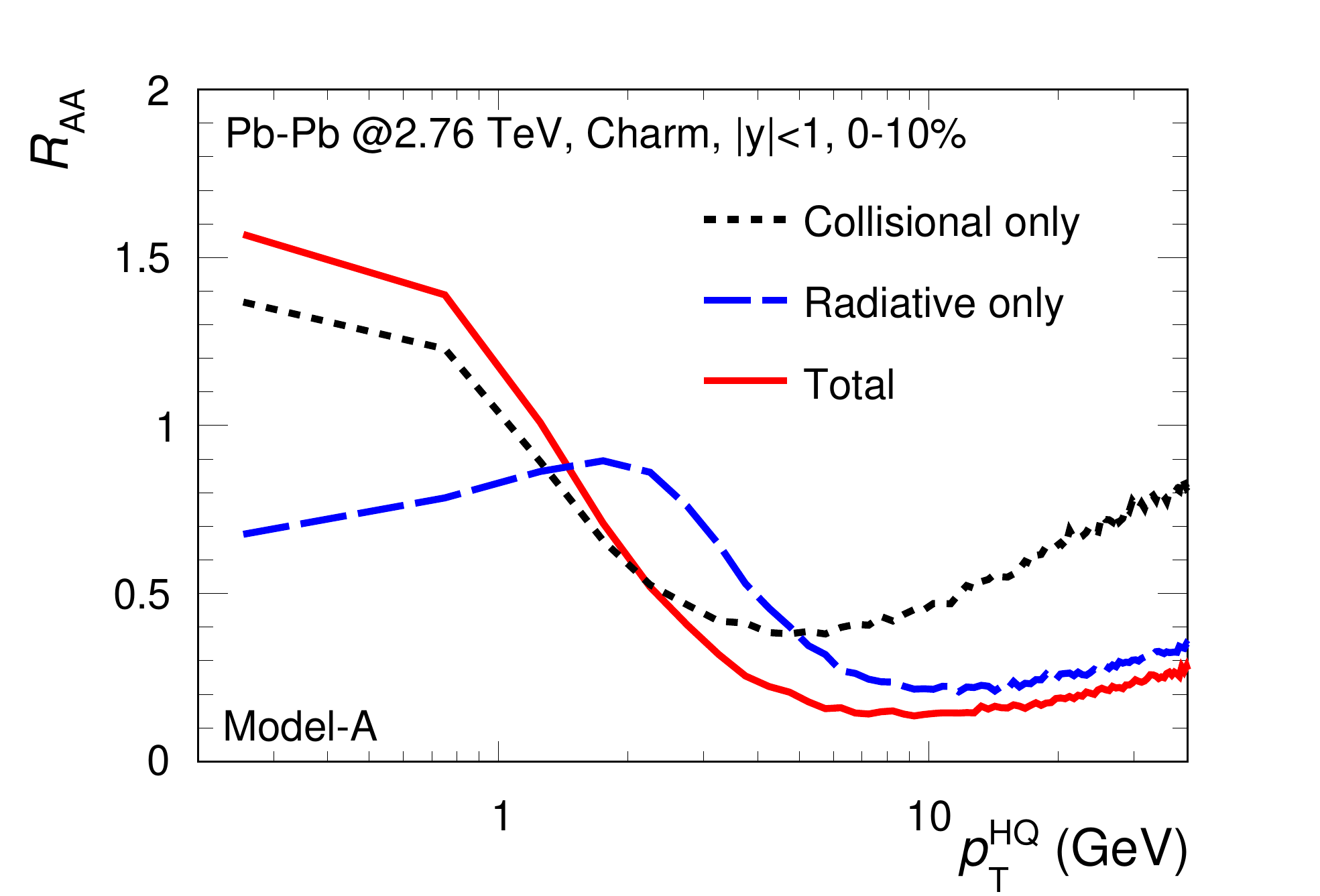}
\caption{(Color online) Nuclear modification factor $\raa$ for charm quark obtained
by considering separately the collisional (dashed black curve) and radiative (long dashed blue curve) energy loss mechanisms,
with Model-A approach in semi-central ($30-50\%$) Pb--Pb collisions at $\snn=2.76~{\rm TeV}$.
The results including both collisional and radiative contributions (solid red curve) are shown as well.}
\label{fig:RAAElssPbPb2760Cent1ModelA}
\end{center}
\end{figure}

\begin{figure*}[!htbp]
\begin{center}
\vspace{-1.0em}
\setlength{\abovecaptionskip}{-0.1mm}
\setlength{\belowcaptionskip}{-1.5em}
\includegraphics[width=.35\textwidth]{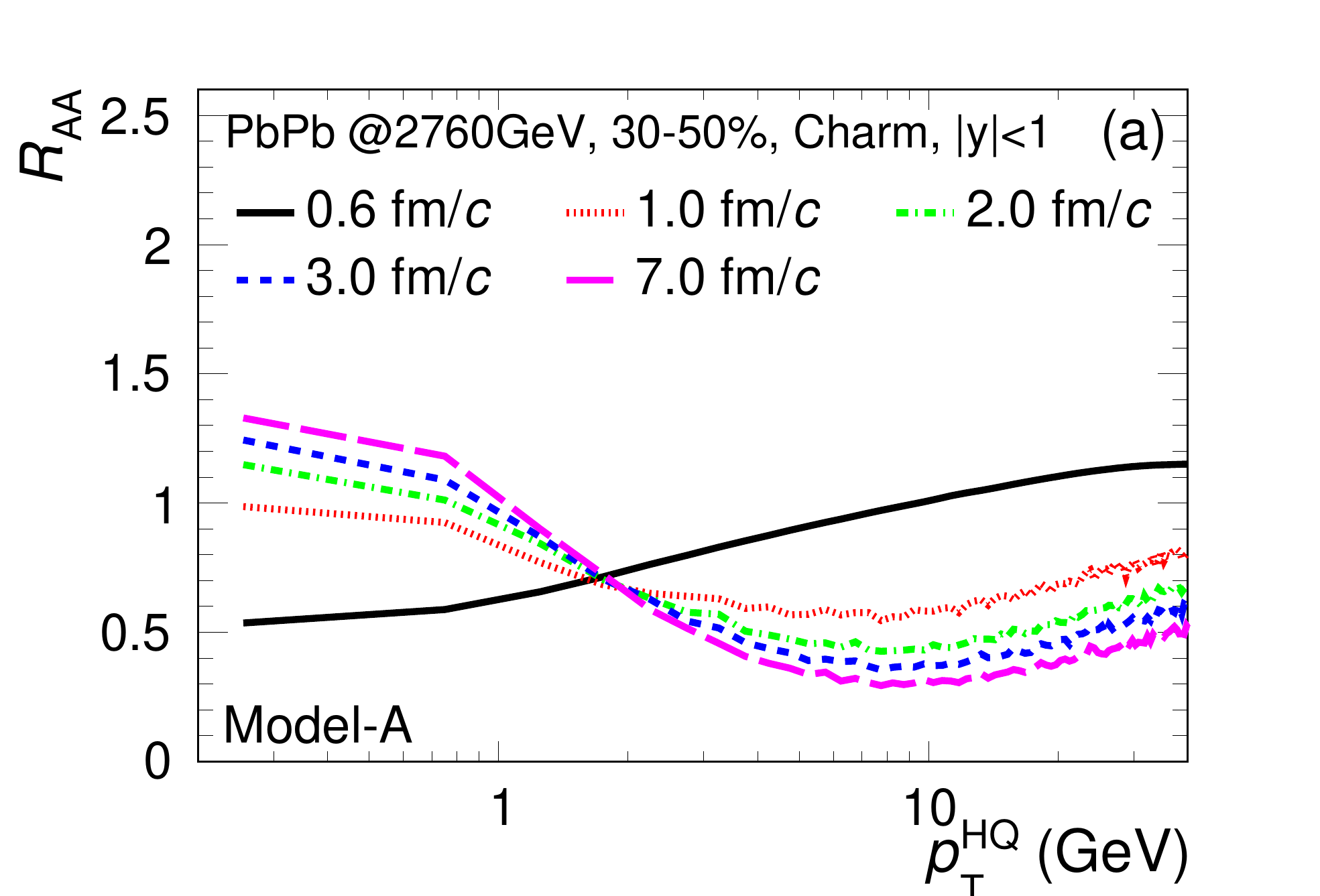}
\hspace{-2.40em}
\vspace{-0.10em}
\includegraphics[width=.35\textwidth]{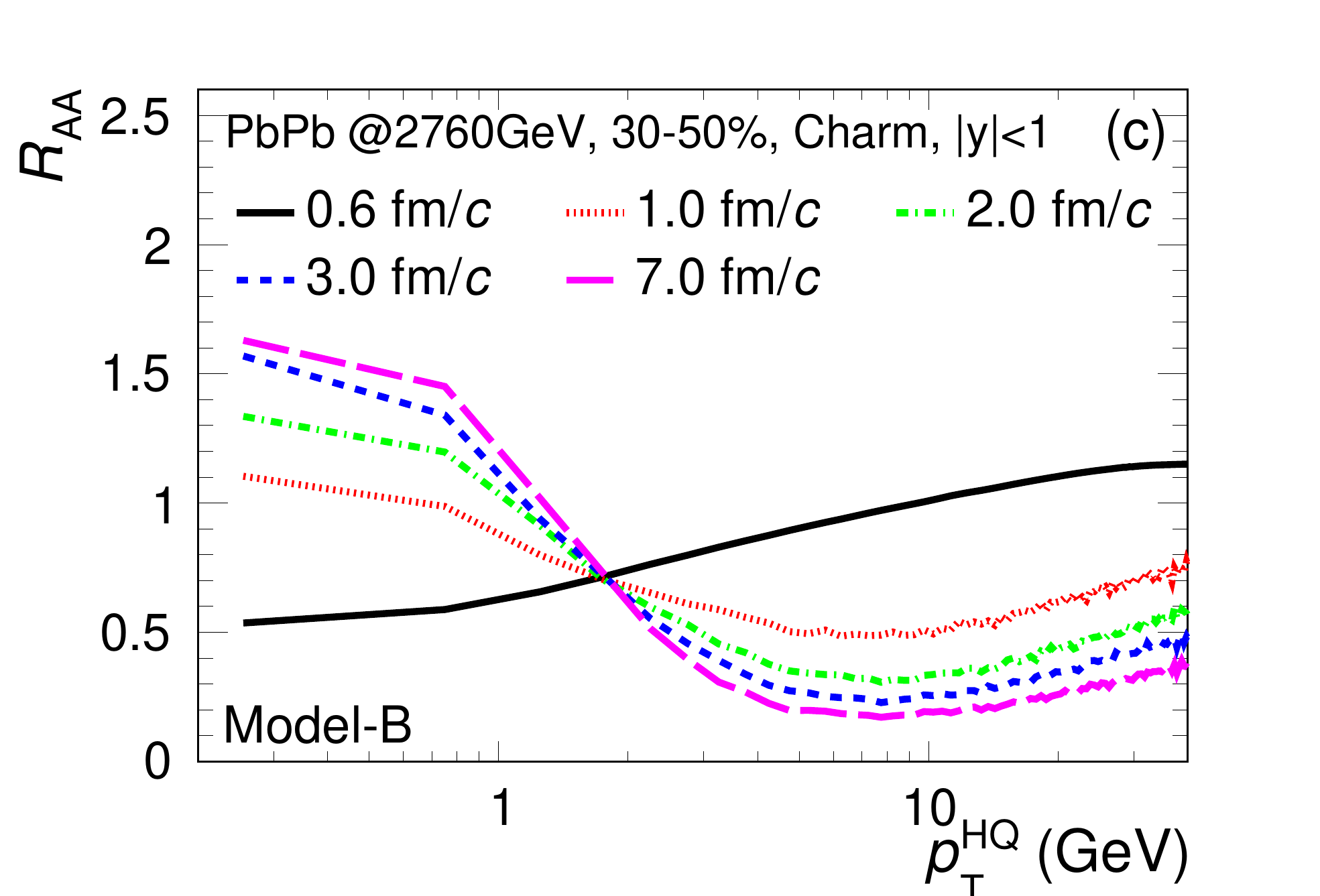}
\hspace{-2.40em}
\vspace{-0.10em}
\includegraphics[width=.35\textwidth]{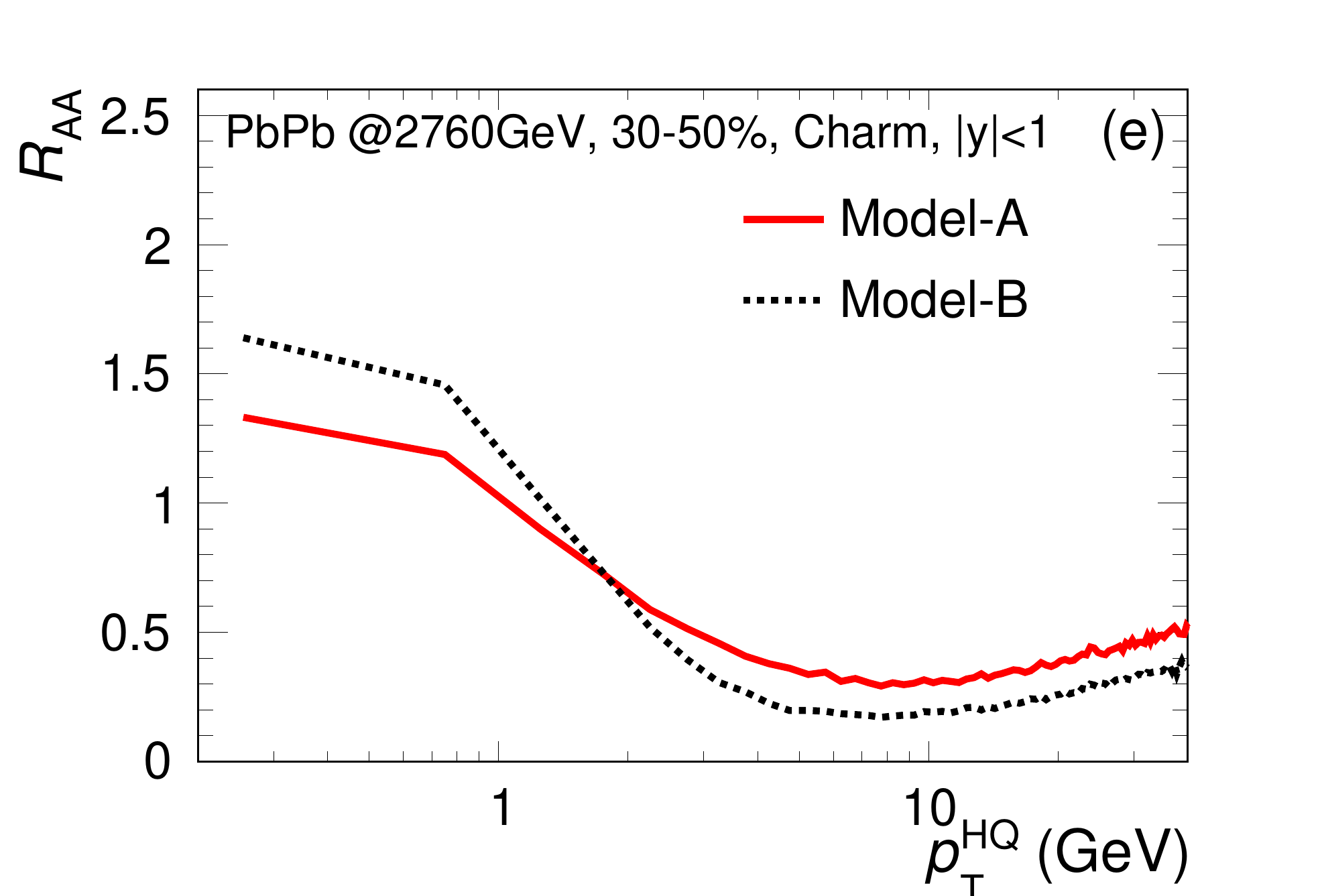}
\vspace{-0.10em}
\includegraphics[width=.35\textwidth]{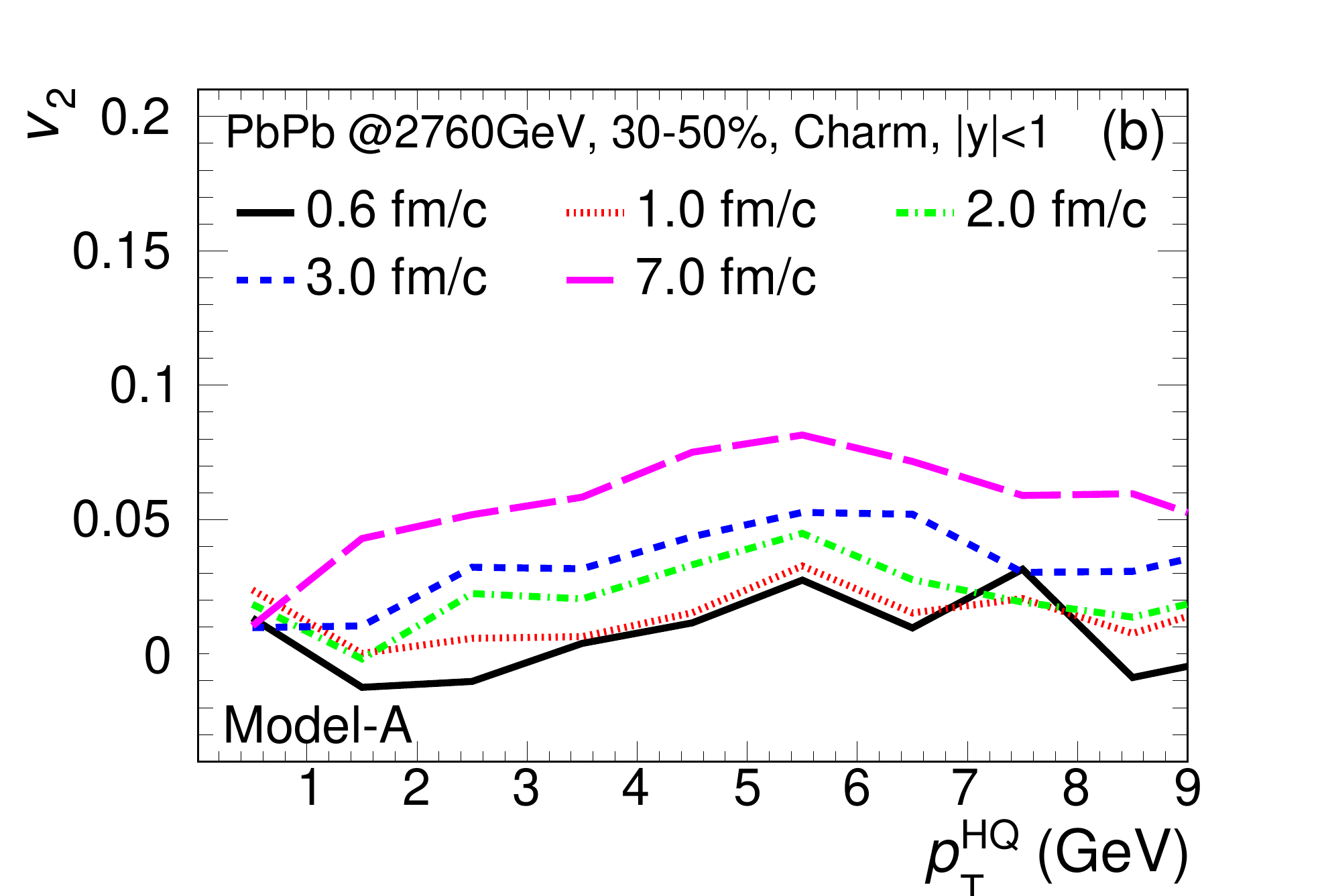}
\hspace{-2.40em}
\includegraphics[width=.35\textwidth]{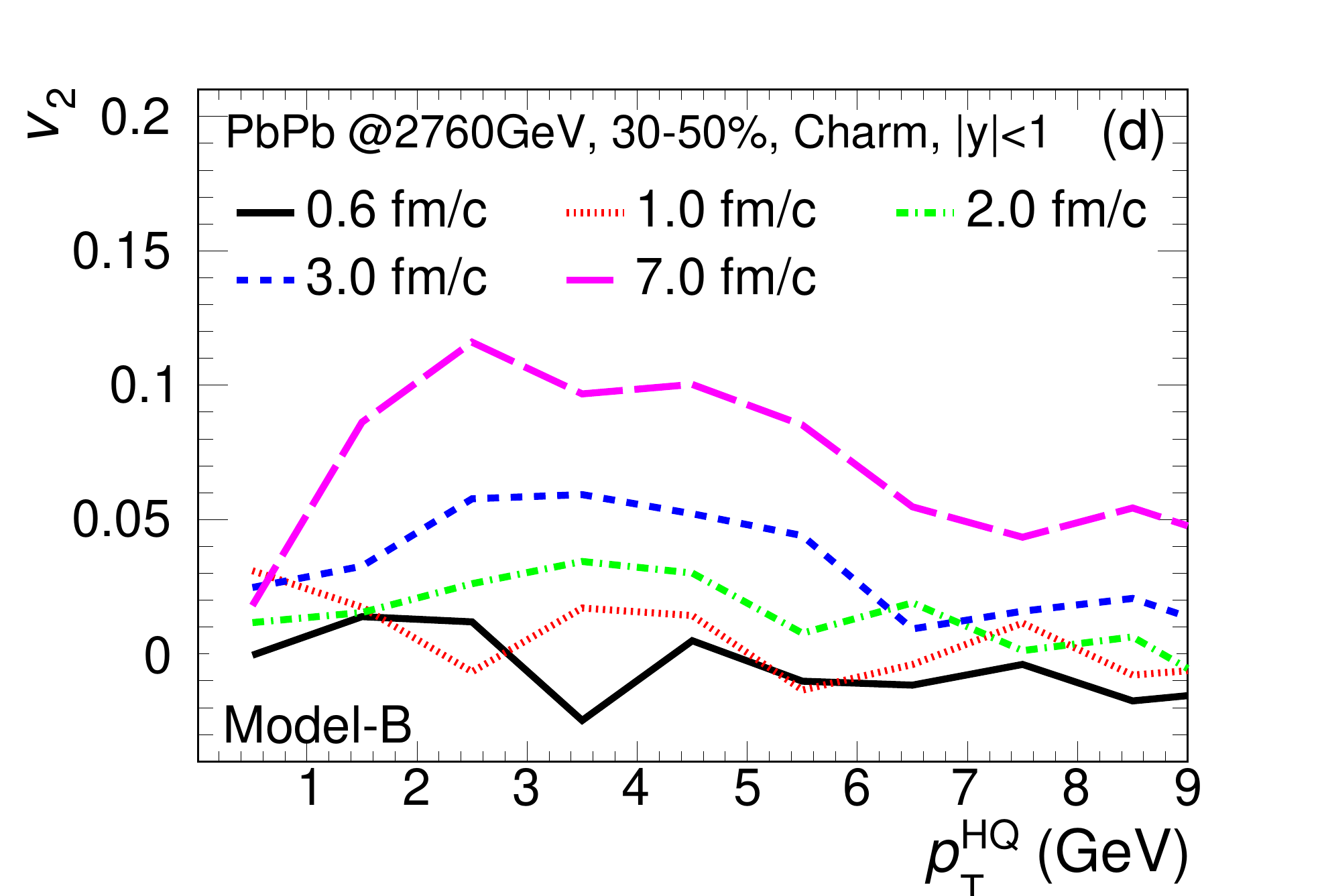}
\hspace{-2.40em}
\includegraphics[width=.35\textwidth]{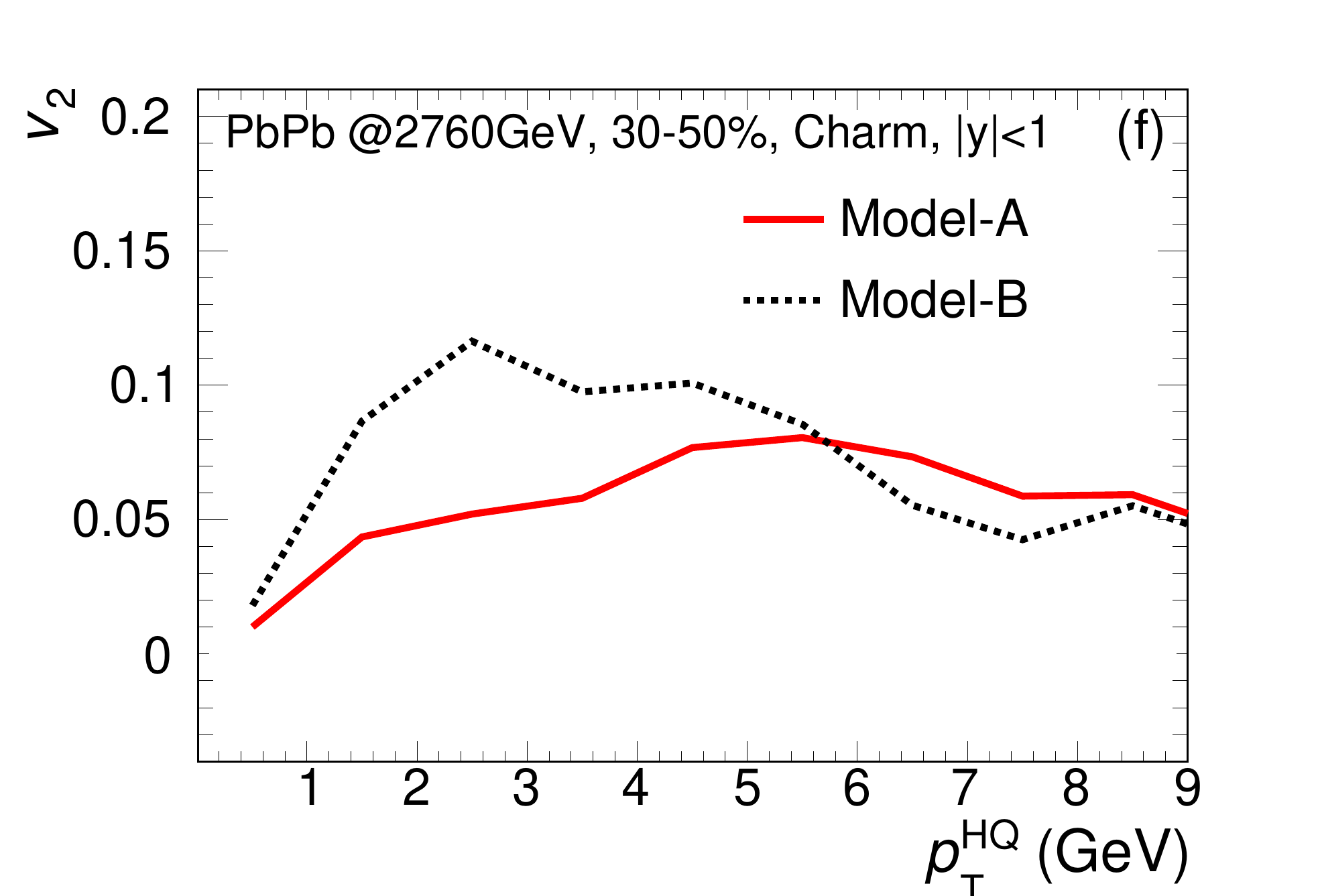}
\caption{(Color online) Upper: charm quark $\raa$ based on (a) Model-A and (c) Model-B
at different times during the hydrodynamical evolution of the medium (see legend for details)
in semi-central ($30-50\%$) Pb--Pb collisions at $\snn=2.76~{\rm TeV}$.
The (e) comparison between Model-A and Model-B results is shown in the upper right panel.
Bottom: same as the above panels but for (b, d and f) charm quark $\vtwo$.}
\label{fig:RAAPbPb2760Cent2ModelAB}
\end{center}
\end{figure*}

\subsubsection{Nuclear modification factor and elliptic flow}\label{subsubsec:HQRAAV2}
Figure~\ref{fig:RAAElssPbPb2760Cent1ModelA} shows the nuclear modification factor $\raa$
of charm quarks obtained by considering only the collisional (dashed black curve) and radiative (long dashed blue curve) energy loss mechanisms,
with Model-A approach in semi-central ($30-50\%$) Pb--Pb collisions at $\snn=2.76~{\rm TeV}$.
The result including the two effects (solid red curve) is closer to the one with only collisional energy loss at low $\pt$,
while it is closer to the radiative curve at high $\pt$.
This is consistent with what we discussed above in Fig.~\ref{fig:ElossPbPb2760Cent1ModelAB},
which indicated that collisional energy loss is the dominant contribution at low momentum/energy of the charm quark,
while at high $\pt$ radiative processes dominate.
Similar behaviour can be found for the results with Model-B.

In Fig.~\ref{fig:RAAPbPb2760Cent2ModelAB} the charm quark $\raa$ (panel-a, c and e) and $\vtwo$ (panel-b, d and f)
are calculated, including both the collisional and radiative energy loss mechanisms,
at various times during the hydrodynamic evolution of the medium.
At the starting time $\tau_{0}=0.6~{\rm fm/{\it c}}$,
$\raa$ (panel-a; upper left) is determined by the (anti-)shadowing effect (see also Fig.~\ref{fig:EPS09Comparison}),
and the corresponding $\vtwo$ (panel-b; bottom left) is close to zero (even though the statistics is limited)
for both Model-A (panel-a and b; left two panels) and Model-B (panel-c and d; middle two panels).
During the evolution up to $\tau=7~{\rm fm/{\it c}}$, $\raa$ rises in the low $\pt$ region, while it drops at high $\pt$,
because the initial drag term is dominant with respect to the diffusion term, as discussed in Fig.~\ref{fig:ElossPbPb2760Cent1ModelAB}.
The variation between neighboring time-windows
exhibits a decreasing trend, and the modification of $\raa$ is less pronounced after $\tau=7~{\rm fm/{\it c}}$.
This may be induced by the late stage collective flow,
which allows to transport the HQ from low momentum to high momentum,
as mentioned in Ref.~\cite{LowPtRAARHICAndrew14}.
This means that the competition between the initial drag and the subsequent collective flow tends to restrict the time dependence of $\raa$.
This can be confirmed by studying the time evolution of $\vtwo$, as displayed in the panel-b (bottom left).
It clearly shows that $\vtwo$ develops mostly at late times, reaching the maximum at $\tau\sim7~{\rm fm/{\it c}}$.
The results of Model-A and Model-B are qualitatively similar.
A quantitative comparision of $\raa$ and $\vtwo$ at $\tau\sim7~{\rm fm/{\it c}}$
is shown in the bottom panel of Fig.~\ref{fig:RAAPbPb2760Cent2ModelAB}.
$\raa$ (panel-e; upper right) is enhanced (suppressed) at low (high) $\pt$ in Model-B compared to Model-A,
while $\vtwo$ (panel-f; bottom right) is significantly enhanced at intermediate $\pt$ ($2\lesssim\pt\lesssim4~{\rm GeV}$).

%%-----------------------------
%%-----------------------------
\subsection{Results for open charmed mesons}\label{subsec:DSpectra}
\subsubsection{Production cross section}\label{subsubsec:DProd}
\begin{figure}[!htbp]
\begin{center}
\vspace{-1.0em}
\setlength{\abovecaptionskip}{-0.1mm}
\setlength{\belowcaptionskip}{-1.0em}
\includegraphics[width=.42\textwidth]{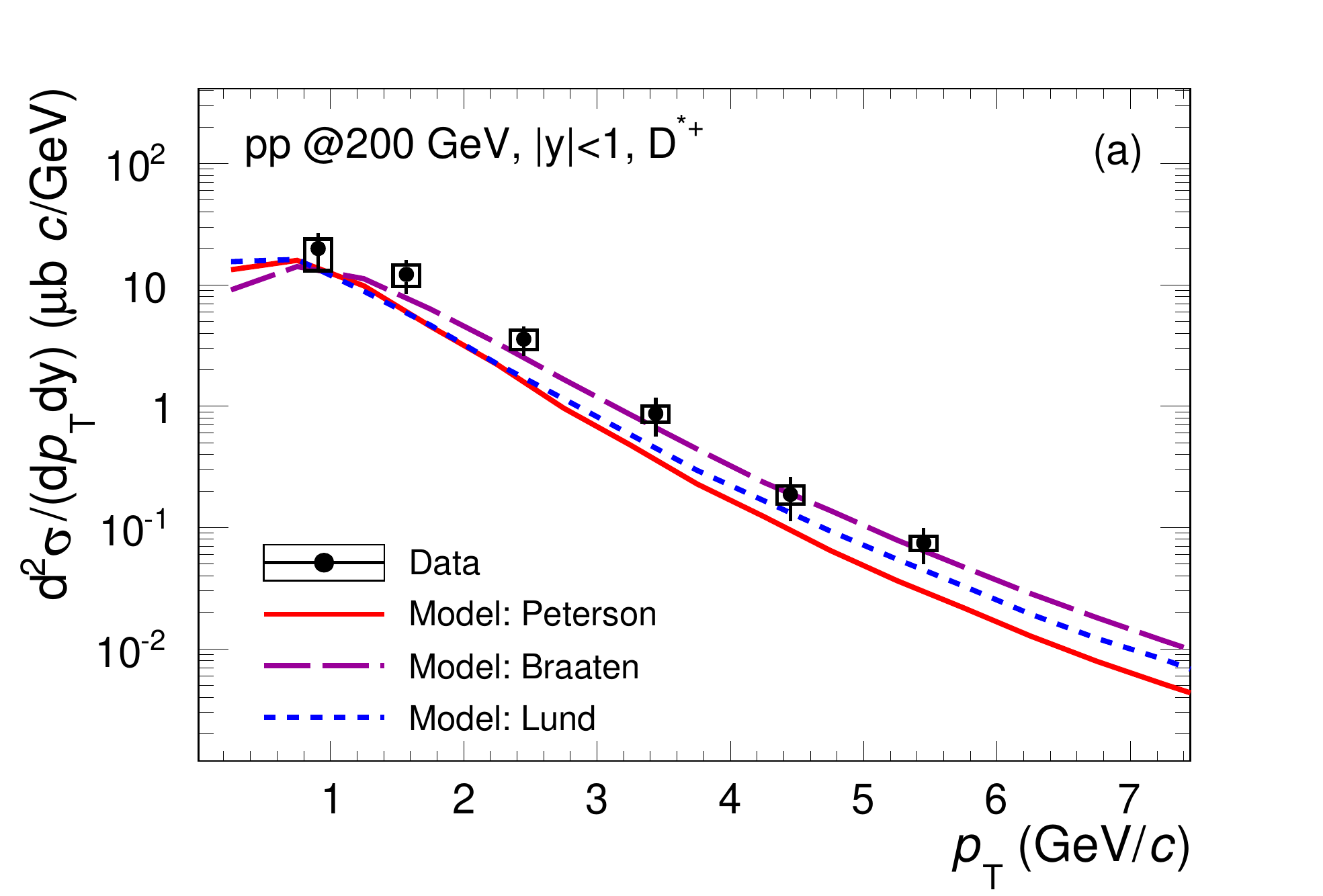}
\includegraphics[width=.42\textwidth]{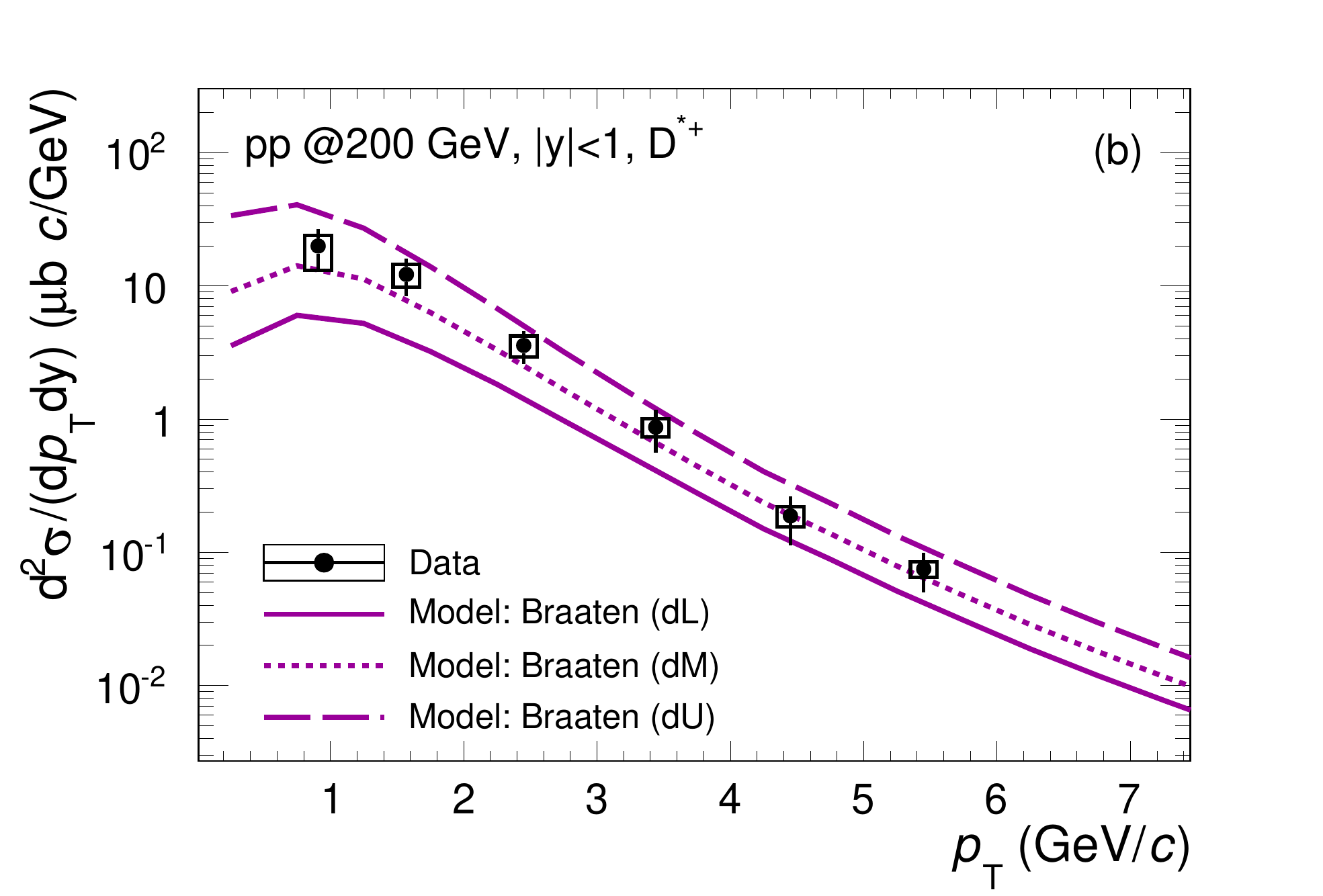}
\caption{(Color online) Upper (a): comparison of the central values of the
$\pt$-differential production cross section for $D^{*+}$ mesons at mid-rapidity ($|y|<1$) in pp collisions at $\s=200~{\rm GeV}$
obtained using different fragmentation models:
Peterson (solid red curve), Braaten (long dashed purple curve) and Lund-Pythia (dashed blue curve).
Bottom (b): $\pt$-differential production cross section for $D^{*+}$ mesons with Braaten fragmentation function,
including the theoretical uncertainties.
Experimental data arederived from Ref.~\cite{RHICPP12}.}
\label{fig:Sigma200_Dstar+}
\end{center}
\end{figure}

The $\pt$-differential production cross section of $D^{*+}$ mesons,
in the range $|y|<1$ and $\pt\lesssim7~{\rm GeV}$ for pp collisions at $\s=200~{\rm GeV}$
is shown in the panel-a (upper) of Fig.~\ref{fig:Sigma200_Dstar+}.
The curves in different styles are the model calculations, for the central values (Eq.~\ref{eq:FONLLCentForm}),
relying on various fragmentation functions: Lund-Pythia (dashed blue curve), Peterson (solid red curve) and Braaten (long dashed purple curve).
As discussed in Sec.~\ref{subsec:hybridModel_Had}, the spectrum with the Braaten fragmentation function
is found to be harder than that with with Peterson function.
The experimental data (black boxes) are shown as well for comparison,
which is obtained by scaling the $c\bar{c}$ production cross section
reported by the STAR experiment~\cite{RHICPP12} by the factor $f(c\rightarrow D^{\ast +})=0.230$.
Within the experimential uncertainties,
the measured $\pt$ differential cross section is better described by the central prediction with the Braaten fragmentation function.

However, one should consider simultaneously the theoretical uncertainties on the
FONLL calculation due to the perturbative QCD scales and the heavy quark mass (Eq.~\ref{eq:FONLLErr}).
The resulting $D^{*+}$ cross section is displayed in the panel-b (bottom) of Fig.~\ref{fig:Sigma200_Dstar+}.
The curves in solid, dotted and long dashed curves denote the lower, central and upper bands of the model calculations, respectively.
Within both the theoretical and the experimental uncertainties,
the results based on the different fragmentation functions can provide a good description of
the measured D meson corss section in the whole $\pt$ region~\cite{RHICPP12}.
Same conclusions can be drawn for pp collisions at $\s=~2.76$ and 7 ${\rm TeV}$.
Hereafter, all the results are based on the $Braaten$ fragmentation function.

\begin{figure}[!htbp]
\begin{center}
\vspace{-1.0em}
\setlength{\abovecaptionskip}{-0.1mm}
\setlength{\belowcaptionskip}{-1.5em}
\includegraphics[width=.42\textwidth]{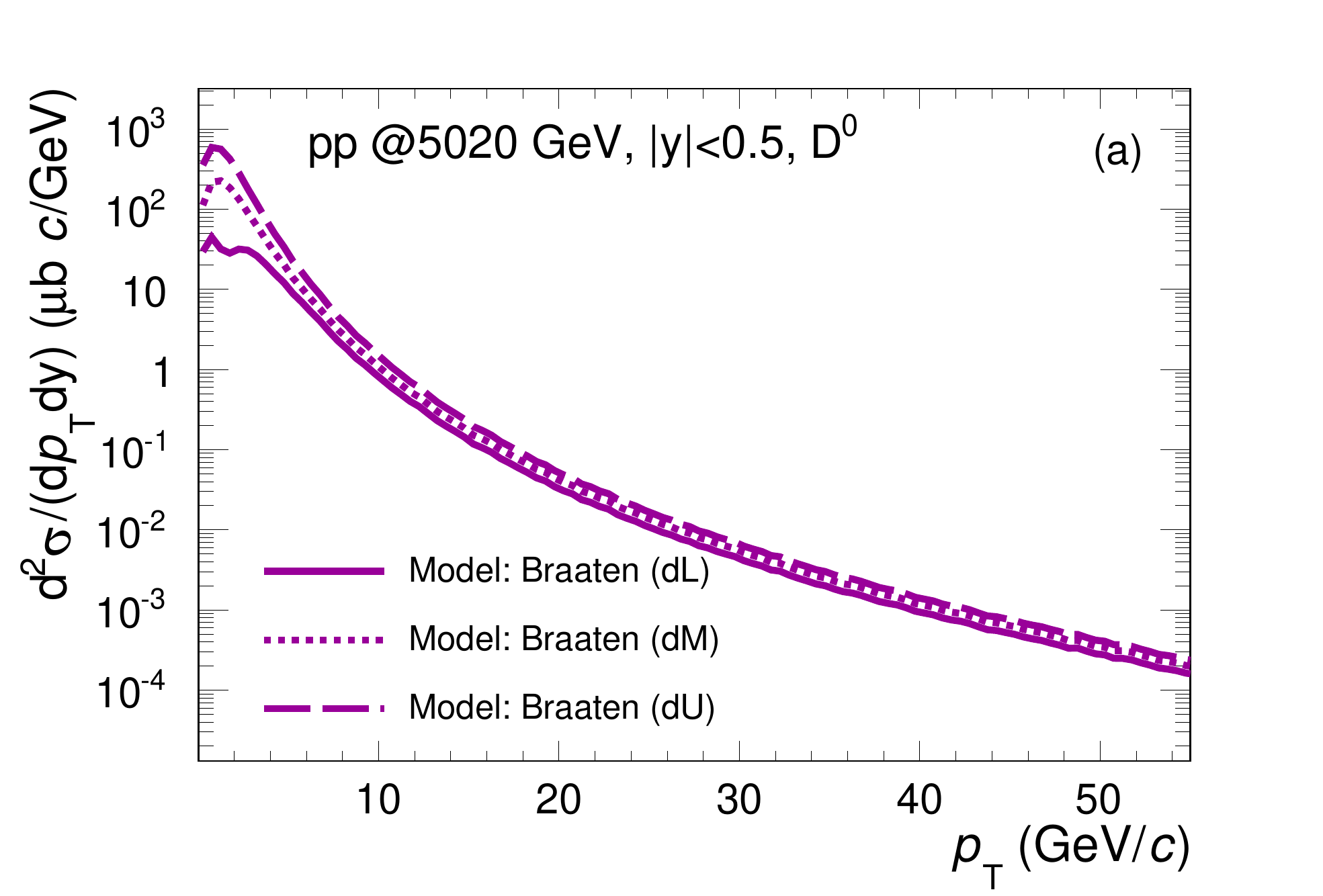}
\includegraphics[width=.42\textwidth]{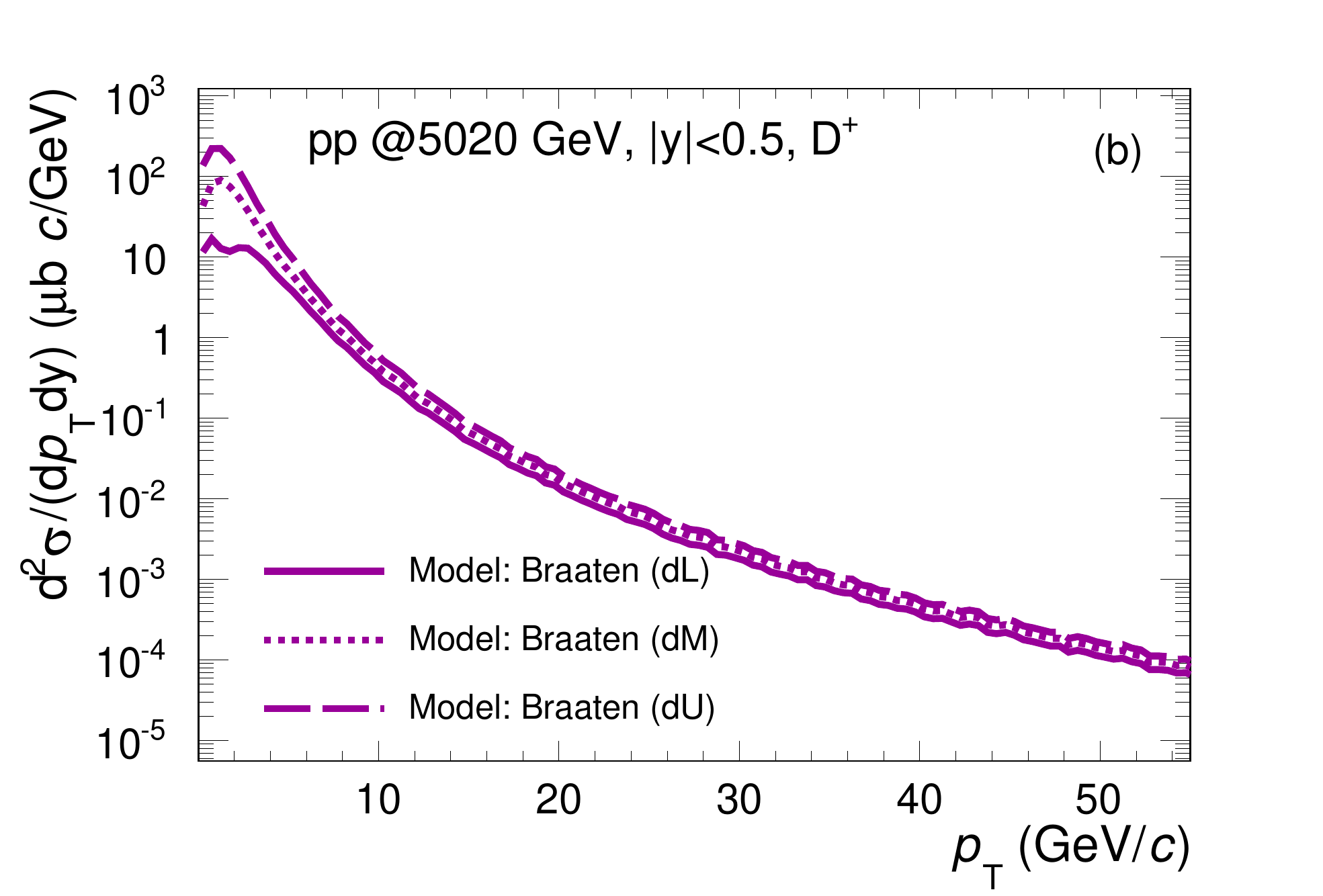}
\includegraphics[width=.42\textwidth]{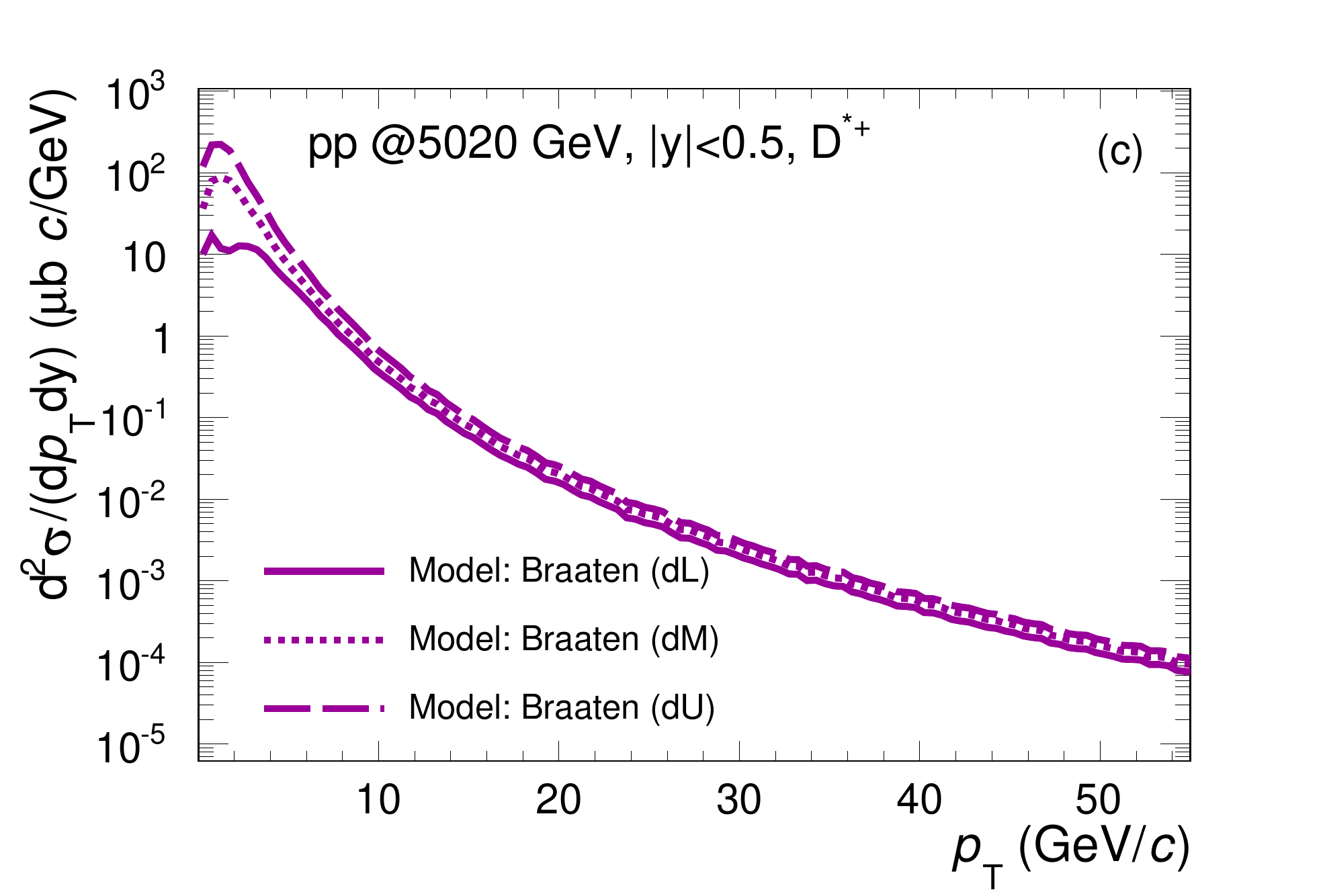}
\caption{(Color online) $\pt$-differential production cross section for (a) $D^{0}$, (b) $D^{+}$ and
(c) $D^{*+}$ mesons at mid-rapidity in pp collisions at $\s=5.02~{\rm TeV}$,
obtained using the Braaten fragmentation function.}
\label{fig:Sigma5020_BK}
\end{center}
\end{figure}
In Fig.~\ref{fig:Sigma5020_BK} the results of the calculations of the $\pt$-differential production cross section,
of $D^{0}$ (panel-a; upper), $D^{+}$ (panel-b; middle) and $D^{*+}$ (panel-c; bottom) mesons
at mid-rapidity ($|y|<0.5$) in pp collisions at $\s=5.02~{\rm TeV}$ are shown.
They can be compared with the upcoming measurements at the LHC.

\subsubsection{Correlation in relative azimuthal angle}\label{subsubsec:DDeltaPhi}
\begin{figure}[!htbp]
\begin{center}
\vspace{-1.0em}
\setlength{\abovecaptionskip}{-0.1mm}
\setlength{\belowcaptionskip}{-1.5em}
\includegraphics[width=.42\textwidth]{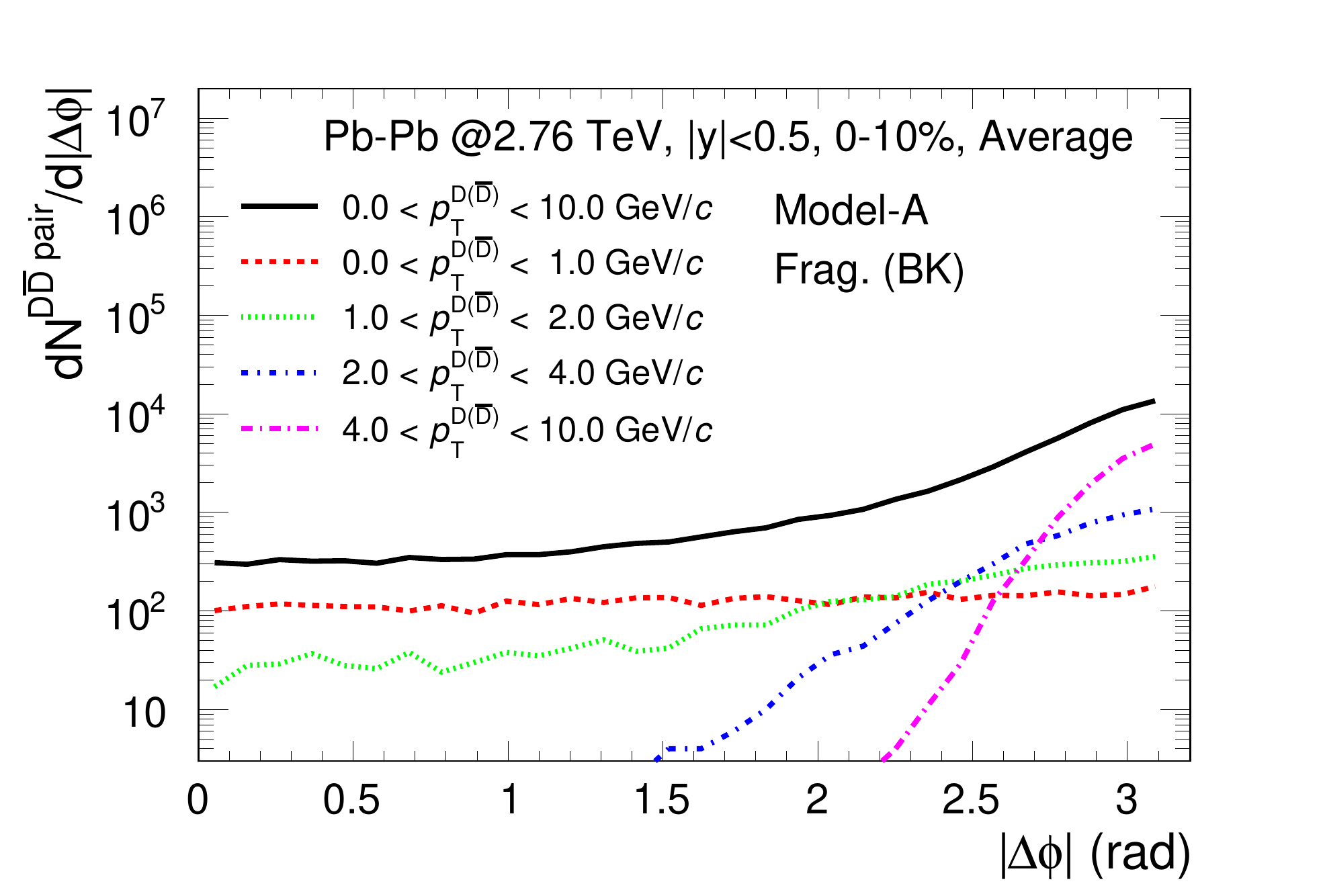}
\caption{(Color online) Relative azimuthal angle dependence for $D\bar{D}$ pairs generated by the initially back-to-back $c\bar{c}$ pairs,
with Model-A approach in central ($0-10\%$) Pb--Pb collisions at $\snn=2.76~{\rm TeV}$.
The results in different $\pt$ intervals are shown as curves with different styles (see legend for details).
The hadronization is carried on considering only quark fragmentation with the Braaten fragmentation functin.}
\label{fig:DmesonDeltaPhi}
\end{center}
\end{figure}

Figure~\ref{fig:DmesonDeltaPhi} shows the (raw) yields of $D\bar{D}$ pairs,
produced from the initially back-to-back generated $c\bar{c}$ pairs,
as function of the relative azimuthal angle $|\Delta\phi|$,
obtained with the Model-A approach for central ($0-10\%$) Pb--Pb collisions at $\snn=2.76~{\rm TeV}$.
Note that: (1) both the in-medium energy loss and nuclear (anti-)shadowing effects are included;
(2) the heavy-light coalescence (Sec.~\ref{subsubsec:CoalMod}) is not considered and all D mesons are produced via fragmentation.
The broadening observed at hadron level is similar to the one observed at quark level (Fig.~\ref{fig:CharmDeltaPhi}).
A significant broadening is observed at low transverse momentum and it decreases with increasing transverse momentum.
A qualitatively similar trend is found with Model-B, but with a more pronounced broadening.

\subsubsection{Nuclear modification factor and elliptic flow}\label{subsubsec:DRAAV2}
\begin{figure}[!tbp]
\begin{center}
\vspace{-1.0em}
\setlength{\abovecaptionskip}{-0.1mm}
\setlength{\belowcaptionskip}{-1.5em}
\includegraphics[width=.42\textwidth]{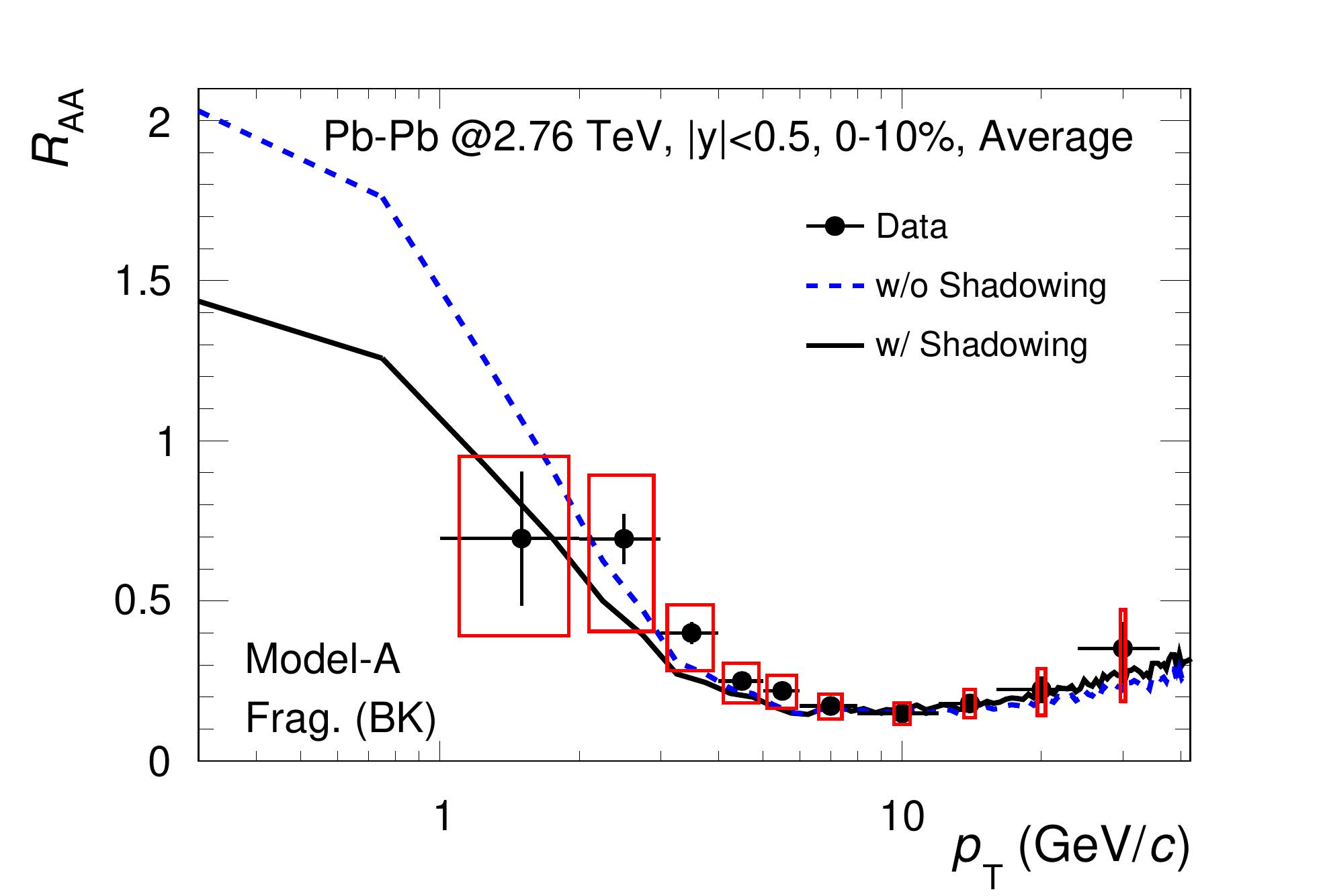}
\caption{(Color online) Comparison of the nuclear modification factor $\raa$ of non-strange D mesons ($D^{0}$, $D^{+}$ and $D^{*+}$)
with and without including the nuclear shadowing effect in central ($0-10\%$) Pb--Pb collisions at $\snn=2.76~{\rm TeV}$.
Experimental data taken from Ref.~\cite{ALICEDesonPbPb2760RAA}.
The used fragmentation model is Braaten.}
\label{fig:RAA2760_Shad_Avg_c0-c10_ModelA}
\end{center}
\end{figure}

Figure~\ref{fig:RAA2760_Shad_Avg_c0-c10_ModelA} shows the nuclear modification factor $\raa$
of non-strange D mesons ($D^{0}$, $D^{+}$ and $D^{*+}$) as a function of $\pt$ in central ($0-10\%$) Pb--Pb collisions at $\snn=2.76~{\rm TeV}$.
The heavy-light coalescence (Sec.~\ref{subsubsec:CoalMod}) is not included.
The solid black and dashed blue curves display the central values of the calculations with and without including the nuclear shadowing effect, respectively.
It is found that the relative ratio (=``with/without'' nPDFs) between them is about 0.7 (1.1) at $\pt$=1 (10) GeV.
This behaviour is consistent with the observation made for charm quarks when discussing the results shown in Fig.~\ref{fig:EPS09Comparison}.
When comparing the model calculations with the corresponding measurements,
we can see that, within the experimental uncertainties,
the data can be described by the results including shadowing in the whole $\pt$ region,
even without taking into account the heavy-light coalescence effect.
The measurements with higher precision are needed to quantify this effect, in particular in the intermediate $\pt$ region.

\begin{figure}[!tbp]
\begin{center}
\vspace{-1.0em}
\includegraphics[width=.42\textwidth]{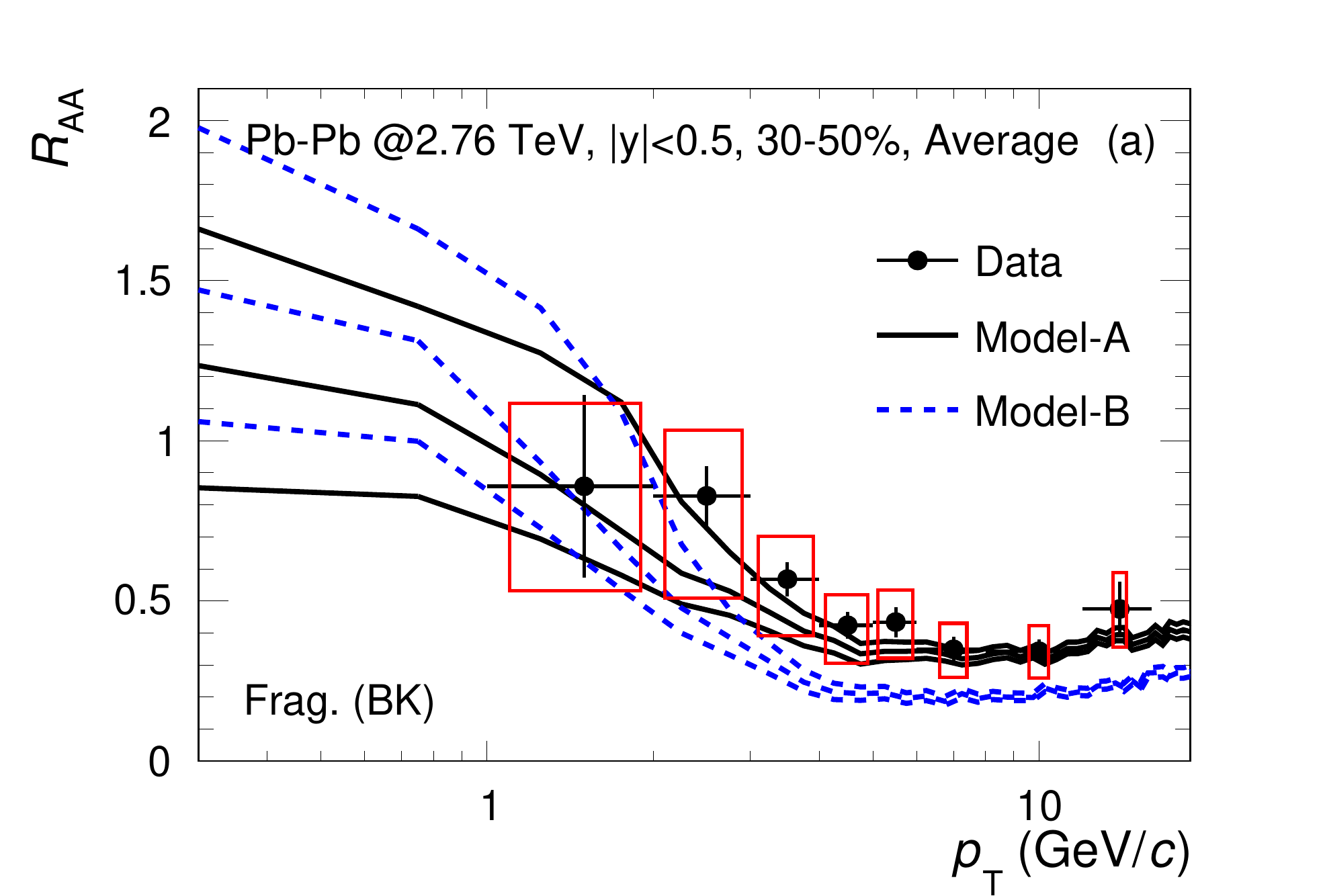}
\includegraphics[width=.42\textwidth]{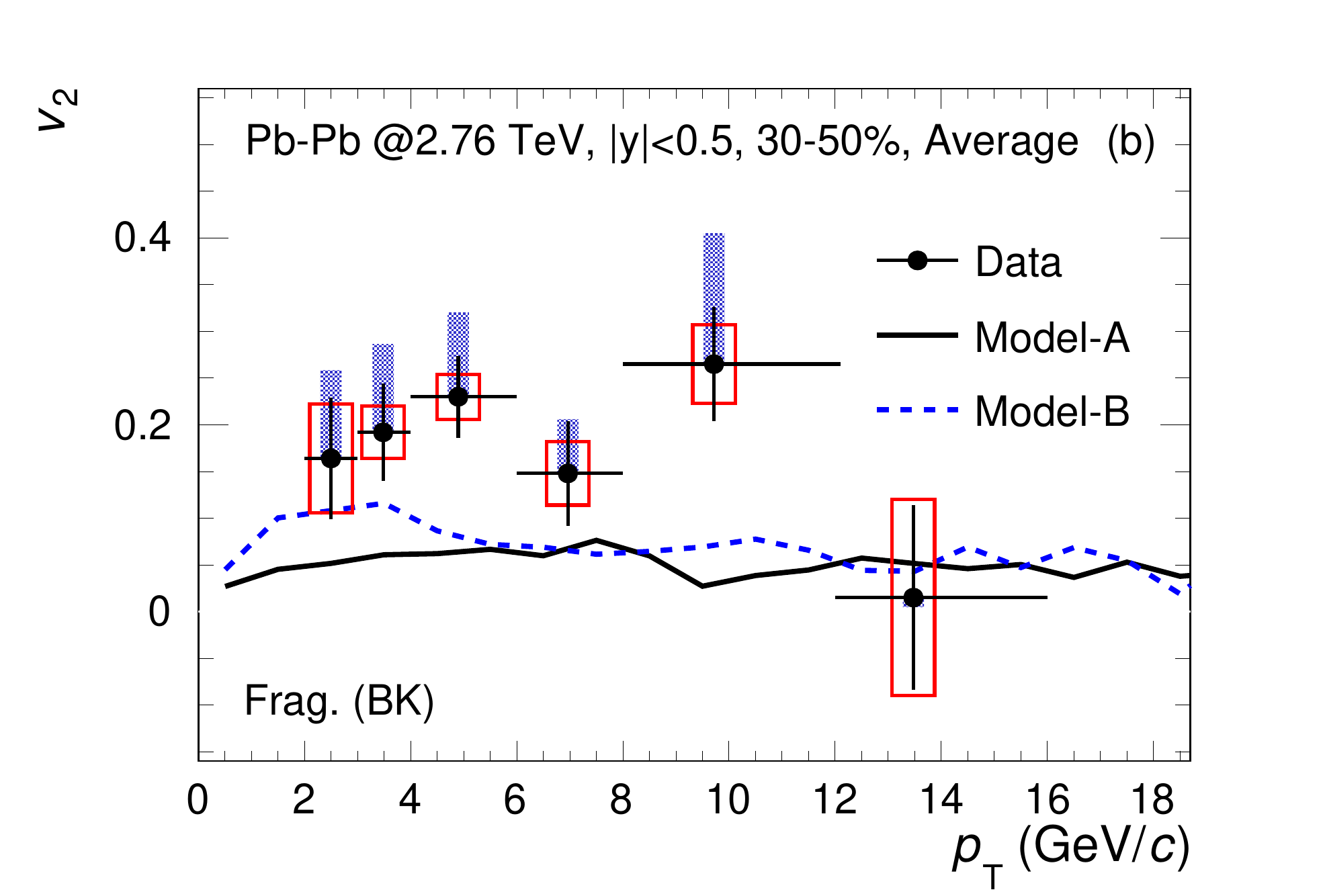}
\caption{(Color online) Comparison of D-meson (a) $\raa$ and (b) $\vtwo$ based on Model-A (solid black curves)
and Model-B (dashed blue curves) calculations as a function of $\pt$ with the measured values,
for Pb--Pb collisions at $\snn=2.76~{\rm TeV}$ in the centrality range $30-50\%$.
The measured $\raa$ and $\vtwo$ taken from Ref.~\cite{ALICEDesonPbPb2760RAA} and Ref.~\cite{ALICEDesonPbPb2760V2}, respectively.
The fragmentation function is the Braaten one.}
\label{fig:RAAV2_2760GeV_BK_Avg_c30_c50_ModelAB}
\end{center}
\end{figure}
After propagating the theoretical uncertainties on the pp reference and on the nuclear (anti-)shadowing to the nuclear modification factor,
we find that the uncertainty in the pp reference is significant at low $\pt$ ($\pt\lesssim3~{\rm GeV/{\it c}}$),
while the one one the nuclear PDFs is dominated at higher $\pt$.
Figure~\ref{fig:RAAV2_2760GeV_BK_Avg_c30_c50_ModelAB} presents the D-meson $\raa$ (panel-a; with full uncertainties)
and $\vtwo$ (panel-b; only central value) for semi-central ($30-50\%$)
Pb--Pb collisions at $\snn=2.76~{\rm TeV}$ for Model-A (solid black curves) and Model-B (dashed blue curves).
The D-meson $\raa$ with Model-B is enhanced (suppressed) at low (high) $\pt$ as compared to Model-A,
while $\vtwo$ is significantly higher at moderate $\pt$.
This is due to the different temperature-dependence of the spatial diffusion coefficient,
as is was also pointed out when discussing the results at parton level (panel-b of Fig.~\ref{fig:RAAPbPb2760Cent2ModelAB}).
The calculations with Model-A seem to give a better description of the measured $\raa$~\cite{ALICEDesonPbPb2760RAA}
as compared to those with Model-B, in particular in the range $\pt\gtrsim4~{\rm GeV}$ even though the
theoretical uncertainties are large.
On the other hands, D-meson $\vtwo$ calculated with Model-B approach
is closer to the available data~\cite{ALICEDesonPbPb2760V2} at $\pt\lesssim5~{\rm GeV}$.
The comparison of $\raa$ and $\vtwo$ gives the opposite indications about Model-A and Model-B,
confirming that it is challenging to describe well $\raa$ and $\vtwo$ simultaneously.
A similar behaviour was observed in Ref.~\cite{Das15}.

\begin{figure}[!htbp]
\begin{center}
\vspace{-1.0em}
\setlength{\abovecaptionskip}{-0.1mm}
\setlength{\belowcaptionskip}{-1.5em}
\includegraphics[width=.42\textwidth]{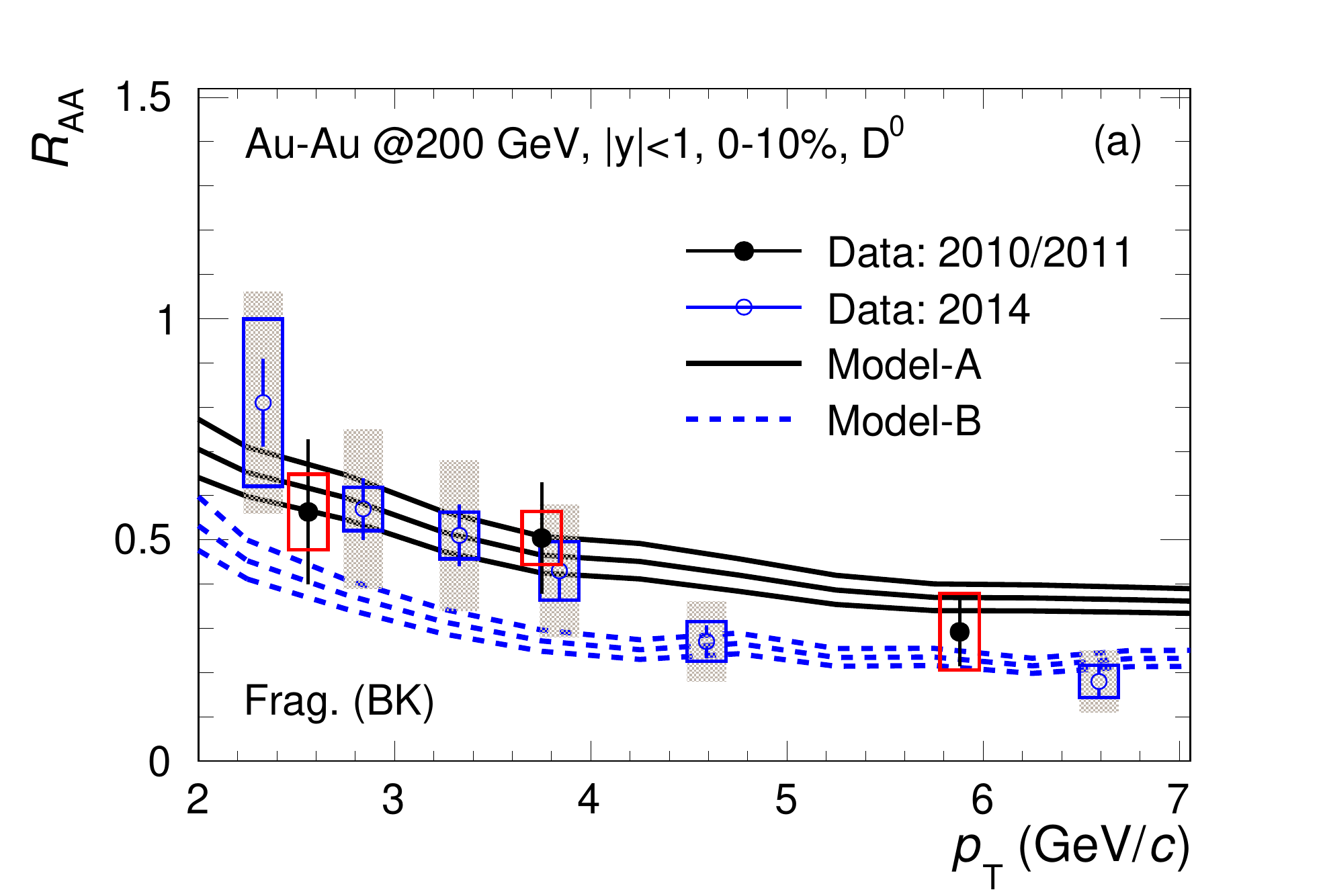}
\includegraphics[width=.42\textwidth]{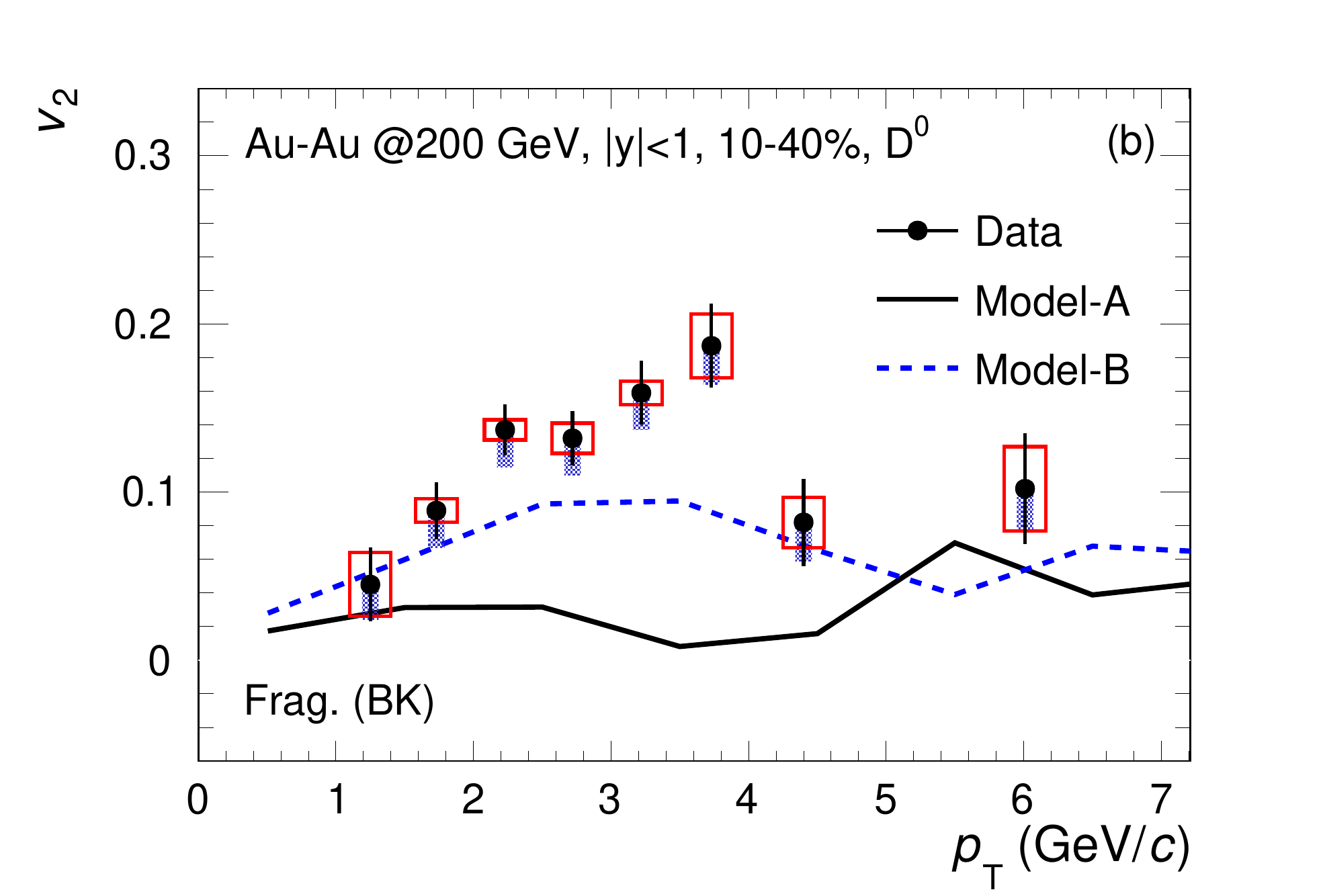}
\caption{Same as Fig.~\ref{fig:RAAV2_2760GeV_BK_Avg_c30_c50_ModelAB}
but for $D^{0}$ $\raa$ and $\vtwo$ at RHIC energy (see legend for details).
The measured $\raa$ and $\vtwo$ are taken from Ref.~\cite{STARD0RAA14,STARD0RAA17} and Ref.~\cite{STARD0V217}.}
\label{fig:RAA_200GeV_BK_D0_c10_c40_ModelAB}
\end{center}
\end{figure}

\begin{figure}[!htbp]
\begin{center}
\vspace{-1.0em}
\setlength{\abovecaptionskip}{-0.1mm}
\setlength{\belowcaptionskip}{-1.5em}
\includegraphics[width=.42\textwidth]{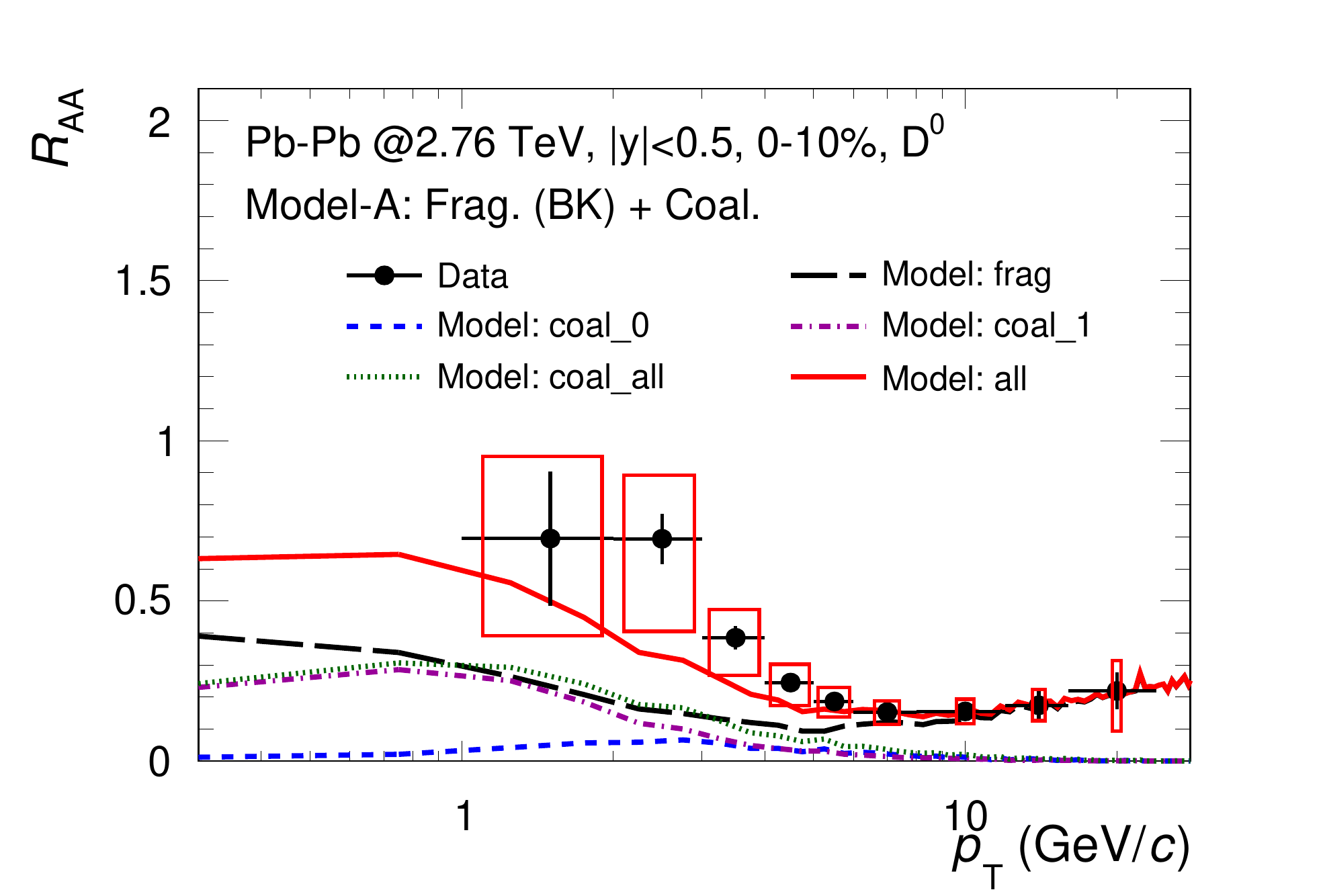}
\caption{(Color online) Comparison of the central predictions for the nuclear modification factor,
of $D^{0}$ mesons at mid-rapidity ($|y|<0.5$) in Pb--Pb collisions at $\snn=2.76~{\rm TeV}$.
The contributions of the different hadronization mechanism are displayed separately:
fragmentation (Braaten) as long dashed black curve and coalescence as dotted green curve.
For the coalescence, the contributions of the ground states (dashed blue curve)
and first exciting states (dot-dashed purple curve) are displayed separately.
Experimental data are taken from Ref.~\cite{ALICEDesonPbPb2760RAA}}.
\label{fig:CoalEff_2760_c0_c10_ModelA}
\end{center}
\end{figure}

The results at RHIC energy are displayed in Fig.~\ref{fig:RAA_200GeV_BK_D0_c10_c40_ModelAB}.
Within the experimental uncertainties, the measured $D^{0}$ $\raa$ but the data samples collected in 2010/2011~\cite{STARD0RAA14}
and 2014~\cite{STARD0RAA17}, for Au--Au collisions at $\snn=200~{\rm GeV}$ in the centrality class $0-10\%$,
can be fairly described by the model predictions with both the Model-A (solid black curves) and Model-B approaches (dashed blue curves).
The results with Model-A are closer to the measurements in the range $2<\pt<4~{\rm GeV}$.
The same conclusion can be drawn for the $D^{+}$ meson $\raa$~\cite{STARD0RAA17}.
The data-to-model comparison can be improved with the future high precision measurements.
As observed at LHC energy (panel-b of Fig.~\ref{fig:RAAV2_2760GeV_BK_Avg_c30_c50_ModelAB}),
the temperature dependent coupling strength of Model-B allows to improve the description of the measured D-meson $\vtwo$.

%%--------------------------------------------------- Results including heavy-light coalescence effect
%%---------------------------------------------------
%%---------------------------------------------------
Figure~\ref{fig:CoalEff_2760_c0_c10_ModelA} shows the $D^{0}$ meson $\raa$ obtained with Model-A approach
at mid-rapidity ($|y|<0.5$) in central ($0-10\%$) Pb--Pb collisions at $\snn=2.76~{\rm TeV}$.
Both the fragmentation and the coalescence mechanisms are considered in the hadronization model.
The fragmentation (long dashed black curve) is dominant in the range $\pt\gtrsim6~{\rm GeV}$,
while the coalescence contribution (dotted green curve) is significant in $1\lesssim\pt\lesssim5~{\rm GeV}$.
Concerning the different components of the coalescence mechanism,
the contributions due to the ground states (dashed blue curve) cannot be neglected in $2\lesssim\pt\lesssim4~{\rm GeV}$,
while the excited states contribution (dot-dashed purple curve) is dominant in the range $\pt\lesssim2~{\rm GeV}$.
This is due to the associated coalescence probabilities (Fig.~\ref{fig:CoalProbPbPb2760}),
as well as the related degeneracy factors ($g_{\rm M}$ in Eq.~\ref{eq:MesonCoal}).

%%--------------------------------------------------- Results predicted for Pb-Pb @ 5.02 TeV 
%%---------------------------------------------------
%%---------------------------------------------------
\begin{figure}[!htbp]
\begin{center}
\vspace{-1.0em}
\includegraphics[width=.42\textwidth]{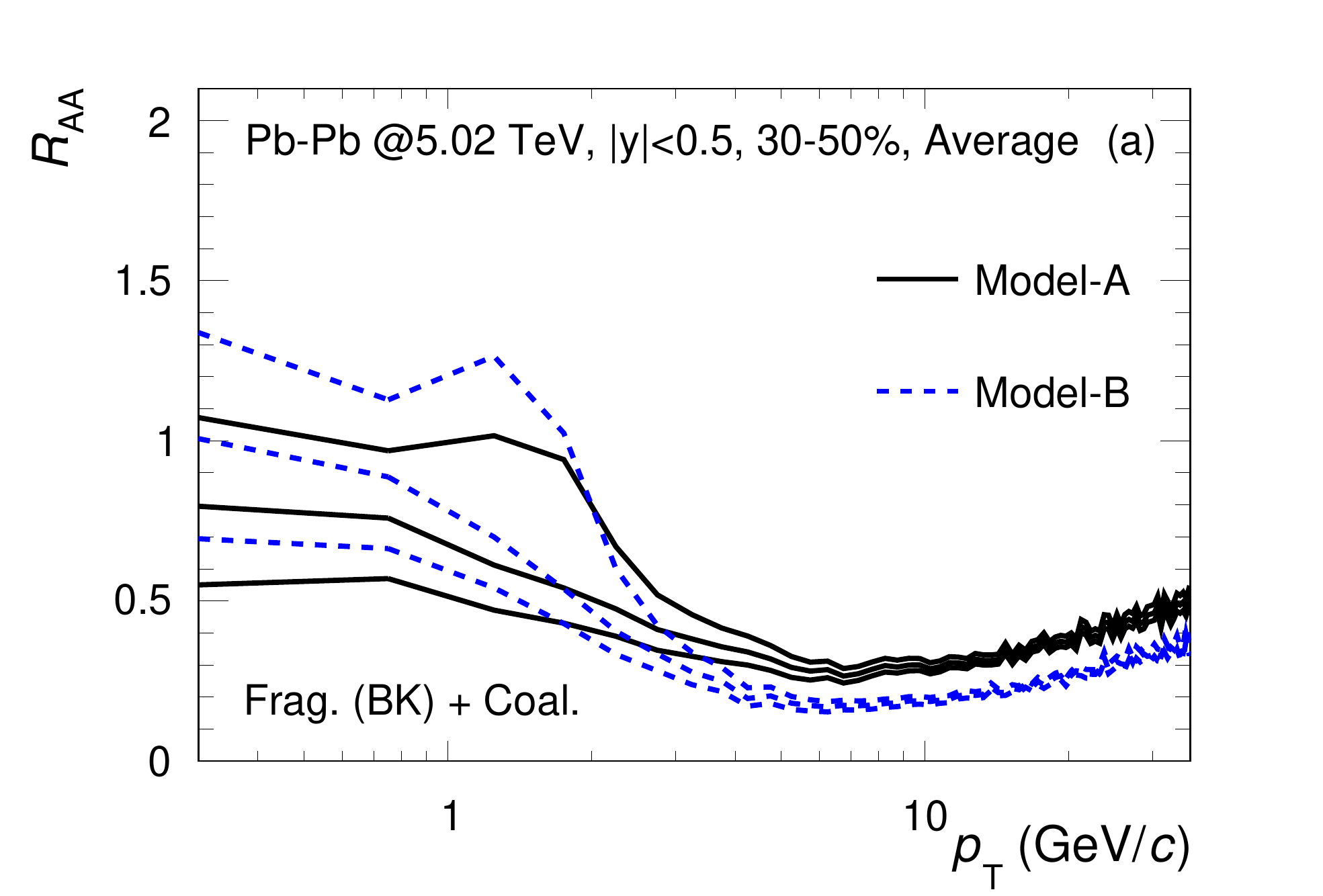}
\includegraphics[width=.42\textwidth]{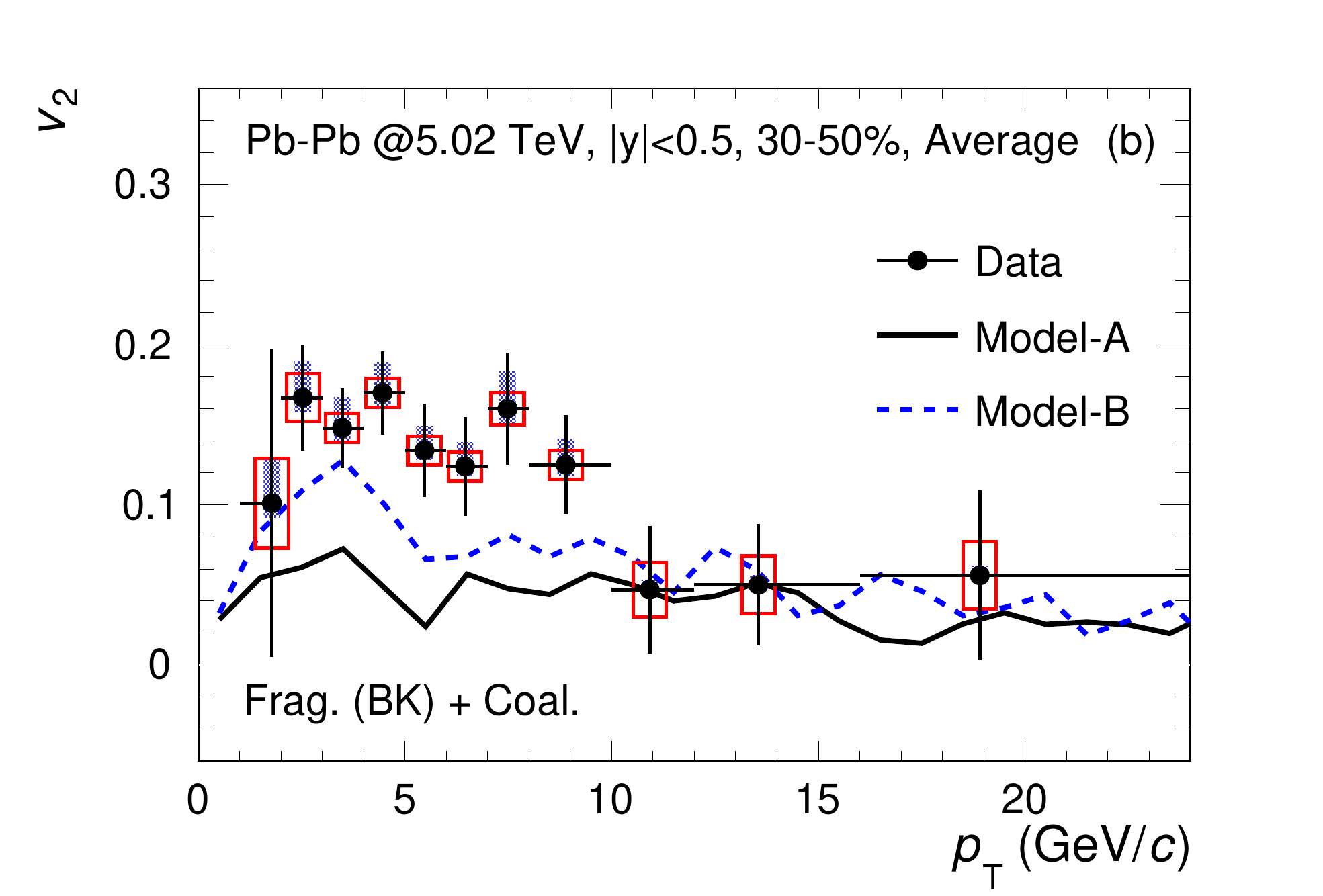}
\caption{Same as Fig.~\ref{fig:RAAV2_2760GeV_BK_Avg_c30_c50_ModelAB} but for Pb--Pb collisions at $\snn=5.02~{\rm TeV}$.
The heavy-light coalescence effect is included in the calculations.
$\vtwo$ data points are taken from Ref.~\cite{ALICEDesonPbPb5020V2}.}
\label{fig:RAAV2_5020GeV_BK_Coal_c30_c50_ModelAB}
\end{center}
\end{figure}
The results of the calculations for D meson $\raa$ and $\vtwo$
for semi-central ($30-50\%$) Pb--Pb collisions at $\snn=5.02~{\rm TeV}$,
are presented in the panel-a and panel-b of Fig.~\ref{fig:RAAV2_5020GeV_BK_Coal_c30_c50_ModelAB}, respectively.
The charm quark hadronization is implemented by using the ``dual'' model, including both the fragmentation and the coalescence.
The available data points for $\vtwo$ (boxes~\cite{ALICEDesonPbPb5020V2}) are shown for comparison.
As already observed at the other collisions energies
(Fig.~\ref{fig:RAAV2_2760GeV_BK_Avg_c30_c50_ModelAB} and~\ref{fig:RAA_200GeV_BK_D0_c10_c40_ModelAB}),
the $\vtwo$ results with Model-B (dashed blue curve) give a better description of the data.

%%==============================================
\section{Summary and Conclusions}\label{sec:conclusion}
In this analysis, we investigated the charm quark evolution through the QGP
together with the relevant open charmed meson observables in relativistic heavy-ion collisions.
We tried to study the temperature dependence of the coupling strength of the charm quark and the medium,
as well as its effects on the nuclear modification factor $\raa$ and the elliptic flow $\vtwo$ of open charmed mesons in Au--Au collisions at $\snn=200~{\rm GeV}$
and Pb--Pb collisions at $\snn=2.76$ and 5.02 ${\rm TeV}$.
We built a theoretical framework to achieve this goal, and all the adjustable parameters were tuned according to the
comprehensive sets of available data at RHIC and LHC energies.
The coupling strength for charm quark $2\pi TD_{s}$, is determined according to lattice QCD calculations
using two different assumptions: $2\pi TD_{s}=const.$ (\textbf{Model-A}) and $2\pi TD_{s}=1.3 + (T/T_{c})^2$ (\textbf{Model-B}).
It is found that:
\begin{enumerate}
\item[(1)] charm quark in-medium energy loss due to gluon radiation is dominant at high $\pt$,
while the quasi-elastic scattering is significant at low $\pt$;
the energy loss is stronger with Model-B approach because the relevant $2\pi TD_{s}$ is smaller,
and therefore the initial drag term is larger, resulting in stronger interactions.
Hence, charm quarks will lose more energy while traversing the QGP;
%%%
\item[(2)] the azimuthal angle ($|\Delta\phi|$) distribution of the initially back-to-back generated $c\bar{c}$ pairs
presents a broadening behaviour, which is mainly due to quark pairs with small initial $\pt$;
%similar trends observed at hadron level;
this broadening effect is more pronounced with Model-B due to the larger drag force,
which is more powerful to pull $c\bar{c}$ pairs from high momentum to low momentum;
%%%
\item[(3)] the charm quark $\raa$ is mostly determined by interactions occuring in the time window $0.6\lesssim\tau\lesssim7~{\rm fm/{\it c}}$,
due to the competition between initial drag and subsequent collective effect;
%%%
\item[(4)] hadronization due to fragmentation is dominant at high $\pt$,
while the coalescence is significant at moderate $\pt$;
the coalescence probability induced by the higher state component is relevant at moderate-low $\pt$,
resulting in an enhancement of D-meson yield in this region;
%%%
\item[(5)] the theoretical uncertainty on the D-meson $\raa$ is dominated by the pp reference uncertainty at $\pt\lesssim3~{\rm GeV/{\it c}}$,
and by the nuclear (anti-)shadowing parameterization at higher $\pt$;
%%%
\item[(6)] model-to-data comparisons for D-meson $\raa$ favor Model-A assumption for the dependence of $2\pi TD_{s}$ on temperature,
while the measured $\vtwo$ prefer Model-B.
This conclusion holds true for all the available measurements at RHIC and LHC energy,
suggesting the need for a temperature dependent $2\pi TD_{s}$, as well as a possible momentum dependent $2\pi TD_{s}$
to describe simultaneously $\raa$ and $\vtwo$.
\end{enumerate}

Some effects such as pre-equilibrium interactions and hadronic rescatterings are missing in this model.
More detailed checks and results on this developments will be discussed in forthcoming publications.

\begin{acknowledgments}
The authors thank Matteo Cacciari, Shanshan Cao, Kyong Chol Han, Yuriy Karpenko, Chun Shen and Taesoo Song
for their kind help and useful discussions.
A big thanks goes to Francesco Prino for carefully reading the manuscript and valuable suggestions.
S.~Li is supported by the CTGU No.1910103, QLPL2018P01, B2018023 and NSFC No.11447023.
C.~W.~Wang acknowledges the support from the NSFHB No.2012FFA085.
X.~B.~Yuan is supported by the NSCF No.11247021.
S.~Q.~Feng is supported by the NSCF No.11747115 and No.11475068.
\end{acknowledgments}

%%==============================================
%\bibliography{Shuang_2017}

\end{document}